\RequirePackage{fix-cm} % Fix LaTeX2e bugs.
\documentclass[a4paper, twoside, reqno, 12pt]{amsart}
\usepackage{fixltx2e}   % Fix LaTeX2e bugs.

\usepackage{etex}

\usepackage[latin1]{inputenc}
\usepackage[T1]{fontenc}

\usepackage{esint}
\usepackage{dsfont}
\usepackage{xspace}
\usepackage{amsgen}
\usepackage{amsthm}
\usepackage{amssymb}
\usepackage{amsmath}
\usepackage{wasysym}
\usepackage{upgreek}
\usepackage{amsfonts}
\usepackage{stmaryrd}
\usepackage{mathtools}

\usepackage[mathcal, mathscr]{euscript}

\usepackage{mathrsfs}
\DeclareMathAlphabet{\mathscrbf}{OMS}{mdugm}{b}{n}

\usepackage[ugly]{nicefrac}

\usepackage[lined, boxed, linesnumbered, english]{algorithm2e}

\usepackage{multicol}
\usepackage{multirow}

\usepackage{arydshln}
\usepackage{tabularx}
\usepackage{colortbl}
\usepackage{multicol}
\usepackage{multirow}

\usepackage{a4wide}

\usepackage[dvipsnames, table]{xcolor}
\definecolor{bckg}{RGB}{20.8, 20.8, 20.8}
\definecolor{oneblue}{rgb}{0.0, 0.0, 0.85}
\definecolor{Lightblue}{RGB}{214, 214, 214}
\definecolor{bluepigment}{rgb}{0.2, 0.2, 0.6}
\definecolor{charcoal}{rgb}{0.21, 0.27, 0.31}
\definecolor{denimblue}{rgb}{0.08, 0.38, 0.74}
\definecolor{Lightgray}{rgb}{0.89, 0.89, 0.89}
\definecolor{darkgrey}{rgb}{0.273, 0.281, 0.30}
\definecolor{darkelectricblue}{rgb}{0.33, 0.41, 0.47}

\usepackage[sort&compress, comma, square, numbers]{natbib}

\usepackage{psfrag}
\usepackage{graphicx}
\usepackage{subfigure}
\usepackage{morefloats}
\usepackage{indentfirst}

\usepackage{acronym}
\usepackage{microtype}
\usepackage[labelsep=period,%
            labelfont={bf,sf,color=bluepigment},%
            textfont=it,%
            justification=raggedright]{caption}

\usepackage[perpage, symbol]{footmisc}

\usepackage[usenames, pdf]{pstricks}
\usepackage{epsfig}
\usepackage{pst-grad} % For gradients
\usepackage{pst-plot} % For axes

\usepackage{url}
\usepackage[colorlinks,
           urlcolor=oneblue,
           linkcolor=denimblue,
           citecolor=NavyBlue,
           bookmarksopen=false,
           pdfpagemode=UseNone,
           pagebackref]{hyperref}

\usepackage[explicit]{titlesec}

\titleformat{\section}[block]
  {\color{NavyBlue}\Large\sffamily\bfseries}
  {}
  {0.0em}
  {\colorbox{bckg!5}{\strut\parbox{\dimexpr\linewidth-2\fboxsep\relax}{\thesection. #1}}}
  [\vspace*{0.33em}]

\titleformat{name=\section,numberless}[block]
  {\color{NavyBlue}\Large\sffamily\bfseries}
  {}
  {0.0em}
  {\colorbox{bckg!5}{\strut\parbox{\dimexpr\linewidth-2\fboxsep\relax}{#1}}}
  [\vspace*{0.33em}]

\titleformat{\subsection}
  {\color{NavyBlue}\large\sffamily\bfseries}
  {}
  {0.0em}
  {\colorbox{bckg!5}{\parbox{\dimexpr\linewidth-2\fboxsep\relax}{\thesubsection. #1}}}
  [\vspace*{0.33em}]

\titleformat{name=\subsection,numberless}
  {\color{NavyBlue}\Large\sffamily\bfseries}
  {}
  {0em}
  {\colorbox{bckg!5}{\parbox{\dimexpr\linewidth-2\fboxsep\relax}{#1}}}
  [\vspace*{0.33em}]

\titleformat{\subsubsection}
  {\color{bluepigment}\sffamily\normalsize\bfseries}
  {\thesubsubsection}
  {0.5em}
  {#1}
  [\vspace*{0.33em}]

\titleformat{\paragraph}[runin]
  {\color{bluepigment}\sffamily\small\bfseries}
  {}
  {0em}
  {#1}

\titlespacing{\section}{0.0em}{1.5em plus 2pt minus 2pt}%
{1.0em plus 2pt minus 2pt}[0em]
\titlespacing{\subsection}{0.5em}{1.5em plus 2pt minus 2pt}%
{1.0em}[0em]
\titlespacing{\subsubsection}{0.5em}{1.5em plus 2pt minus 2pt}%
{1.0em plus 2pt minus 2pt}[0em]

\usepackage{titletoc}

\contentsmargin{0.5em}
\newlength{\tocsep} 
\titlecontents{section}[\tocsep]
  {\addvspace{10pt}\bfseries\sffamily}
  {\contentslabel[\thecontentslabel]{\tocsep}}
  {}
  {\ \titlerule*[0.75pc]{.}\ \thecontentspage}
  []
\titlecontents{subsection}[\tocsep]
  {\addvspace{8pt}\sffamily}
  {\contentslabel[\thecontentslabel]{\tocsep}}
  {}
  {\ \titlerule*[0.5pc]{.}\ \thecontentspage}
  []
\titlecontents*{subsubsection}[\tocsep]
  {\addvspace{2pt}\footnotesize\sffamily}
  {}
  {}
  {\ \titlerule*[0.35pc]{.}\ \thecontentspage}
  [\\*]

\makeatletter
\def\@setauthors{%
  \begingroup
  \def\thanks{\protect\thanks@warning}%
  \trivlist
  \centering\footnotesize \@topsep30\p@\relax
  \advance\@topsep by -\baselineskip
  \item\relax
  \author@andify\authors
  \def\\{\protect\linebreak}%
  \textsc{\normalsize\textcolor{darkelectricblue}{\authors}}%
  \ifx\@empty\contribs
  \else
    ,\penalty-3 \space \@setcontribs
    \@closetoccontribs
  \fi
  \endtrivlist
  \endgroup
}
\def\@settitle{\begin{center}%
  \baselineskip14\p@\relax
    \bfseries
    \textsc{\Large\textcolor{charcoal}{\@title}}
  \end{center}%
}
\makeatother

\usepackage{enumitem}
\setlist[description]{%
  topsep=30pt,               % space before start / after end of list
  itemsep=5pt,               % space between items
  font={\bfseries\sffamily\color{NavyBlue}}, % if colour is needed
}

\usepackage{fancyhdr}
\usepackage{lastpage}

\newcommand*\Title{\textcolor{bluepigment}{A new model for simulating heat, air and moisture transport}}
\newcommand*\Authors{\textcolor{bluepigment}{J.~Berger, D.~Dutykh \etal}}
\newcommand*{\plogo}{\textcolor{gray}{{\texttt{arXiv.org} / \textsc{hal}}}} % Generic publisher logo

\numberwithin{equation}{section}

\newcommand{\ie}{\emph{i.e.}\xspace}
\newcommand{\eg}{\emph{e.g.}\xspace}

\newcommand{\etal}{\emph{et al.}\xspace}

\newcommand*\unit[1]{\bigl[\, \mathsf{#1} \,\bigr]}

\newcommand{\DF}{\textsc{Du$\,$Fort}--\textsc{Frankel}}
\newcommand{\Eu}{\textsc{Euler}}
\newcommand{\RK}{\textsc{Runge}--\textsc{Kutta}}
\newcommand{\SG}{\textsc{Scharfetter}--\textsc{Gummel}}

\newcommand{\Bi}{\mathrm{Bi}}
\newcommand{\Fo}{\mathrm{Fo}}
\newcommand{\K}{\mathrm{Ko}}
\newcommand{\Pe}{\mathrm{Pe}}
\newcommand{\Ps}{P_{\,\mathrm{s}}}

\newcommand{\vit}{\mathsf{v}}
\newcommand{\Vp}{V_{\,\mathrm{p}}}
\newcommand{\w}{\mathsf{w}}
\newcommand{\sigmaT}{\sigma_{_{\,T}}}
\newcommand{\sigmaPv}{\sigma_{_{\,P_{\,1}}}}

\newcommand{\scal}{\boldsymbol{\cdot}}
\newcommand*\od[2]{\frac{\mathrm{d} #1}{\mathrm{d} #2}}
\newcommand*\pd[2]{\frac{\partial #1}{\partial #2}}
\renewcommand{\div}{\grad\scal}
\newcommand{\grad}{\boldsymbol{\nabla}}
\newcommand{\eqdef}{\mathop{\stackrel{\,\mathrm{def}}{:=}\,}}

\newcommand{\half}{\frac{1}{2}} %{{\textstyle{1\over2}}}

\newcommand*\e[1]{\cdot 10^{\,#1}}

\newcommand*\egal{\ = \ }
\newcommand*\plus{\ + \ }
\newcommand*\moins{\ - \ }

\begin{document}

\title[\Title]{A new model for simulating heat, air and moisture transport in porous building materials}

\author[J.~Berger]{Julien Berger$^*$}
\address{\textbf{J.~Berger:} Univ. Grenoble Alpes, Univ. Savoie Mont Blanc, CNRS, LOCIE, 73000 Chamb\'ery, France and LOCIE, UMR 5271 CNRS, Universit\'e Savoie Mont Blanc, Campus Scientifique, F-73376 Le Bourget-du-Lac Cedex, France}
\email{Berger.Julien@univ-smb.fr}
\urladdr{https://www.researchgate.net/profile/Julien\_Berger3/}
\thanks{$^*$ Corresponding author}

\author[D.~Dutykh]{Denys Dutykh}
\address{\textbf{D.~Dutykh:} Univ. Grenoble Alpes, Univ. Savoie Mont Blanc, CNRS, LAMA, 73000 Chamb\'ery, France and LAMA, UMR 5127 CNRS, Universit\'e Savoie Mont Blanc, Campus Scientifique, F-73376 Le Bourget-du-Lac Cedex, France}
\email{Denys.Dutykh@univ-smb.fr}
\urladdr{http://www.denys-dutykh.com/}

\author[N.~Mendes]{Nathan Mendes}
\address{\textbf{N.~Mendes:} Thermal Systems Laboratory, Mechanical Engineering Graduate Program, Pontifical Catholic University of Paran\'a, Rua Imaculada Concei\c{c}\~{a}o, 1155, CEP: 80215-901, Curitiba -- Paran\'a, Brazil}
\email{Nathan.Mendes@pucpr.edu.br}
\urladdr{https://www.researchgate.net/profile/Nathan\_Mendes/}

\author[B.~Rysbaiuly]{Bolatbek Rysbaiuly}
\address{\textbf{B.~Rysbaiuly:} International Information Technology University, Almaty, Manas Str. 34/1, Kazakhstan}
\email{b.rysbaiuly@mail.ru}

\keywords{heat and mass transfer; porous material; Benchmarking with experimental data; \SG ~numerical scheme; \DF ~numerical scheme}

%%% ------------------------------------------------------------------------ %%%

\begin{titlepage}
\thispagestyle{empty} % Remove page numbering on this page
\noindent
{\Large Julien \textsc{Berger}}\\
{\it\textcolor{gray}{LOCIE--CNRS, Universit\'e Savoie Mont Blanc, France}}
\\[0.02\textheight]
{\Large Denys \textsc{Dutykh}}\\
{\it\textcolor{gray}{LAMA--CNRS, Universit\'e Savoie Mont Blanc, France}}
\\[0.02\textheight]
{\Large Nathan \textsc{Mendes}}\\
{\it\textcolor{gray}{Pontifical Catholic University of Paran\'a, Brazil}}
\\[0.02\textheight]
{\Large Bolatbek \textsc{Rysbaiuly}}\\
{\it\textcolor{gray}{IIT University, Almaty, Kazakhstan}}
\\[0.10\textheight]

\colorbox{Lightblue}{
  \parbox[t]{1.0\textwidth}{
    \centering\huge\sc
    \vspace*{0.7cm}
    
    \textcolor{bluepigment}{A new model for simulating heat, air and moisture transport in porous building materials}

    \vspace*{0.7cm}
  }
}

\vfill % Whitespace between the title block and the publisher

\raggedleft     % Right-align all text
{\large \plogo} % Publisher and logo
\end{titlepage}

%%% ------------------------------------------------------------------------ %%%

\newpage
\thispagestyle{empty} % Remove page numbering on this page
\par\vspace*{\fill}   % Whitespace until the bottom
\begin{flushright} % Right-align all text
{\textcolor{denimblue}{\textsc{Last modified:}} \today}
\end{flushright}

%%% ------------------------------------------------------------------------ %%%

\newpage
\maketitle
\thispagestyle{empty}

%%% ------------------------------------------------------------------------ %%%

\begin{abstract}

This work presents a detailed mathematical model combined with an innovative efficient numerical model to predict heat, air and moisture transfer through porous building materials. The model considers the transient effects of air transport and its impact on the heat and moisture transfer. The achievement of the mathematical model is detailed in the continuity of \textsc{Luikov}'s work. A system composed of two advection--diffusion differential equations plus one exclusively diffusion equation is derived. The main issue to take into account the transient air transfer arises in the very small characteristic time of the transfer, implying very fine discretisation. To circumvent these difficulties, the numerical model is based on the \DF ~explicit and unconditionally stable scheme for the exclusively diffusion equation. It is combined with a two--step \RK ~scheme in time with the \SG ~numerical scheme in space for the coupled advection--diffusion equations. At the end, the numerical model enables to relax the stability condition, and, therefore, to save important computational efforts. A validation case is considered to evaluate the efficiency of the model for a nonlinear problem. Results highlight a very accurate solution computed about $16$ times faster than standard approaches. After this numerical validation, the reliability of the mathematical model is evaluated by comparing the numerical predictions to experimental observations. The latter is measured within a multi-layered wall submitted to a sudden increase of vapor pressure on the inner side and driven climate boundary conditions on the outer side. A very satisfactory agreement is noted between the numerical predictions and experimental observations indicating an overall good reliability of the proposed model.

\bigskip
\noindent \textbf{\keywordsname:} heat and mass transfer; porous material; Benchmarking with experimental data; \SG ~numerical scheme; \DF ~numerical scheme \\

% heat and mass transfer; porous material; benchmarking with experimental data;  Scharfetter-Gummel numerical scheme; Dufort-Frankel numerical scheme

\smallskip
\noindent \textbf{MSC:} \subjclass[2010]{ 35R30 (primary), 35K05, 80A20, 65M32 (secondary)}
\smallskip \\
\noindent \textbf{PACS:} \subjclass[2010]{ 44.05.+e (primary), 44.10.+i, 02.60.Cb, 02.70.Bf (secondary)}

\end{abstract}

%%% ------------------------------------------------------------------------ %%%

\newpage
\tableofcontents
\thispagestyle{empty}

%%% ------------------------------------------------------------------------ %%%

\newpage
\section{Introduction}

Moisture is a key factor on durability and performance of buildings. An excessive level compromises the construction quality, impacts the indoor air quality and the thermal comfort, as well as the building energy efficiency \cite{Berger2015a}. As a consequence, a number of models for predicting the impact of moisture on building energy efficiency are proposed in the literature. A primary overview may be consulted in \cite{Mendes2017}. Among the physical phenomena, the air transfer through the porous building media has a crucial impact on the amount of moisture. Diverse studies enhance these effects using both experimental and numerical results \cite{Wang2017, Kalamees2010, Desta2011}.

Several numerical models are proposed in the literature for the prediction of physical phenomena of coupled heat, air and moisture transport through porous building materials. Their physical representations are based on the mass conservation laws for the dry air, vapor and liquid water, as well as the energy conservation law as detailed in the early work of \textsc{Luikov} \cite{Luikov1966}. As the continuity of his work, the numerical models proposed in the literature can be divided into two main groups. The first group considers the three evolution differential equations to compute the temperature, the mass content and the air pressure in the porous media. In \cite{DosSantos2009a}, a model is proposed for the simulation of transfer through hollow porous blocks. It is based on an implicit finite-difference numerical scheme. More recently, in \cite{Abahri2016a, Mnasri2017}, the commercial \texttt{COMSOL\texttrademark} software is used to propose a numerical model for such physical problems. As mentioned by the authors, the scheme is based on an explicit in time finite element approach. The main drawback of these numerical models is their computational cost. The implicit approach requires costly sub-iterations at each time step to handle severe nonlinearities of the problem. The explicit scheme requires very fine time steps to satisfy the so-called \textsc{Courant}--\textsc{Friedrichs}--\textsc{Lewy} (CFL) stability conditions. Indeed, the characteristic time of air transfer is very small compared to the ones for heat and mass transfer.

To handle this computational issue, the second group of models does not consider the transient phenomena for the air transport through the porous matrix. In other words, the evolution differential equation for air transfer is transformed into a simple steady boundary value problem which is solved at prescribed time instances. It enables somehow to relax the stability conditions and, therefore, to save computational efforts comparing to the models from the first group. Some examples can be found in \cite{Tariku2010} or \cite{Langmans2012} for one-dimensional transfer. The former references uses an explicit scheme  provided by the commercial \texttt{COMSOL\texttrademark} software. The latter is based on an implicit scheme based on the generic ODE solver from the \texttt{SUN-DIALS} solver package \cite{Hindmarsh2005}. More recently, a numerical model is proposed in \cite{Belleudy2015, Belleudy2016} for the simulation of two--dimensional transfer in building structures. It is also based on an explicit scheme from \texttt{COMSOL\texttrademark} software. The assumption of neglecting the transient term is justified by the low velocities occurring through the porous matrix. As mentioned by authors, the numerical predictions are reliable only in the context of simulations with standard hourly climate driven boundary conditions. It should be noted that this condition is not often satisfied, particularly for the air pressure surrounding building walls as detailed in \cite{Khanduri1998}. Moreover, even if the restriction on the time discretisation is relaxed, the numerical models are based on standard approaches and still have a high computational cost.

Therefore, the goal is to propose an efficient numerical model considering the three transient equations to improve the reliability of its predictions. It requires to be accurate with a reduced computational cost. To address this issue, this paper proposes to use the \SG ~scheme combined with a two--step \RK ~approach. The \SG ~numerical scheme was proposed in 1969 in \cite{Scharfetter1969} with very recent theoretical results in \cite{Gosse2016, Gosse2017}. In the context of building porous media, it is successfully applied in \cite{Berger2017a} to water transport and then in \cite{Berger2018a} to combined heat and moisture transfer. The contributions of the present paper is two fold. First, the model proposed in \cite{Berger2017a} is extended by including the air transport equation. Then, the extension of the \SG ~approach to a system of three coupled advection--diffusion equations is proposed. The combination with a two-step \RK ~scheme is investigated in order to relax the stability restrictions on the choice of the time discretisation.

The paper is organized as follows. Section~\ref{sec:definition_physical_model} presents the demonstration of the mathematical model to describe the physical phenomena and its dimensionless formulation. Then, Section~\ref{sec:numerical_model} presents the numerical method to solve the system of three differential advection--diffusion equations. In Section~\ref{sec:validation_numerical_model}, one case studie is considered to validate the numerical model. The purpose is to quantify the accuracy and efficiency in terms of computational time and relaxation of the stability condition. For each case, a reference solution is proposed, computed by a numerical pseudo--spectral approach. In the last Section, the reliability of the numerical predictions is evaluated by comparing them to experimental observations. A wall composed of two layers of wood fiberboard is submitted to a controlled environment on the inner side and to climate driven boundary condition on the outer side. Three points of measurements of temperature and vapor pressure within the wall are used for comparison purposes.

%%% ------------------------------------------------------------------------ %%%

\section{Formulation of the physical phenomena}
\label{sec:definition_physical_model}

First, the mathematical model including the governing equations to describe the physical phenomena is presented.

%%% ------------------------------------------------------------------------ %%%

\subsection{Porous medium basics}
\label{sec:porous_material}

The term moisture is used to designate the water vapor, denoted by index~$1$, and the liquid water, denoted by index~$2\,$. The gas mixture (moist air) is composed of water vapor and dry air indexed by $3\,$. The porous matrix of the material is symbolized by index~$0\,$. The \emph{volumetric concentration} $\w_{\,i} \ \unit{kg/m^{\,3}}$ of specie(s) $i\,$, which varies in space and time, is defined by $\w_{\,i} \ \eqdef \ \lim\limits_{V \, \rightarrow \, 0} \dfrac{m_{\,i} \,(\,V\,\bigr)}{V} \,$. We denot by $m_{\,i}\,(\,V\,\bigr) \ \unit{kg}$ the mass of the specie(s) in the representative volume $V \ \unit{m^{\,3}}\,$. The representative volume of material $V$ is composed of the volume of the solid porous matrix $V_{\,0}$ , the volume of water in its liquid form $V_{\,2}$ and the volume of the voids formed by the pores $\Vp\,$. Another important property of a porous material is the so-called \emph{sorption isotherm curve}, which provides a relation between the moisture content $w_{\,12}\,$, in liquid and vapor phases, with the relative humidity:
\begin{align}\label{eq:sorption_curve}
  \w_{\,12} \, \eqdef \, \frac{m_{\,12}}{V} \egal \mathsf{f} \,\bigl(\, \phi \,\bigr) \,,
\end{align}
where $f$ can be a fitted function from experimental data.  By introducing this property, an instantaneous thermodynamic equilibrium is assumed in the material. As reported in \cite{Soudani2016}, most of the models assume the water content provided by the sorption curve corresponds approximately to the measure of the liquid water content, since the mass of the vapor phase is negligible compared to the one of liquid, \ie, $m_{\,1} \ \ll \ m_{\,2}\,$. In the present work, this simplification is not considered. The measure of the sorption curve includes both vapor and liquid water masses. The \emph{saturation rate} $\sigma \ \unit{-}$ corresponds to the volume of liquid water out of the volume of pores. The estimation of the saturation is a difficult task since the weighting of a material only provides the amount of both liquid and vapor within the porous structure. When the assumption $m_{\,12} \ \approx \ m_{\,2}$ is considered, the saturation rate is evaluated directly by the sorption curve. Presently, the saturation rate is given by:
\begin{align*}
  \sigma \egal \frac{1}{\rho_{\,2} \, R_{\,1} \, T \moins P_{\,1}} \Biggl(\, \frac{R_{\,1} \,T}{\Pi} \, \w_{\,12} \moins P_{\,1} \,\Biggr) \,.
\end{align*}
Assuming the perfect gas law, the volumetric vapor and dry air content are given by:
\begin{align*}
  \w_{\,1} & \egal \frac{P_{\,1}}{R_{\,1} \, T} \ \Pi \ \bigl(\,1 \moins \sigma \,\bigr) \,,
  &&
  \w_{\,3} \egal \frac{P_{\,3}}{R_{\,3} \, T} \ \Pi \ \bigl(\,1 \moins \sigma \,\bigr) \,.
\end{align*}
Last, the liquid volumetric mass content is given by:
\begin{align}\label{eq:liquid_volumetric_mass}
  \w_{\,2} \egal \w_{\,12} \moins \w_{\,1} \egal \w_{\,12} \moins \frac{P_{\,1}}{R_{\,1} \, T} \ \Pi \ \bigl(\,1 \moins \sigma \,\bigr)  \,.
\end{align}
Interested readers may consult \cite{Luikov1966} for further information on these definitions. Using \textsc{Whitaker} volume averaging method to link the microscopic and macroscopic approaches \cite{Whitaker1986, Whitaker1986a}, the conservation law for each specie can be written according to the following differential equation:
\begin{align}\label{eq:conservation_law}
  & \pd{\w_{\,i}}{t} \egal - \, \div \, \boldsymbol{j}_{\,c\,,\,i} \plus I_{\,i} \,, && i \ \in \ \Bigl\{\,1 \,,\,2  \,,\, 3 \,\Bigr\} \,,
\end{align}
where $I_{\,i} \ \unit{kg/(m^{\,3}.\,s)}$ is the volumetric term source or sink of specie $i\,$. Since chemical reactions have not been taking into account, $I_{\,3} \egal 0\,$. Moreover, it is assumed that the temperature is always higher than the freezing point so no ice water may appear in the material. Thus, the following relation is stated:
\begin{align*}
  I_{\,1} \plus I_{\,2} \egal 0 \,. 
\end{align*}
The flow of mass is denoted by $\boldsymbol{j}_{\,c\,,\,i} \ \unit{kg/(m^{\,2}.\,s)}\,$. Due to the airflow occurring through the porous medium, the mechanism of mass transfer is driven by both diffusion $\boldsymbol{j}_{\,d}$ and advection $\boldsymbol{j}_{\,a}$ flows: 
\begin{align*}
  \boldsymbol{j}_{\,c} \egal \boldsymbol{j}_{\,d} \plus \boldsymbol{j}_{\,a} \,.
\end{align*}
It is important to mention that the gravity effects are neglected since the transport phenomena will be investigated on a horizontal plane perpendicular to the gravity forces.

%%% ------------------------------------------------------------------------ %%%

\subsection{Moisture mass balance}

By summing Eq.~\eqref{eq:conservation_law} ($i \ \leftarrow \ 1$) and Eq.~\eqref{eq:conservation_law} ($i \ \leftarrow \ 2$), the moisture transfer equation is obtained: 
\begin{align*}
  \pd{}{t} \Bigl(\, \w_{\,1} \plus \w_{\,2} \,\Bigr) \egal - \, \div \, \Bigl(\, \boldsymbol{j}_{\,c\,,\,1} \plus \boldsymbol{j}_{\,c\,,\,2} \,\Bigr) \,.
\end{align*}

The term $\w_{\,1} \plus \w_{\,2}$ is equal to the moisture content $\w_{\,12}\,$. Using the expression of the sorption isotherm curve Eq.~\eqref{eq:sorption_curve}, the partial differential of the moisture content becomes:
\begin{align*}
  \pd{\w_{\,12}}{t} \egal \pd{\w_{\,12}}{\phi} \, \frac{1}{\Ps} \, \pd{P_{\,1}}{t} \,,
\end{align*}
where $\Ps$ is the saturation pressure, depending on temperature, and $\displaystyle \pd{}{\phi}\, \Bigl(\, \w_{\,1} \plus \w_{\,2} \,\Bigr)$ is the derivative of the sorption curve of the material. In this way, the moisture transfer equation becomes:
\begin{align*}
  \pd{\w_{\,12}}{\phi} \, \frac{1}{\Ps}  \, \pd{P_{\,1}}{t} \egal - \, \div \, \Bigl(\, \boldsymbol{j}_{\,c\,,\,1} \plus \boldsymbol{j}_{\,c\,,\,2} \,\Bigr) \,.
\end{align*}

%%% ----------------------------------------------------------------------- %%%

\subsection{Moist air mass balance}
\label{sec:air_transfer}

Since the air content source term vanishes and adding Eq.~\eqref{eq:conservation_law} ($i \ \leftarrow \ 1$) to Eq.~\eqref{eq:conservation_law} ($i \ \leftarrow \ 3$), one obtains:
\begin{align}\label{eq:air_conservation_1}
  \pd{}{t} \Bigl(\, \w_{\,1} \plus \w_{\,3} \,\Bigr) \egal - \, \div \, \Bigl(\, \boldsymbol{j}_{\,c\,,\,1} \plus \boldsymbol{j}_{\,c\,,\,3} \,\Bigr) \plus I_{\,1} \,,
\end{align} 
where $\w_{\,1} \plus \w_{\,3}$ accounts for the amount of mass of dry air and vapor in the whole mixture and can be expressed through the total pressure $P$:
\begin{align}\label{eq:concentration_melange}
  \w_{\,1} \plus \w_{\,3} \egal \frac{P}{R_{\,13} \, T}\, \Pi \, (\,1 \moins \sigma \,)  \,,
\end{align}
where $P$ is the total pressure of the mixture of vapor and dry air. The term $R_{\,13}$ is the gas constant for the mixture of dry air and vapor. It is computed using the following expression:
\begin{align}\label{eq:definition_R13}
  R_{\,13} \egal \frac{R_{\,3}}{1 \moins \frac{P_{\,1}}{P_{\,3}} \, \Bigl(\,1 \moins \nicefrac{R_{\,3}}{R_{\,1}} \Bigr)} \,.
\end{align}
Assuming the porosity to be invariant, the following equation is obtained by computing the differential of Eq.~\eqref{eq:concentration_melange}:
\begin{align}\label{Eq:differentiel_dw13}
  \partial \, \bigl(\, \w_{\,1} \plus \w_{\,3} \,\bigr) \egal \frac{\Pi \, (\,1 \moins \sigma \,)}{R_{\,13} \, T} \, \partial P \plus \frac{\Pi \, P}{R_{\,13} \, T} \, \partial  (\,1 \moins \sigma \,) \moins \frac{\Pi \, (\,1 \moins \sigma \,) \, P}{R_{\,13} \, T^{\,2}} \, \partial T \,.
\end{align}
The variation of the gas mixture mass (composed of vapor and dry air) depends on the variation of the total pressure $P\,$, of the volume gas $1 \moins \sigma$ and of the temperature $T\,$. Thus, Eq.\eqref{eq:air_conservation_1} becomes:
\begin{align*}
  \frac{\Pi \, (\,1 \moins \sigma \,)}{R_{\,13} \, T} \, \pd{P}{t} \egal - \, \div \, \Bigl(\, \boldsymbol{j}_{\,c\,,\,1} \plus \boldsymbol{j}_{\,c\,,\,3} \,\Bigr) \plus I_{\,1} \plus \frac{\Pi \, P}{R_{\,13} \, T} \, \pd{\sigma}{t} \plus \frac{\Pi \, (\,1 \moins \sigma \,) \, P}{R_{\,13} \, T^{\,2}} \, \pd{T}{t}
 \,.
\end{align*}
The variation of the pressure of the mixture depends on the divergence of the fluxes, on the vapor source due to vaporization/condensation and on the variation of the gas volume and the temperature.

%%% ----------------------------------------------------------------------- %%%

\subsection{Energy balance}

The energy balance equation is derived from the first law of thermodynamics, which states the internal energy variation in time is due to the balance of energy across the volume control, \ie, that the volumetric concentration of the enthalpy $h \ \unit{J/kg}$ equals the divergence of the conduction and enthalpy flux:
\begin{align}\label{eq:conservation_enthalpy}
  \pd{}{t} \biggl(\, h_{\,0} \, \rho_{\,0} \plus \sum_{i=1}^{3} \, h_{\,i} \, \w_{\,i} \,\biggr) \egal - \, \div \, \Bigl(\, \boldsymbol{j}_{\,q} \plus \sum_{i=1}^{3} \, h_{\,i} \, \boldsymbol{j}_{\,c\,,\,i} \,\Bigr) \,,
\end{align}
where $\rho_{\,0} \ \unit{kg/m^{\,3}}$ is the dry material density and $c_{\,i} \ \unit{J/(kg.K)}$ the specific heat of each species. The expression of the volumetric water content $\w_{\,i}$ is detailed in Section~\ref{sec:porous_material}. By assuming a constant volume, Eq~\eqref{eq:conservation_enthalpy} becomes:
\begin{multline}\label{eq:conservation_enthalpy2}
  \biggl(\, c_{\,0} \, \rho_{\,0} \plus \sum_{i=1}^{3} \, c_{\,i} \, \w_{\,i} \,\biggr) \pd{T}{t} \plus \sum_{i=1}^{3} \, h_{\,i} \, \pd{\w_{\,i}}{t} \egal - \, \div \, \boldsymbol{j}_{\,q} \\ 
  \moins \sum_{i=1}^{3} \, h_{\,i} \, \div \, \boldsymbol{j}_{\,c\,,\,i} \moins \sum_{i=1}^{3} \,  \grad \, \bigl(\, h_{\,i} \,\bigr) \scal \boldsymbol{j}_{\,c\,,\,i} \,.
\end{multline}
Since $I_{\,3} \egal 0$, by summing Eq.~\eqref{eq:conservation_law} ($i \ \leftarrow \ 1$), multiplied by $h_{\,1}$, Eq.~\eqref{eq:conservation_law} ($i \ \leftarrow \ 2$), multiplied by $h_{\,2}$  and Eq.~\eqref{eq:conservation_law} ($i \ \leftarrow \ 3$), multiplied by $h_{\,3}$, we obtain that:
\begin{align*}
  \sum_{i=1}^{3} \, h_{\,i} \, \pd{\w_{\,i}}{t} \egal \- \sum_{i=1}^{3} \, h_{\,i} \, \div \, \boldsymbol{j}_{\,c\,,\,i}  \plus h_{\,1} \, I_{\,1} \plus h_{\,2} \, I_{\,2} \,.
\end{align*}
Thus, Eq.~\eqref{eq:conservation_enthalpy2} becomes:
\begin{align*}
  \biggl(\, c_{\,0} \, \rho_{\,0} \plus \sum_{i=1}^{3} \, c_{\,i} \, \w_{\,i} \,\biggr) \pd{T}{t} \egal - \, \div \, \boldsymbol{j}_{\,q} \moins h_{\,1} \, I_{\,1} \moins h_{\,2} \, I_{\,2} \moins \sum_{i=1}^{3} \,  \grad \, \bigl(\, c_{\,i} \, T \,\bigr) \scal \boldsymbol{j}_{\,c\,,\,i} \,.
\end{align*}
We denote by $r_{\,12} \eqdef h_{\,1} \moins h_{\,2} \ \unit{J/kg}$ as the latent heat of vaporization. According to \textsc{Kelvin}'s law, it is defined as \cite{Soudani2016}:
\begin{align}\label{eq:definition_r12}
  r_{\,12} \egal r_{\,12}^{\,\circ} \plus \bigl(\, c_{\,1} \moins c_{\,2} \, \bigr) \cdot \bigl(\,T \moins T_{\,\mathrm{ref}} \,\bigr) \moins R_{\,1} \, T \, \ln \, \phi \,.
\end{align}
Finally, the energy conservation equation yields:
\begin{align*}
  \biggl(\, c_{\,0} \, \rho_{\,0} \plus \sum_{i=1}^{3} \, c_{\,i} \, \w_{\,i} \,\biggr) \pd{T}{t} \egal - \, \div \, \boldsymbol{j}_{\,q} \moins r_{\,12} \ I_{\,1} \moins \sum_{i=1}^{3} \,  \grad \, \bigl(\, c_{\,i} \, T \,\bigr) \scal \boldsymbol{j}_{\,c\,,\,i} \,.
\end{align*}
The temperature variations in the material depend on the variation of the thermal flux, on the energy due to vaporization and the enthalpy transport of each species by advection.

%%% ----------------------------------------------------------------------- %%%

\subsection{Volumetric vapor source}
\label{sec:source_I2}

In order to write the system of governing differential equations, it is important to express the volumetric vapor source $I_{\,1}\,$. For that, several possibilities are suggested in literature. In its original work \cite{Luikov1966}, \textsc{Luikov} neglected the time variation of the vapor mass and expressed the source term as
\begin{align}\label{eq:source_term1}
  I_{\,1} \egal \div \, j_{\,c\,,\,1} \,.
\end{align}
Similar approach has been adopted in \cite{Kuenzel1995, Tariku2010}. This assumption was discussed in \cite{Soudani2016}, for the case of rammed earth material, by comparing the results using expression~\eqref{eq:source_term1} and the following one:
\begin{align*}
  I_{\,1} \egal - \, \biggl(\, \pd{\w_{\,2}}{t} \plus \div \, j_{\,c\,,\,2} \,\biggr) \,.
\end{align*}
The latter expression assumes that the moisture concentration $\w_{\,2}$ is obtained by the sorption isotherm $w_{\,12} \ \approx \ w_{\,2} \,$. This assumption is not considered in this work, a third expression of the source is proposed: 
\begin{align*}
  I_{\,1} \egal \pd{\w_{\,1}}{t} \plus \div \, j_{\,c\,,\,1} \,,
\end{align*}
where the vapor concentration $\w_{\,1}$ is expressed as:
\begin{align*}
  \w_{\,1} \egal \frac{\Pi \, \, \bigl(\, 1 \moins \sigma \, \bigr)}{R_{\,1} \, T} \,.
\end{align*}
The differential of $\w_{\,1}$ gives:
\begin{align}\label{Eq:differentiel_dw1}
  \partial \, \w_{\,1} \egal \frac{\Pi \, (\,1 \moins \sigma \,)}{R_{\,1} \, T} \, \partial P_{\,1} \plus \frac{\Pi \, P_{\,1}}{R_{\,1} \, T} \, \partial  (\,1 \moins \sigma \,) \moins \frac{\Pi \, (\,1 \moins \sigma \,) \, P_{\,1}}{R_{\,1} \, T^{\,2}} \, \partial T \,.
\end{align}
Then, we obtain the final expression of the volumetric source term $I_{\,1}\,$:
\begin{align*}
  I_{\,1} \egal \frac{\Pi \, (\,1 \moins \sigma \,)}{R_{\,1} \, T} \,\pd{P_{\,1}}{t} \moins \frac{\Pi \, P_{\,1}}{R_{\,1} \, T} \, \pd{\sigma}{t} \moins \frac{\Pi \, (\,1 \moins \sigma \,) \, P_{\,1}}{R_{\,1} \, T^{\,2}} \, \pd{T}{t} \plus \div \,  \, j_{\,c\,,\,1} \,,
\end{align*} 
which depends on the vapor pressure, the amount of liquid in the pores, the temperature and the variations of the vapor fluxes.

%%% ----------------------------------------------------------------------- %%%

\subsection{Expression of the fluxes}

A filtration of the dry air occurs through the porous material. The vapor velocity and the air velocity are assumed to be the same, $\boldsymbol{\vit_{\,3}} \egal \boldsymbol{\vit_{\,2}} \egal \boldsymbol{\vit}\,$, and can be expressed by the \textsc{Darcy} law, remembering that the gravity effects are neglected:
\begin{align}\label{eq:Darcy_law}
  \boldsymbol{\vit} \egal - \, \frac{k_{\,13}}{\mu} \, \grad P \,,
\end{align}
where $k_{\,13} \ \unit{m^{\,2}}$ is the intrinsic permeability of the material and $\mu$ is the dynamic viscosity of the fluid. The diffusion of vapor and air is small compared to the filtration transfer. Thus, the total flux of the vapor and dry air is written as:
\begin{align*}
  \boldsymbol{j}_{\,c\,,\,1} \plus \boldsymbol{j}_{\,c\,,\,3} \egal \frac{\w_{\,1} \plus \w_{\,3}}{\Pi \, \bigl(\, 1 \moins \sigma \, \bigr)} \, \boldsymbol{\vit} \,,
\end{align*}
which can be rewritten using Eq.~\eqref{eq:Darcy_law} for the velocity:
\begin{align*}
  \boldsymbol{j}_{\,c\,,\,1} \plus \boldsymbol{j}_{\,c\,,\,3} \egal - \, \frac{\bigl(\,\w_{\,1} \plus \w_{\,3}\, \bigr)}{\Pi \, \bigl(\, 1 \moins \sigma \, \bigr)} \, \frac{k_{\,13}}{\mu} \, \grad P \,.
\end{align*}
The vapor flow is driven by diffusion and advection:
\begin{align*}
  \boldsymbol{j}_{\,c\,,\,1} \egal - \, k_{\,1} \, \grad P_{\,1} \plus \frac{\w_{\,1}}{\Pi \, \bigl(\, 1 \moins \sigma \, \bigr)}  \, \boldsymbol{\vit} \egal \moins k_{\,1} \, \grad P_{\,1} \plus \frac{P_{\,1}}{R_{\,1} \, T} \, \boldsymbol{\vit} \,,
\end{align*}
where $k_{\,1}$ is the vapor permeability of the material and $R_{\,1} \ \unit{J/(kg.K)}$ the vapor gas constant. The air filtration is sufficiently slow to have any influence on the liquid water. Thus, the total moisture flow is written as:
\begin{align*}
  \boldsymbol{j}_{\,c\,,\,1} \plus \boldsymbol{j}_{\,c\,,\,2} \egal - \, k_{\,2} \, \grad P_{\,2} \moins k_{\,1} \, \grad P_{\,1} \plus \frac{P_{\,1}}{R_{\,1} \, T} \, \boldsymbol{\vit} \,.
\end{align*}
It has been shown that the diffusion of moisture can be written using the vapor pressure $P_{\,1}\,$, introducing the global moisture permeability $k_{\,m}$ \cite{Berger2018a}. Thus, the total moisture flux yields:
\begin{align*}
  \boldsymbol{j}_{\,c\,,\,1} \plus \boldsymbol{j}_{\,c\,,\,2} \egal - \, k_{\,m} \, \grad P_{\,1} \plus \frac{P_{\,1}}{R_{\,1} \, T} \, \boldsymbol{\vit} \,.
\end{align*}
The heat flux is expressed as:
\begin{align*}
  j_{\,q} \egal - \, k_{\,q} \, \grad T \plus \bigl(\, \w_{\,1} \, c_{\,1} \plus \w_{\,3} \, c_{\,3} \, \bigr) \, \, \frac{T }{\Pi \, \bigl(\, 1 \moins \sigma \, \bigr)} \, \boldsymbol{\vit} \,.
\end{align*}
It is driven by conduction and advection of the moist air.

%%% ----------------------------------------------------------------------- %%%

\subsection{Governing Equations}

For the sake of clarity, we introduce the following advection coefficients:
\begin{align*}
  & \boldsymbol{a}_{\,q} \ \eqdef \ \frac{\bigl(\, \w_{\,1} \, c_{\,1} \plus \w_{\,3} \, c_{\,3} \, \bigr)}{\Pi \, \bigl(\, 1 \moins \sigma \, \bigr)}\,\boldsymbol{\vit} \,, && \boldsymbol{a}_{\,v} \ \eqdef  \frac{\boldsymbol{\vit}}{R_{\,1} \, T} \,,
\end{align*}
along with the storage coefficients:
\begin{align*}
  & c_{\,a} \ \eqdef \ \frac{\Pi \, (\,1 \moins \sigma \,)}{R_{\,13} \, T} \,, 
  && c_{\,as} \ \eqdef \ \frac{\Pi}{T} \, \biggl(\, \frac{P}{R_{\,13}} \moins \frac{P_{\,1}}{R_{\,1}} \,\biggr) \,, 
  && c_{\,at} \ \eqdef \ \frac{\Pi \, (\,1 \moins \sigma \,) }{T^{\,2}} \, \biggl(\, \frac{P}{R_{\,13}} \moins \frac{P_{\,1}}{R_{\,1}} \,\biggr) \,,  \\[4pt]
  & c_{\,av} \ \eqdef \ \frac{\Pi \, (\,1 \moins \sigma \,) }{R_{\,1} \, T } \,,
  && c_{\,qv} \ \eqdef \ \frac{\Pi \, (\,1 \moins \sigma \,)}{R_{\,1} \, T} \,, 
  && c_{\,qs} \ \eqdef \ \frac{\Pi \, P_{\,1}}{R_{\,1} \, T} \,, \\[4pt]
  & c_{\,m} \ \eqdef \ \pd{w_{\,12}}{\phi} \, \frac{1}{\Ps} \,,
  && c_{\,q} \ \eqdef \ \biggl(\, c_{\,0} \, \rho_{\,0} \plus \sum_{i=1}^{3} \, c_{\,i} \, \w_{\,i} \moins \frac{\Pi \, P_{\,1} \, (\,1 \moins \sigma \,) }{R_{\,1} \, T^{\,2}} \, r_{\,12} \,\biggr) \,, \span\span
\end{align*}
and the permeability coefficients:
\begin{align*}
  & k_{\,a} \ \eqdef \ \frac{\bigl(\,\w_{\,1} \plus \w_{\,3}\, \bigr)}{\Pi \, \bigl(\, 1 \moins \sigma \, \bigr)} \, \frac{k_{\,13}}{\mu} \,, 
  && k_{\,v} \ \eqdef \ k_{\,1} \,.
\end{align*}
Finally, the set of partial differential governing equations for the physical problem of heat, air and moisture transfer through porous building material becomes:
\begin{subequations}\label{eq:phys_dim_HAM}
\begin{align}\label{eq:phys_dim_M} 
  c_{\,m} \, \pd{P_{\,1}}{t} & \egal \div \, \Bigl(\, k_{\,m} \, \grad P_{\,1} \moins \boldsymbol{a}_{\,v} \, P_{\,1} \, \Bigr) \,, \\[4pt]
  \label{eq:phys_dim_H} 
  c_{\,q} \, \pd{T}{t} & \egal \div \, \Bigl(\, k_{\,q} \, \grad T \moins \boldsymbol{a}_{\,q} \, T \, \Bigr) \plus r_{\,12} \, \div \, \Bigl(\, k_{\,v} \, \grad P_{\,1} \moins \boldsymbol{a}_{\,v} \, P_{\,1} \, \Bigr)  \\[4pt]
  & \qquad \moins \sum_{i=1}^{3} \,  \grad \, \bigl(\, c_{\,i} \, T \,\bigr) \scal \boldsymbol{j}_{\,c\,,\,i} \moins r_{\,12} \, c_{\,qv} \, \pd{P_{\,1}}{t} \plus r_{\,12} \, c_{\,qs} \, \pd{\sigma}{t} \,, \nonumber \\[4pt]
  \label{eq:phys_dim_A}
  c_{\,a} \, \pd{P}{t} & \egal \div \, \Bigl(\, k_{\,a} \, \grad P \, \Bigr) \moins \div \, \Bigl(\, k_{\,v} \, \grad P_{\,1} \moins \boldsymbol{a}_{\,v} \, P_{\,1} \, \Bigr)  \\[4pt]
  & \qquad \plus c_{\,av} \, \pd{P_{\,1}}{t} \plus c_{\,at}  \, \pd{T}{t} \plus c_{\,as} \, \pd{\sigma}{t} \,.  \nonumber
\end{align}
\end{subequations}

%%% ----------------------------------------------------------------------- %%%

\subsection{Boundary, interface and initial conditions}
\label{sec:BC_In_IC_cond}

For the set of governing equations~\eqref{eq:phys_dim_M}, \eqref{eq:phys_dim_H} and \eqref{eq:phys_dim_A}, boundary, interface and initial conditions are provided for a one-dimensional problem. At the interface between the material and the ambient air, the liquid flow is imposed:
\begin{align*}
  \boldsymbol{j}_{\,c\,,\,2} \egal \boldsymbol{g}_{\,m\,,\infty} \,,
\end{align*}
where $g_{\,m\,,\infty} \ \unit{kg/(m^{\,2}.s)}$ is the liquid water flux due to wind driven rain that crosses the porous surface. For the vapor phase, the total vapor flow is proportional to the surface and ambient conditions:
\begin{align*}
  \boldsymbol{j}_{\,c\,,\,1} \egal -\, \alpha_{\,m} \, \Bigl(\, P_{\,1} \moins P_{\,1\,,\, \infty}  \, \Bigr) \cdot \boldsymbol{n} \,,
\end{align*}
where $\alpha_{\,m} \ \unit{s/m}$ is the surface vapor transfer coefficient and $\boldsymbol{n}$ is the surface normal unitary vector. Therefore, for the moisture transfer equation, the following boundary conditions are stated:
\begin{align*}
  \boldsymbol{j}_{\,c\,,\,1} \plus \boldsymbol{j}_{\,c\,,\,2} \egal -\, \alpha_{\,m} \, \Bigl(\, P_{\,1} \moins P_{\,1\,,\, \infty} \Bigr) \cdot \boldsymbol{n} \plus \boldsymbol{g}_{\,m\,,\,\infty} \,.
\end{align*}
Similarly, for the heat transfer, the flow is proportional to the surface and ambient conditions. An additional heat flux $\boldsymbol{g}_{\,q\,,\,\infty}$ is introduced to take into account the radiative and convective heat transfer: 
\begin{align*}
  \boldsymbol{j}_{\,c\,,\,q} \egal -\, \alpha_{\,q} \, \Bigl(\, T  \moins T_{\, \infty} \, \Bigr) \cdot \boldsymbol{n} \plus \boldsymbol{g}_{\,q\,,\,\infty} \,.
\end{align*}
Therefore, the boundary condition for the heat transfer equation can be written as: 
\begin{multline*}
  \Bigl(\, k_{\,q} \, \grad T \moins \boldsymbol{a}_{\,q} \, T \, \Bigr) \plus r_{\,12} \, \div \, \Bigl(\, k_{\,v} \, \grad P_{\,1} \moins \boldsymbol{a}_{\,v} \, P_{\,1} \, \Bigr) \egal -\, \alpha_{\,q} \, \Bigl(\, T  \moins T_{\, \infty} \, \Bigr) \cdot \boldsymbol{n} \plus \\ 
  \boldsymbol{g}_{\,q\,,\,\infty} \moins r_{\,12} \, \alpha_{\,m} \, \Bigl(\, P_{\,1} \moins P_{\,1\,,\, \infty} \Bigr) \cdot \boldsymbol{n} \plus r_{\,12} \, \boldsymbol{g}_{\,m\,,\,\infty} \,.
\end{multline*}
Finally, the air pressure at the interface between the material and the ambient air is imposed:
\begin{align*}
  P \egal P_{\,\infty} \,.
\end{align*}
At the interface $x_{\,i}$ between two materials $A$ and $B\,$, a perfect contact is assumed so that air, moisture and heat fluxes are assumed to be continuous across the interfaces \cite{DeFreitas1996, Guimaraes2018}:
\begin{align*}
  \boldsymbol{j}_{\,c\,,\,\mathrm{A}} \,(\,x_{\,i}\,) \egal \boldsymbol{j}_{\,c\,,\,\mathrm{B}} \,(\,x_{\,i}\,) \,.
\end{align*}
As initial conditions, the fields may depend on the space coordinate $x\,$: 
\begin{align*}
  & T \egal T_{\,i}\,(\,x\,) \,, & & P \egal P_{\,i}\,(\,x\,) \,, & & P_{\,1} \egal P_{\,1\,,\, i}\,(\,x\,) \,.
\end{align*}
It is important to note that at $t \egal 0\,$, the boundary and initial conditions must be consistent, \emph{i.e.} the initial condition must verify the boundary conditions. The numerical model to solve the mathematical problem is presented in Section~\ref{sec:numerical_model} based on a dimensionless formulation.

%%% ----------------------------------------------------------------------- %%%

\section{Elaborating an efficient numerical model}
\label{sec:numerical_model}

Since the physical phenomena have been described, the second part in the elaboration of a numerical model consists in detailing the numerical method to solve the physical problem. For this, it is of major importance to define the strategy of building a numerical model that reduces the computational effort and maximize the accuracy of the solution. First, since we have a nonlinear problem, explicit scheme is preferred than implicit approaches to avoid costly subiterations to treat the nonlinearities at each time step. Then, an important observation can be made on the physical problem. In most cases, the air transfer in the porous material is faster than moisture and heat ones. The following inequalities can be set for the \textsc{Fourier} numbers:
\begin{align*}
  \Fo_{\,a} \ \gg \ \max \ \Bigl\{\, \Fo_{\,v} \,,\, \Fo_{\,m} \,,\, \Fo_{\,q} \, \Bigr\} \,.
\end{align*}
Thus, when using an \Eu ~explicit scheme, since we have a system of equations coupled through the advection coefficient $a_{\,v}^{\,\star}\,$, the stability condition will be imposed by the air transfer equation. To circumvent this restriction, the diffusive air transfer equation is solved using the \DF ~scheme. It provides an efficient highly stable stable scheme. The moisture and heat equations are advection--diffusion types. The \SG ~approach has shown great efficiency in preliminary studies for a single equation \cite{Berger2017a} and a system of two coupled equations \cite{Berger2018a}. Therefore it will be used for the spatial discretisation of the moisture and heat equations. Since this scheme has a stability condition, a two-step \RK ~will be used for the time discretisation to extend the stability region of the numerical scheme \cite{Chollom2003}.

A uniform discretisation is considered for space and time intervals. The discretisation parameters are denoted using $\Delta x$ for the space and $\Delta t$ for the time. The spatial cell $\mathcal{C} \egal \bigl[\,x_{\,j \moins \half} \,,\, x_{\,j\plus\half} \,\bigr]$ is illustrated in Figure~\ref{fig:SG_stencil}. The discrete values of function $u \, (\,x\,,\,t\,)$ are denoted by $u_{\,j}^{\,n} \ \eqdef \ u(\,x_{\,j}\,,\,t^{\,n}\,)$ with $j \ \in \ \bigl\{\, 1 \,, \ldots \,, N \,\bigr\}$ and $n \ \in \ \bigl\{\, 1 \,, \ldots \,, N_{\,t} \,\bigr\} \,$.

For the sake of clarity to explain the numerical method, the system Eqs.~\eqref{eq:phys_dim_HAM} will be written using a simpler notation and considering a linear one-dimensional problem for $x \ \in \ \bigl[\, 0 \,,\, 1\,\bigr]$ and $ t \ > \ 0\,$:
\begin{subequations}\label{eq:systeme_uvw}
\begin{align}
  \pd{u}{t} & \egal \pd{f}{x} \,, \label{eq:systeme_u} \\[4pt]
  \pd{v}{t} & \egal \pd{g}{x} \plus S\,, \label{eq:systeme_v}\\[4pt]
  \pd{w}{t} & \egal \pd{h}{x} \plus \Sigma \,, \label{eq:systeme_w}
\end{align}
\end{subequations}
where 
\begin{align*}
  & f \egal k_{\,1} \, \pd{u}{x} \moins a_{\,1} \, u \,, \\
  & g \egal k_{\,22} \, \pd{v}{x} \moins a_{\,22} \, v \plus k_{\,21} \, \pd{u}{x} \moins a_{\,21} \, u\,, 
  && S \egal - \, c_{\,21} \, \pd{u}{t}\plus c_{\,2s} \pd{\sigma}{t}   \,, \\[4pt]
  & h \egal k_{\,33} \, \pd{w}{x} \moins k_{\,31} \, \pd{u}{x} \plus a_{\,31} \, u\,, 
  && \Sigma \egal c_{\,31} \, \pd{u}{t} \plus  c_{\,32} \, \pd{v}{t} \plus c_{\,3s} \, \pd{\sigma}{t} \,.
\end{align*}
The term $\displaystyle \sum_{i=1}^{3} \,  \grad \, \bigl(\, c_{\,i} \, T \,\bigr) \scal \boldsymbol{j}_{\,c\,,\,i}$ in Eq.~\eqref{eq:phys_dim_H} is first omitted for the description of the numerical method. Its inclusion is detailed in Section~\ref{sec:extension_nonlinear}.

%%% ----------------------------------------------------------------------- %%%

\subsection{Spatial discretisation}
\label{sec:spat_discretisation}

\subsubsection{Equation~\eqref{eq:systeme_u}, moisture mass balance}
\label{sec:spat_discretisation_u}

The discretisation of Eq.~\eqref{eq:systeme_u} gives the following semi-discrete difference relation: 
\begin{align*}
  \od{u_{\,j}}{t} \egal \frac{1}{\Delta x} \, \biggl[\, f_{\,j+\half} \moins f_{\,j-\half} \,\biggr] \,.
\end{align*}

The \SG ~scheme was first proposed in \cite{Scharfetter1969}. It assumes that the numerical flux is constant on the dual cell $\mathcal{C}^{\,\star} \egal \Bigl[\, x_{\,j} \,,\, x_{\,j+1} \,\Bigr]\,$, which is illustrated in Figure~\ref{fig:SG_stencil}. Additional theoretical results have been proposed recently in \cite{Gosse2013, Gosse2017}. The description of the scheme for advection--diffusion transport in porous material are provided in \cite{Berger2017a, Berger2018a}. Thus, for the sake of compactness, interested readers may refer to the above mentioned references. The semi-discrete difference relation for Eq.~\eqref{eq:systeme_u} is directly given by:
\begin{align}\label{eq:ode_uj}
  \od{u_{\,j}}{t} \egal \frac{d_{\,1}}{\Delta x} \Biggl[\,\mathcal{B} \bigl(\, \Theta \,\bigr) \, u_{\,j+1} \moins \biggl(\, \mathcal{B} \bigl(\, \Theta \,\bigr) \plus \mathcal{B} \bigl(\, - \, \Theta \,\bigr)  \,\biggr)  \, u_{\,j} \plus \mathcal{B} \bigl(\, - \, \Theta \,\bigr) \, u_{\,j-1}\, \Biggr] \,.
\end{align}
As mentioned in \cite{Berger2018a}, one important advantage of the \SG ~scheme is the possibility of computing the flux $f_{\,j+\half}$ as wells as the exact solution of $u\,$:
\begin{align}\label{eq:solu}
  & u \,(\,x\,) \egal - \, \frac{1}{a_{\,1}} \, f_{\,j+\half} \plus C_{\,j} \, \mathrm{e}^{\,\dfrac{a_{\,1}\,x}{k_{\,1}} } \,, && \forall \, x \, \in \, \mathcal{C}^{\,\star}  \,,
\end{align}
where 
\begin{align*}
  C_{\,j}\ \eqdef \ \nicefrac{ \Bigl(\, u_{\,j} \moins u_{\,j+1} \, \Bigr)}{1 \moins \mathrm{e}^{\,\Theta}} \; \mathrm{e}^{\,-\,\dfrac{a_{\,1} \, x_{\,j}}{k_{\,1}}} 
\end{align*}
is a defined constant.

%%% ----------------------------------------------------------------------- %%%

\subsubsection{Equation~\eqref{eq:systeme_v}, energy balance}

For the spatial discretisation of Eq.~\eqref{eq:systeme_v}, the semi-discrete difference relation yields:
\begin{align*}
  \od{v_{\,j}}{t} \egal \frac{1}{\Delta x} \, \biggl[\, g_{\,j+\half} \moins g_{\,j-\half} \,\biggr] \plus S_{\,j} \,.
\end{align*}
The source term $S_{\,j}$ is evaluated using Eq.~\eqref{eq:ode_uj}:
\begin{align*}
  S_{\,j} \egal -\, c_{\,21} \, \od{u_{\,j}}{t} \,.
\end{align*}
By analogy, using the \SG ~approach, the numerical flux $g_{\,j+\half}$ is assumed constant on the dual cell $\mathcal{C}^{\,\star}\,$. The exact solution of $u \, (\,x\,)$ is already known for $x \, \in \, \mathcal{C}^{\,\star}$ so that the numerical flux $g_{\,j+\half}$ in the dual cell can be computed. It is important to remark that the solution considers a fully coupled approach between Equations~\eqref{eq:systeme_u} and \eqref{eq:systeme_v}. For the sake of notation compactness, the computation is provided in the supplementary \texttt{Maple\texttrademark} file. It should be noted that the exact solution $v \, (\,x\,) \, \in \, \mathcal{C}^{\,\star}$ can also be computed.

%%% ----------------------------------------------------------------------- %%%

\subsubsection{Equation~\eqref{eq:systeme_w}, moist air mass balance}

The following semi-discrete difference relation stands for Eq.~\eqref{eq:systeme_w}:
\begin{align*}
\od{w_{\,j}}{t} \egal \frac{1}{\Delta x} \, \biggl[\, h_{\,j+\half} \moins h_{\,j-\half} \,\biggr] \plus \Sigma_{\,j}\,,
\end{align*}
where the numerical flux $h_{\,j+\half}$ is given by:
\begin{align}
\label{eq:flux_h}
h_{\,j+\half} \egal 
k_{\,33} \, \pd{w}{x}\,\biggl|_{\,j+\half}
\moins k_{\,31} \, \pd{u}{x}\,\biggl|_{\,j+\half}  \plus a_{\,31} \, u_{\,j+\half} \,.
\end{align}
The exact solution of $u \, (\,x\,)$ is already known so that it is possible to compute the second and the third right-hand side terms of Eq.~\eqref{eq:flux_h}. The hypothesis of the \SG ~approach, assuming the flux to be constant through the dual cell $\mathcal{C}^{\,\star}$, is not considered for the diffusion term. This term is approximated using central finite difference scheme. At the end, the semi-discrete difference relation for Eq.~\eqref{eq:systeme_w} is:
\begin{align}\label{eq:ode_w}
  \od{w_{\,j}}{t} \egal & \frac{k_{\,33}}{\Delta x^{\,2}} \, \biggl(\, w_{\,j+1} \moins 2 \, w_{\,j}  \plus  w_{\,j-1}\,\biggr) \\[4pt]
  & \moins \frac{k_{\,31} \; a_{\,1}}{\Delta x \; k_{\,1}} \Biggl(\,C_{\,j} \, \mathrm{e}^{\,\dfrac{a_{\,1}\,x_{\,j+\half}}{k_{\,1}} } \moins C_{\,j-1} \, \mathrm{e}^{\,\dfrac{a_{\,1}\,x_{\,j-\half}}{k_{\,1}} }\,\Biggr) \\[4pt]
  & \moins \frac{a_{\,31}}{a_{\,1}} \, \biggl(\, f_{\,j+\half} \moins f_{\,j-\half}\,\biggr) \plus a_{\,31} \, \Biggl(\, C_{\,j}\, \mathrm{e}^{\,\dfrac{a_{\,1}\,x_{\,j+\half}}{k_{\,1}} } \moins C_{\,j-1} \, \mathrm{e}^{\,\dfrac{a_{\,1}\,x_{\,j-\half}}{k_{\,1}} } \,\Biggr) \plus \Sigma_{\,j} \,,
\end{align}
where the source term 
\begin{align*}
  \Sigma_{\,j} \egal c_{\,31} \, \od{u_{\,j}}{t} 
\end{align*}
is evaluated using Eq.~\eqref{eq:ode_uj}. For the treatment of the boundary conditions, interested reader is invited to consult \cite{Berger2018a} for the details.

\begin{figure}
  \centering
  \subfigure[\label{fig:SG_stencil}]{\includegraphics[width=.45\textwidth]{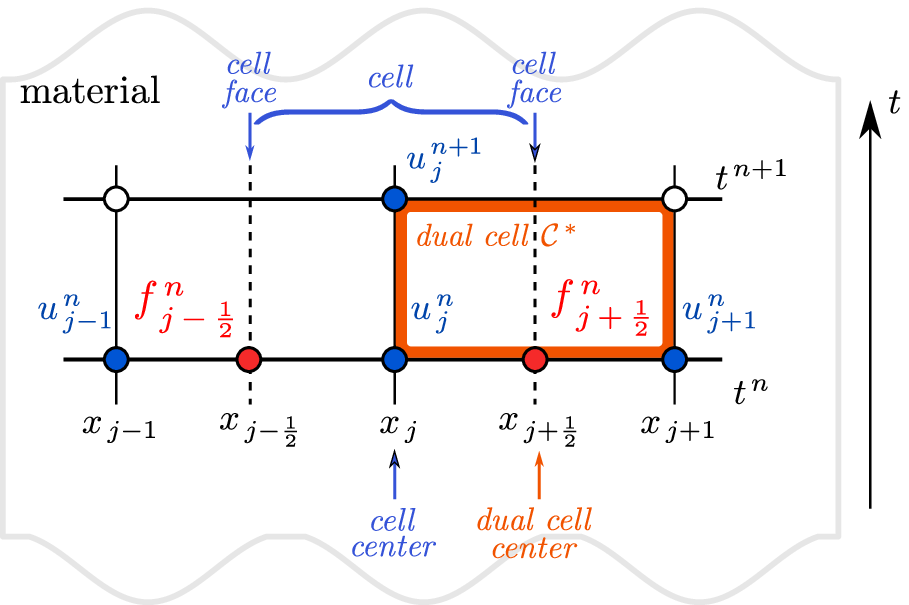}}
  \subfigure[\label{fig:DF_stencil}]{\includegraphics[width=.45\textwidth]{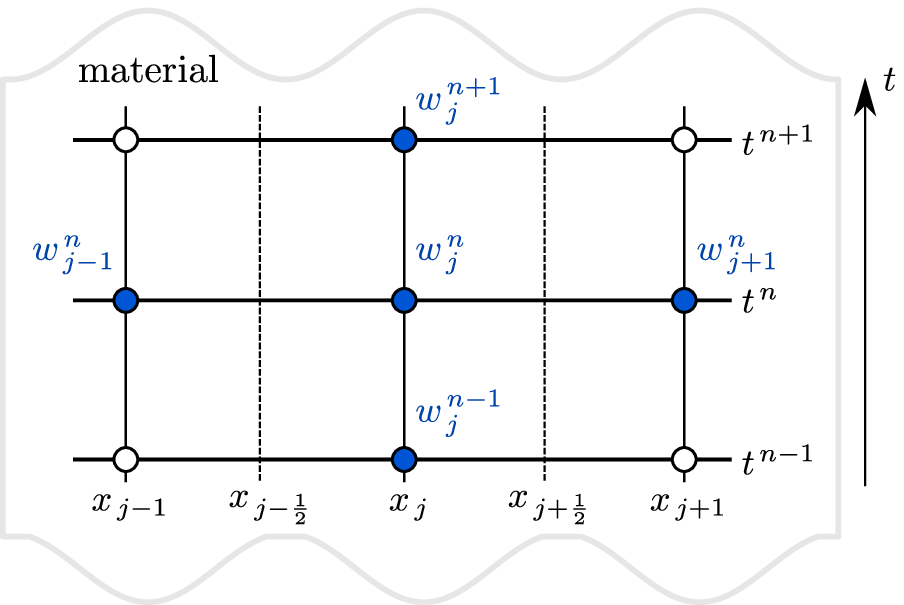}}
  \caption{Stencils of the \SG (a) and \DF (b) ~schemes.}
\end{figure}

%%% ----------------------------------------------------------------------- %%%

\subsection{Temporal discretisation}

The purpose is to relax the stability restriction of the system of differential equations~\eqref{eq:systeme_uvw}. The main restriction comes from Eq.~\eqref{eq:systeme_w} and will be softened by using the \DF ~scheme presented in Section~\ref{sec:temp_discretisation_air_transfer}. Nevertheless, it is also of major importance to relax the stability condition coming from Eqs.~\eqref{eq:systeme_u} and \eqref{eq:systeme_v}. For this purpose, the two--step \RK ~methods can be used since it enables to provide with fewer stages the same order of accuracy than standard one-step \RK ~approaches \cite{Jackiewicz1991}. Moreover, it is possible to extend the regions of stability of the scheme. In addition, the numerical scheme is written in an explicit way, avoiding costly sub--iterations to treat the nonlinearities observed when using implicit approaches. Thus, the two--step \RK ~scheme provides a competitive option compared to the classical explicit \Eu, one--step \RK ~or implicit methods.

%%% ----------------------------------------------------------------------- %%%

\subsubsection{Equations~\eqref{eq:systeme_u} and \eqref{eq:systeme_v},  moisture mass and energy balances}
\label{sec:temp_discretisation_uv}

For the description of the temporal discretisation, Eqs.~\eqref{eq:systeme_u} and \eqref{eq:systeme_v} are written in the following form:
\begin{align*}
  & \pd{u}{t} \egal F(\,u\,) \,, && F(\,u\,) = \pd{f}{x} \,, \\[4pt]
  & \pd{v}{t} \egal G(\,v\,) \,, && G(\,v\,) = \pd{g}{x} \plus S \,.
\end{align*}
The two-step \RK (TSRK) methods give the following fully discrete scheme \cite{Chollom2003}:
\begin{align*}
  u^{\,n+1} & \egal \theta \, u^{\,n-1} \plus \bigl(\, 1 \moins \theta \,\bigr) \, u^{\,n} \plus \Delta t \, \sum_{\,k=1}^{\,s} \, \Bigl[\,\nu_{\,k} \, F \bigl(\, \,U_{\,k}^{\,n-1} \, \bigr) \plus \mu_{\,k} \, F \bigl(\, \,U_{\,k}^{\,n} \, \bigr) \,\Bigr] \,, \\[4pt]
  U_{\,k}^{\,n} & \egal \lambda_{\,k} \, u^{\,n-1} \plus \bigl(\, 1 \moins \lambda_{\,k} \,\bigr) \, u^{\,n} \plus \Delta t \, \sum_{\,p=1}^{\,s}  \, \Bigl[\,a_{\,k\,p} \, F \bigl(\, \,U_{\,p}^{\,n-1} \, \bigr) \plus b_{\,k\,p} \, F \bigl(\, \,U_{\,p}^{\,n} \, \bigr) \,\Bigr] 
\end{align*}
and similarly for $v\,$:
\begin{align*}
  v^{\,n+1} & \egal \theta \, v^{\,n-1} \plus \bigl(\, 1 \moins \theta \,\bigr) \, v^{\,n} \plus \Delta t \, \sum_{\,k=1}^{\,s} \, \Bigl[\,\nu_{\,k} \, G \bigl(\, \,V_{\,k}^{\,n-1} \, \bigr) \plus \mu_{\,k} \, G \bigl(\, \,V_{\,k}^{\,n} \, \bigr) \,\Bigr] \,, \\[4pt]
  V_{\,k}^{\,n} & \egal \lambda_{\,k} \, v^{\,n-1} \plus \bigl(\, 1 \moins \lambda_{\,k} \,\bigr) \, v^{\,n} \plus \Delta t \, \sum_{\,p=1}^{\,s}  \, \Bigl[\,a_{\,k\,p} \, G \bigl(\, \,V_{\,p}^{\,n-1} \, \bigr) \plus b_{\,k\,p} \, G \bigl(\, \,V_{\,p}^{\,n} \, \bigr) \,\Bigr] \,,
\end{align*}
where $\theta\,$, $\nu_{\,k}\,$, $\mu_{\,k}\,$, $\lambda_{\,k}\,$, $a_{\,kp}\,$ and $b_{\,kp}$ are numerical coefficients provided in \cite{Chollom2003} depending on the value of the number of stage $s\,$. According to the vocabulary in \cite{Chollom2003}, the approach has two--steps since it computes the fields $\Bigl(\,\,u^{\,n+1} \,,\, v^{\,n+1}\,\Bigr)$ using the previous values $\Bigl(\,\,u^{\,n} \,,\, v^{\,n}\,\Bigr)$ and $\Bigl(\,\,u^{\,n-1} \,,\, v^{\,n-1}\,\Bigr)$ at $t^{\,n}$ and $t^{\,n-1}\,$, respectively.

%%% ----------------------------------------------------------------------- %%%

\subsubsection{Equation~\eqref{eq:systeme_w}, moist air mass balance}
\label{sec:temp_discretisation_air_transfer}

For the description, Eq.~\eqref{eq:ode_w}, resulting from the semi-discrete spatial discretisation of Eq.~\eqref{eq:systeme_w} with the method of horizontal lines \cite{Kreiss1992}, is written as:
\begin{align}\label{eq:ode_w2}
  \od{w_{\,j}}{t} \egal & \frac{k_{\,33}}{\Delta x^{\,2}} \, \biggl(\, w_{\,j+1} \moins 2 \, w_{\,j}  \plus  w_{\,j-1}\,\biggr) \plus H_{\,j} \,,
\end{align}
where 
\begin{align*}
  H_{\,j} \egal & \, \frac{k_{\,31} \; a_{\,1}}{\Delta x \; k_{\,1}}\Biggl(\,C_{\,j} \, \mathrm{e}^{\,\dfrac{a_{\,1}\,x_{\,j+\half}}{k_{\,1}} } \moins C_{\,j-1} \, \mathrm{e}^{\,\dfrac{a_{\,1}\,x_{\,j-\half}}{k_{\,1}} }\,\Biggr) \\[4pt]
  & \moins \frac{a_{\,31}}{a_{\,1}} \, \biggl(\, f_{\,j+\half} \moins f_{\,j-\half}\,\biggr) \plus a_{\,31} \, \Biggl(\, C_{\,j}\, \mathrm{e}^{\,\dfrac{a_{\,1}\,x_{\,j+\half}}{k_{\,1}} } \moins C_{\,j-1} \, \mathrm{e}^{\,\dfrac{a_{\,1}\,x_{\,j-\half}}{k_{\,1}} } \,\Biggr) \plus \Sigma_{\,j} \,.
\end{align*}

Using the \DF ~approach \cite{DuFort1953}, the numerical scheme given for Eq.~\eqref{eq:ode_w2} is expressed as: 
\begin{align}\label{eq:DF_w1}
  \frac{w_{\,j}^{\,n+1} \moins w_{\,j}^{\,n-1}}{2 \, \Delta t} \egal \frac{k_{\,33}}{\Delta x^{\,2}} \, \biggl(\, w_{\,j+1}^{\,n} \moins \bigl(\,w_{\,j}^{\,n-1} \plus w_{\,j}^{\,n+1} \,\bigr)  \plus  w_{\,j-1}^{\,n}\,\biggr) \plus H_{\,j}^{\,n} \,,
\end{align}
where the term $2 \, w_{\,j}^{\,n}$ has been replaced by $w_{\,j}^{\,n+1} + w_{\,j}^{\,n-1}\,$. The stencil is illustrated in Figure~\ref{fig:DF_stencil}. Re-arranging Eq.~\eqref{eq:DF_w1} to compute $w_{\,j}^{\,n+1}\,$, it gives the following discrete system:
\begin{align*}
  w_{\,j}^{\,n+1} \egal \frac{1 \moins \delta}{1 \plus \delta} \; w_{\,j}^{\,n-1} \plus \frac{\delta}{1 \plus \delta} \, \biggl(\,u_{\,j+1}^{\,n} \plus w_{\,j-1}^{\,n} \,\biggr) \plus \frac{2 \, \Delta t}{1 \plus \delta} \, H_{\,j}^{\,n} \,,
\end{align*}
where
\begin{align*}
  \delta \ \eqdef \ \frac{2 \, \Delta t}{\Delta x^{\,2}} \, k_{\,33} \,.
\end{align*}

According to the standard \textsc{von Neumann} stability analysis, the \DF ~scheme is unconditionally stable \cite{Gasparin2017, Taylor1970}. Further details and examples of applications of this approach are presented in \cite{Gasparin2017, Gasparin2017b}.

%%% ----------------------------------------------------------------------- %%%

\subsection{Extension to nonlinear coefficients}
\label{sec:extension_nonlinear}

The physical problem~\eqref{eq:phys_dim_HAM} involves coefficients $c\,$, $k\,$ and $a$ varying with the fields $u\,$, $v\,$ and/or $w\,$. To extend the numerical method for nonlinear coefficients, the assumption of frozen coefficients on the dual cell $\mathcal{C}^{\,\star}\,$ is adopted. Thus, the flux $f_{\,j\plus \half}$ is written as:
\begin{align*}
  f_{\,j+\half} & \egal k_{\,1\,,\,j+\half} \; \pd{u}{x} \moins a_{\,1\,,\,j+\half} \, u \,, 
\end{align*}
where the coefficients are computed as:
\begin{align*}
  & k_{\,1\,,\,j+\half} \egal k_{\,1}\, \biggl(\,u_{\,j+\half} \,,\, v_{\,j+\half} \,,\, w_{\,j+\half} \,\biggr) \,,
  && a_{\,1\,,\,j+\half} \egal a_{\,1}\, \biggl(\,u_{\,j+\half}\,,\, v_{\,j+\half} \,,\, w_{\,j+\half}\,\biggr) \,,
\end{align*}
with
\begin{align*}
  & u_{\,j+\half} \egal \frac{1}{2} \, \biggl(\,u_{\,j+1} \plus u_{\,j} \,\biggr) \,,
  && v_{\,j+\half} \egal \frac{1}{2} \, \biggl(\,v_{\,j+1} \plus v_{\,j} \,\biggr) \,,
  && w_{\,j+\half} \egal \frac{1}{2} \, \biggl(\,w_{\,j+1} \plus w_{\,j} \,\biggr) \,.
\end{align*}
Then, the numerical flux $f_{\,j\pm\half}$ is computed using the approach described in Section~\ref{sec:spat_discretisation_u}. By analogy, the approach is extended for the computation of the fluxes $g_{\,j\pm\half}$ and $h_{\,j\pm\half}\,$.

The term $\displaystyle \sum_{i=1}^{3} \; \beta_{\,i} \; j_{\,c\,,\,i} \; \pd{v}{x}$ is taken into account by freezing the flux $j_{\,c\,,\,i}^{\,n}$ at each time step $t^{\,n}\,$:
\begin{align*}
  \biggl(\, \beta_{\,i} \; j_{\,c\,,\,i} \; \pd{v}{x} \,\biggr)_{\,t \,=\, t^{\,n}} \egal \beta_{\,i} \; j_{\,c\,,\,i}^{\,n} \; \pd{v}{x} \,.
\end{align*}
Then, the flux $g$ in Equation~\eqref{eq:systeme_v} is written as:
\begin{align*}
  g \egal k_{\,22} \, \pd{v}{x} \moins \biggl(\, a_{\,22} \plus \sum_{i=1}^{3} \; \beta_{\,i} \; j_{\,c\,,\,i}^{\,n} \,\biggr) \, v \plus k_{\,21} \, \pd{u}{x} \moins a_{\,21} \, u\,, 
\end{align*}
and the whole description of the \SG ~approach for the spatial discretisation can be applied.

%%% ----------------------------------------------------------------------- %%%

\subsection{Important features of the numerical scheme}
\label{sec:important_features_num_scheme}

A synthesis of the schemes used for building the numerical model is provided in Table~\ref{tb:synthesis_num_scheme}. Some important features of the numerical scheme should be highlighted. First, the \SG ~scheme is well balanced and asymptotically preserving as detailed in \cite{Berger2018a}. When the advection coefficient is much greater than the diffusion one, the expression of the numerical flux tends to the so-called upwind scheme. Inversely, when the diffusion coefficient is more important, the flux is approximated by central finite difference. Therefore, the flux is correct independently from the grid parameters. This feature applies also to for the numerical flux $g$ from Eq.~\eqref{eq:systeme_v}.

Another important point is that we have an explicit expression for each governing equations of system~\eqref{eq:systeme_uvw}. Thus, when dealing with nonlinear coefficients, the computational efforts are preserved. No sub-iterations are required at each time step, contrarily to implicit numerical schemes.

It is of major importance to comment the stability condition of the system~\eqref{eq:systeme_uvw}. First of all, with an explicit \Eu ~scheme and central--finite differences, the stability condition of the system~\eqref{eq:systeme_uvw} is:
\begin{align}\label{eq:CFL_standard}
  \Delta t \ \leqslant\ \frac{\Delta x^{\,2}}{2} \; \min_{1 \, \leqslant \, k \,,\, l \, \leqslant \, 2} \ \biggl\{\, \min_{1 \, \leqslant \, j \, \leqslant \, N} \ \frac{1}{k_{\,kl\,,\,j}} \,\biggr\} \,.
\end{align}
Using the \DF ~approach, we have an unconditionally stable explicit scheme to solve Eq.~\eqref{eq:systeme_w}. Thus, the discretisation parameters for this equation need only to be chosen in terms of the characteristic time of the physical phenomena. The scheme ensure to compute a bounded solution. However, for Eqs.~\eqref{eq:systeme_u} and \eqref{eq:systeme_v}, the stability condition of the \SG ~scheme combined with \Eu ~explicit approach is given \cite{Gosse2016}:
\begin{align}\label{eq:CFL_SG}
  \Delta t \, \max_{1 \, \leqslant \, k \,,\, l \, \leqslant \, 2} \ \Biggl\{\,\max_{1 \, \leqslant \, j \, \leqslant \, N} k_{\,kl\,,\,j} \ \max_{1 \, \leqslant \, j \, \leqslant \, N} \biggl\{\, \frac{a_{\,kl\,,\,j}}{k_{\,kl\,,\,j}} \, \tanh \, \biggl(\, \dfrac{a_{\,kl\,,\,j} \, \Delta x}{2 \, k_{\,kl\,,\,j}} \,\biggr)^{\,-1} \,\biggr\}\,\Biggr\}\ \leqslant\ \Delta x \,.
\end{align}
The stability condition~\eqref{eq:CFL_SG} is nonlinear in $\Delta x \,$. It can be observed that for large space discretisation $\Delta x\,$, the conditions is equivalent to $\Delta t \, \leqslant \ C_{\,1} \, \Delta x \,$ \cite{Berger2018a}, $C_{\,1}$ being a defined constant. This condition is much less restrictive than the standard approach, which is given by the standard \textsc{Courant}--\textsc{Friedrichs}--\textsc{Lewy} (CFL) conditions~\eqref{eq:CFL_standard}: $\Delta t \, \leqslant \, C_{\,2} \, \Delta x^{\,2}$ for parabolic equations, with $C_{\,2}$ as a defined constant. Thus, it is not necessary to use fine spatial discretisation while using the \SG ~scheme for Eqs.~\eqref{eq:systeme_u} and \eqref{eq:systeme_v}. Moreover, since the \SG ~scheme provides an exact solution of the fields $u$ and $v$ in the dual cell $\mathcal{C}^{\,\star}$. Thus, no interpolation is required, ensuring an accurate computation of $u$ and $v$ in any point of the spatial domain. It should be noted that for the field $w$, since the \SG ~approach is not used, an interpolation using $w_{\,j}$ and $w_{\,j+1}$ is required to compute $w(\,x\,) \,, \forall \ x \ \in \ \bigl[\, x_{\,j} \,,\, x_{\,j+1}\,\bigr]\,$.

As described in Section~\eqref{sec:temp_discretisation_uv}, a two-step \RK ~scheme is used for the temporal discretisation of Eqs.~\eqref{eq:systeme_u} and \eqref{eq:systeme_v}. This scheme enables to extend the stability region compared to the classic \Eu ~approach. It can be expected a less restrictive stability condition than the one given by Eq.~\eqref{eq:CFL_SG} is obtained.

Summarizing, the System~\eqref{eq:systeme_uvw} is composed of three coupled parabolic equations. One diffusion driven equation solved using the \DF ~approach providing an unconditionally stable explicit scheme. Two advection--diffusion equations solved by means of the \SG ~approach, by implementing an explicit well balanced and asymptotically preserved numerical scheme. The stability condition $\Delta t \ \sim \ \Delta t\,$, for large spatial discretisation, is relaxed using an efficient two--step \RK ~approach for the temporal discretisation of these two equations. It should be noted that the \SG ~scheme cannot be used for the diffusion-type Equation~\eqref{eq:systeme_w}.

\begin{table}
\centering\small
\caption{Synthesis of the numerical scheme used to build the numerical model.}
\label{tb:synthesis_num_scheme}
\smallskip
\begin{tabular}{l|c|c}
\hline
\hline
\textit{Equation} 
& Eqs.~\eqref{eq:systeme_u} and \eqref{eq:systeme_v}
& Eq.~\eqref{eq:systeme_w} \\ 
\hline
\hline
Type
& Advection--diffusion
& Diffusion \\
Space discr.
& \SG 
& Central Finite--Differences \\ 
Time discr.
& Two--step \RK
& \DF\\ 
Features
& Explicit, extended stability conditions $\Delta t\ \sim \ \Delta x$
& Explicit, unconditionally stable\\
\hline
\hline
\end{tabular}
\end{table}

%%% ----------------------------------------------------------------------- %%%

\subsection{Efficiency and reliability of a numerical model}
\label{sec:efficiency_of_num_models}

To compare the efficiency of the numerical model several criteria are considered. First, the error with a reference solution $u^{\, \mathrm{ref}}\, (\,x \,, t \,)$ is evaluated. According to the definition provided in Section~\ref{sec:definition_physical_model}, it enables to evaluate the numerical approximations introduced by the numerical model compared to the mathematical model. The error between the solution, obtained by the numerical model, and the reference one is computed as a function of $x$ using:
\begin{align*}
  \varepsilon_{\,2}\, (\, x\,)\ &\ \eqdef \ \sqrt{\,\frac{1}{N_{\,t}} \, \sum_{n\, =\, 1}^{N_{\,t}} \, \Bigl( \, u\, (\,x \,, t^{\,n} \,) \moins u^{\mathrm{\, ref}}\, (\,x \,, t^{\,n} \,) \, \Bigr)^{\,2}}\,,
\end{align*}
where $N_{\,t}$ is the number of temporal steps. The global uniform error is measured using the norm of the space
$$
\mathcal{L}_{\,\infty} \, \biggl(\, \mathcal{L}_{\,2} \Bigl(\, \bigl[\, 0 \,,\, \tau \,\bigr] \,\Bigr) \,,\, \bigl[\, 0 \,,\, L \,\bigr] \,\biggr)
$$
is given by the maximum value of $\varepsilon_{\,2}\, (\, x\,)\,$:
\begin{align*}
  \varepsilon_{\, \infty}\ &\ \eqdef \ \sup_{x \ \in \ \bigl[\, 0 \,,\, L \,\bigr]} \, \varepsilon_{\,2}\, (\, x\,) \,,
\end{align*}
where $\tau$ is the time horizon of simulation. Another criterion is the significative correct digits of the solution, computed according to \cite{Soderlind2006a}:
\begin{align*}
  \mathrm{scd}\, (\, u\,) &\ \eqdef \moins \log_{\,10} \, \biggl|\biggl|\,\frac{u\, (\,x \,, \tau \,) \moins u^{\, \mathrm{ref}}\, (\,x \,, \tau \,)}{u^{\, \mathrm{ref}}\, (\,x \,, \tau \,)}\, \biggr|\biggr|_{\,\infty} \,.
\end{align*}
The last criterion is the computational (CPU) time required by the numerical model to compute the solution. It is measured using \texttt{Matlab\texttrademark} platform on a computer using Intel i7 CPU and $32\,$GB of RAM.

For the purposes of comparing the numerical predictions of a model with experimental observations to evaluate the reliability of the model, the relative error for temperature and vapor pressure are defined as:
\begin{align*}
  \varepsilon_{\,r}^{\,T} \, (\, x_{\,0} \,, t\,)\ &\ \eqdef \ \frac{ \displaystyle  T^{\mathrm{\, num}}\, (\,x_{\,0} \,, t \,) \moins T^{\mathrm{\, obs}}\, (\,x_{\,0} \,, t \,) }{\displaystyle \max_{\,t} T^{\mathrm{\, obs}}\, (\,x_{\,0} \,, t \,) \moins  \min_{\,t} T^{\mathrm{\, obs}}\, (\,x_{\,0} \,, t \,)}\,, \\[4pt]
  \varepsilon_{\,r}^{\,P_{\,1}} \, (\, x_{\,0} \,, t\,)\ &\ \eqdef \ \frac{ \displaystyle  P_{\,1}^{\mathrm{\, num}}\, (\,x_{\,0} \,, t \,) \moins P_{\,1}^{\mathrm{\, obs}}\, (\,x_{\,0} \,, t \,) }{P_{\,1}^{\mathrm{\, obs}}\, (\,x_{\,0} \,, t \,)}\,,
\end{align*}
where super script $^{\mathrm{\, num.}}$ stands for the output field obtained by the numerical model and  $^{\mathrm{\, obs.}}$  for the experimental observation of the field obtained at the sensor location $x_{\,0}\,$. The relative error is computed in a such way to avoid scaling problems due to the temperature scale.

Last, some physical hypotheses assumed in the mathematical model can be discussed after obtaining the numerical predictions. For this, the relative $\mathcal{L}_{\, 2}$ error is computed according to:
\begin{align*}
\varepsilon_{\,2\,,r} \bigl[\, q \,\bigr] (\, t\,)\ &\ \eqdef \ \sqrt{\, 
\frac{ \displaystyle \sum_{j\, =\, 1}^{N_{\,x}} \, \Bigl( \, q^{\mathrm{\, app}}\, (\,x_{\, j} \,, t \,) \moins q^{\mathrm{\, ref}}\, (\,x_{\, j} \,, t \,) \, \Bigr)^{\,2}}{
 \displaystyle \sum_{j\, =\, 1}^{N_{\,x}} \, \Bigl(\, q^{\mathrm{\, ref}}\, (\,x_{\, j} \,, t \,) \,\Bigr)^{\,2}}
}\,.
\end{align*}
Since the hypothesis can be computed on any physical quantity $q$ (such as the enthalpy or volumetric moisture source among others), it is important to specify it in the definition of the error $\varepsilon_{\,2\,,r}\,$. The quantity $q^{\mathrm{\, app}}$ designates the quantity approximated by the hypothesis and $q^{\mathrm{\, ref}}$ the reference one.

%%% ----------------------------------------------------------------------- %%%

\section{Validation of the numerical model}
\label{sec:validation_numerical_model}

After presenting the numerical model, a case study with nonlinear coefficients and \textsc{Robin}--type boundary conditions is considered to validate its implementation. The reference solution is computed using a numerical pseudo--spectral approach obtained with the \texttt{Matlab\texttrademark} open source toolbox \texttt{Chebfun} \cite{Driscoll2014}.

%%% ----------------------------------------------------------------------- %%%

\subsection{Description of the case}
\label{sec:case3}

The dimensionless problem is considered for the validation: 
\begin{align*}
  c_{\,m}^{\,\star} \; \pd{u}{t^{\,\star}} \egal & \Fo_{\,m} \; \pd{}{x^{\,\star}} \biggl(\, k_{\,m}^{\,\star} \; \pd{u}{x^{\,\star}} \moins \Pe_{\,m} \; a_{\,m}^{\,\star} \; u \,\biggr)\,, \\ [4pt]
  c_{\,q}^{\,\star} \; \pd{v}{t^{\,\star}} \egal & \Fo_{\,q} \; \pd{}{x^{\,\star}} \biggl(\, k_{\,q}^{\,\star} \; \pd{v}{x^{\,\star}} \moins \Pe_{\,q} \; a_{\,q}^{\,\star}  \; v \,\biggr) \plus \Fo_{\,q} \; \gamma \; r^{\,\star} \; \pd{}{x^{\,\star}} \biggl(\, k_{\,v}^{\,\star} \; \pd{u}{x^{\,\star}} \moins \Pe_{\,v} \; a_{\,v}^{\,\star} \; u \,\biggr) \\[4pt] 
  & \moins \K_{\,qv} \; c_{\,qv}^{\,\star} \; r^{\,\star} \, \pd{u}{t^{\,\star}} \plus \K_{\,qs} \; c_{\,qs}^{\,\star} \; r^{\,\star} \, \pd{\sigma}{t^{\,\star}} \moins \Fo_{\,q} \; \sum_{i = 1}^{3} \; \beta_{\,i} \;\pd{v}{x}  \; j_{\,c\,,\,i}\,,  \nonumber  \\[4pt]
  c_{\,a}^{\,\star} \; \pd{w}{t^{\,\star}} \egal & \Fo_{\,a} \; \pd{}{x^{\,\star}} \biggl(\, k_{\,a}^{\,\star} \; \pd{w}{x^{\,\star}} \,\biggr) \moins \Fo_{\,a} \; \delta \; \pd{}{x^{\,\star}} \biggl(\, k_{\,v}^{\,\star} \; \pd{u}{x^{\,\star}} \moins \Pe_{\,v} \; a_{\,v}^{\,\star} \; u \,\biggr) \\[4pt]
  & \plus \K_{\,av} \; c_{\,av}^{\,\star} \, \pd{u}{t^{\,\star}} \plus \K_{\,as} \; c_{\,as}^{\,\star} \, \pd{\sigma}{t^{\,\star}} \plus \K_{\,at} \; c_{\,at}^{\,\star} \, \pd{v}{t^{\,\star}}\,, \nonumber 
\end{align*}
with the following boundary conditions at the air--material interface:
\begin{align*}
  k_{\,m}^{\,\star} \; \pd{u}{x^{\,\star}} \moins \Pe_{\,m} \; a_{\,m}^{\,\star} \; u & \egal  \Bi_{\,m} \; \eta \; \Bigl(\,  u_{\,\infty} \moins u \,\Bigr) \plus \eta \; g_{\,m\,,\,\infty}^{\,\star} \,, \\[4pt]
  k_{\,q}^{\,\star} \, \pd{v}{x^{\,\star}} - \Pe_{\,q} \; a_{\,q}^{\,\star}  \, v + \gamma \, r^{\,\star} \, \biggl(\, k_{\,v}^{\,\star} \, \pd{u}{x^{\,\star}} - \Pe_{\,v} \; a_{\,v}^{\,\star} \; u \,\biggr) & \egal  \Bi_{\,q} \; \eta \; \Bigl(\, v_{\,\infty} \moins v \,\Bigr) \\[4pt]
  & \qquad \plus \gamma \, \Bi_{\,v} \, r^{\,\star} \; \eta \; \Bigl(\, u_{\,\infty} - u \,\Bigr) + \eta \; g_{\,q\,,\,\infty}^{\,\star} \,, \\ 
  w & \egal w_{\,\infty}  \,,
\end{align*}
where $\eta$ corresponds to the projection of the surface normal unity vector $\boldsymbol{n}$ on the axis $x\,$. The simulation is performed for $x^{\,\star} \ \in \ \bigl[\,0 \,,\, 1 \,\bigr]$ and $t^{\,\star} \ \in \ \bigl[\,0 \,,\, 24 \,\bigr]\,$. At the initial state, all fields are set to zero. The boundary conditions are defined as:
\begin{align*}
  & u_{\,\infty\,,\,L} \egal 0.6 \, \Biggl( \sin \bigl(\, \pi \, t\,\bigr)^{\,2} \plus \sin \biggl(\, \frac{2 \, \pi}{24}\, t\,\biggr)\,\Biggr) \,, && u_{\,\infty\,,\,R} \egal 0.9 \, \sin \biggl(\, \frac{2 \, \pi}{6}\, t\,\biggr)^{\,2} \,,  \\[4pt]
  & v_{\,\infty\,,\,L} \egal 1.2 \, \Biggl(\, \sin \biggl(\, \frac{2 \, \pi}{5}\, t\,\biggr)^{\,2} \plus \sin \biggl(\, \frac{2 \, \pi}{24}\, t\,\biggr)\,\Biggr) \,, && v_{\,\infty\,,\,R} \egal 0.5 \, \sin \biggl(\, \frac{2 \, \pi}{3}\, t\,\biggr)^{\,2} \,,
\end{align*}
\begin{equation*}
  w(\,0\,,\,t\,)  \egal 1.3 \, \sin \biggl(\, \frac{2 \, \pi}{6}\, t\,\biggr)^{\,2} \,, \qquad w(\,1\,,\,t\,) \egal 0.7 \, \Biggl(\, \sin \biggl(\, \frac{2 \, \pi}{4}\, t\,\biggr)^{\,2} \plus \sin \biggl(\, \frac{2 \, \pi}{24}\, t\,\biggr) \,\Biggr) \,,
\end{equation*}
with the following numerical values for the \textsc{Biot} numbers: 
\begin{align*}
  & \text{for } x^{\,\star} \egal 0 \,, && \Bi_{\,m} \egal 200 \,, && \Bi_{\,q} \egal 30 \,, && \Bi_{\,v} \egal 10^{\,4} \,,  \\[4pt]
  & \text{for } x^{\,\star} \egal 1 && \Bi_{\,m} \egal 125 \,, && \Bi_{\,q} \egal 22.5 \,, && \Bi_{\,v} \egal 5 \cdot 10^{\,3} \,.
\end{align*}
The numerical values of other dimensionless numbers are:
\begin{align*}
  & \Fo_{\,m} = 4 \cdot 10^{\,-3} \,, && \Fo_{\,q} = 4 \cdot 10^{\,-2} \,, && \Fo_{\,a} = 4 \cdot 10^{\,-1} \,,  && \Pe_{\,m} = 2 \cdot 10^{\,-3} \,, && \Pe_{\,v} = 3 \cdot 10^{\,-3} \,, \\[4pt]
  & \Pe_{\,q} = 5 \cdot 10^{\,-3} \,, && \K_{\,qv} = 1 \cdot 10^{\,-2} \,, &&\K_{\,av} = 2 \cdot 10^{\,-2} \,, && \gamma = 1.5 \cdot 10^{\,-3} \,, && \delta = 3 \cdot 10^{\,-4} \,, 
\end{align*}
with the following functions for the parameters $c^{\,\star}\,$, $k^{\,\star}$ and $a^{\,\star}\,$:
\begin{align*}
  k_{\,q}^{\,\star} & \egal 1 \plus 0.05 \cdot u \plus 0.03 \cdot v \plus 0.01 \cdot w \,, && a_{\,q}^{\,\star} \egal 1 \plus 0.01 \cdot v \,, \\[4pt]
  c_{\,q}^{\,\star}  & \egal 1 \plus 0.6 \cdot u^{\,2} \plus 0.1 \cdot v \plus 0.3 \cdot w \,, && a_{\,v}^{\,\star} \egal 1 \plus 0.01 \cdot u \,, \\[4pt]
  k_{\,v}^{\,\star} & \egal 1 \plus 0.1 \cdot u \plus 0.01 \cdot u^{\,2} \plus 0.02 \cdot v \plus 0.02 \cdot w \,, && c_{\,qv}^{\,\star} \egal c_{\,av}^{\,\star} \egal 1 \plus 0.06 \cdot u \,,\\[4pt]
  c_{\,m}^{\,\star} & \egal 1 \plus 0.05 \cdot u^{\,2} \plus 0.03 \cdot v^{\,2} \plus 0.01 \cdot w^{\,2} \,, &&  a_{\,m}^{\,\star} \egal 1 \plus 0.03 \cdot u \,, \\[4pt]
  k_{\,m}^{\,\star} & \egal 1 \plus 0.6 \cdot u \plus 0.1 \cdot v \plus 0.3 \cdot w \,, \\[4pt]
  k_{\,a}^{\,\star} &  \egal 1 \plus 0.01 \cdot u \plus 0.02 \cdot u^{\,2}\plus 0.07 \cdot v^{\,2} \plus 0.04 \cdot w^{\,2} \,, \\[4pt]
  c_{\,a}^{\,\star} & \egal 1 \plus 0.04 \cdot u \plus 0.03 \cdot v \plus 0.01 \cdot w \,,
\end{align*}
Other parameters are equal to zero and $r^{\,\star}\,$ is set to one. 

The efficiency of the numerical model proposed in Section~\ref{sec:numerical_model} is evaluated according to the criteria defined in Section~\ref{sec:efficiency_of_num_models}. The model is tested for different values of $s \in \bigl\{\,2\,,\,3\,\Bigl\}\,$ for the two--step \RK ~method, which coefficients are given in the following tables \cite{Chollom2003}:
\begin{subequations}\label{eq:coefficients_TSRK}
  \begin{align}
  & \begin{array}{c|c|c}
  \lambda_{\,k} & a_{\,kp} & b_{\,kp} \\ \hline
  \Theta & \nu_{\,k} & \mu_{\,k}\\
  \multicolumn{3}{c}{s}
   \end{array} \,, \\[7pt]
  & \begin{array}{c|cc|cc}
  0.308343 & 0.154172 & 0.154172 & & \\
  -0.113988 & -0.556994 & 0.00838564 & 1.43462 & \\ \hline
  0 & -0.526458 & 0.0787465 & 1.47417 & -0.0264581 \\
  \multicolumn{5}{c}{s \egal 2}
   \end{array} \,,  \\[7pt]
  & \begin{array}{c|ccc|ccc}
  0.0213802 & 0.264446 & -0.507512 & 0.264446 & 0 & 0 & 0 \\
  0.0991119 & 0.240292 & -0.631471 & 0.597963 & 0.392328 & 0 & 0 \\
  0.353437 & 0.0969341 & -0.578148 & 0.98944 & 0.582927 & 0.262283 & 0 \\ \hline
  0 & 0.197972 & -0.543131 & 0.521515 & 0.0867149 & 0.621047 & 0.115883 \\
  \multicolumn{7}{c}{s \egal 3}
   \end{array} \,.
  \end{align}
\end{subequations}

The proposed numerical model will be compared to the reference solution. The abbreviation NM~$3$ corresponds to the proposed one for different values of parameter $s\,$: NM~$3$b stands for $s \egal 2$ and NM~$3$c for $s \egal 3\,$. The following discretisation parameters are set $\Delta x^{\,\star} \egal 0.01$ and $\Delta t^{\,\star} \egal 2 \cdot 10^{\,-3}\,$. The numerical model NM~$1$, based on an \Eu ~explicit approach for the temporal discretisation, is also used with $\Delta x^{\,\star} \egal 0.01$ and $\Delta t^{\,\star} \egal 10^{\,-4}\,$, which is consistent with the CFL condition. It can be noted that for higher time discretisation (with $\Delta t^{\,\star} \egal 0.01$), the model NM~$1$ does not compute a bounded solution. The reference solution is the pseudo--spectral one obtained with the \texttt{Matlab\texttrademark} open source toolbox \texttt{Chebfun} \cite{Driscoll2014}.

%%% ----------------------------------------------------------------------- %%%

\subsection{Results}

\begin{figure}
  \centering
  \subfigure[\label{fig:Case3_u_ft}]{\includegraphics[width=.45\textwidth]{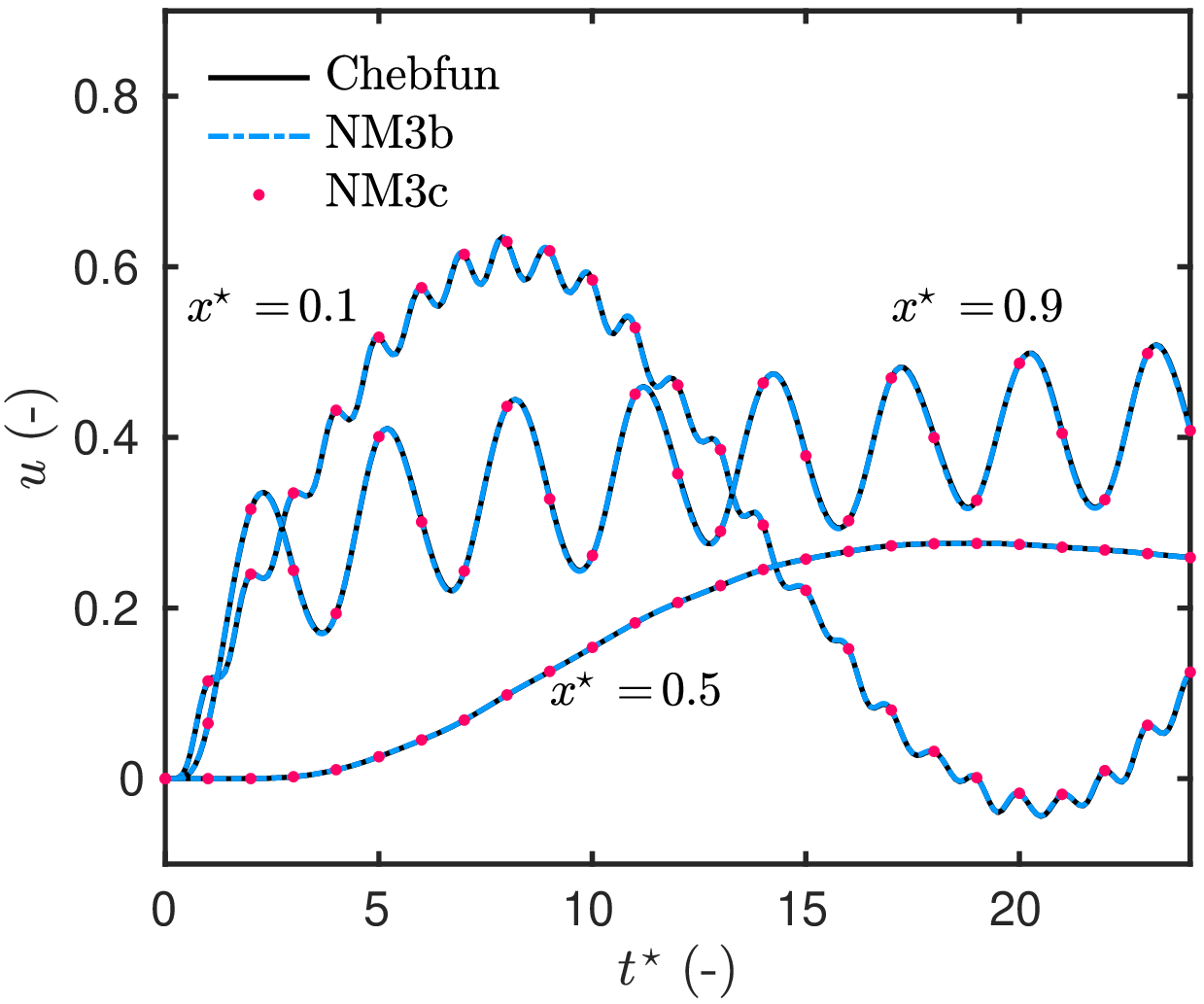}}
  \hspace{0.2cm}
  \subfigure[\label{fig:Case3_u_fx}]{\includegraphics[width=.45\textwidth]{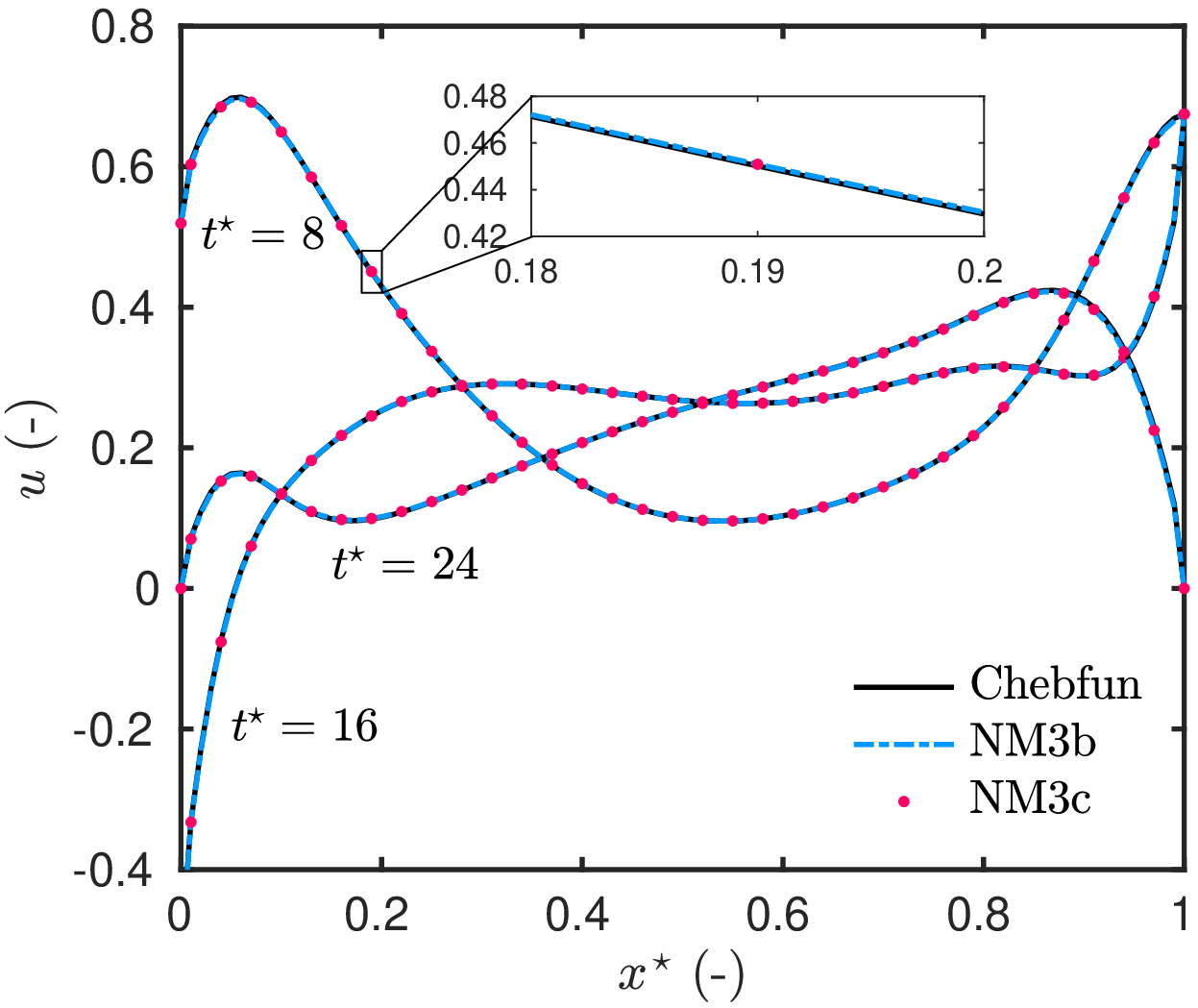}}
  \\
  \subfigure[\label{fig:Case3_v_ft}]{\includegraphics[width=.45\textwidth]{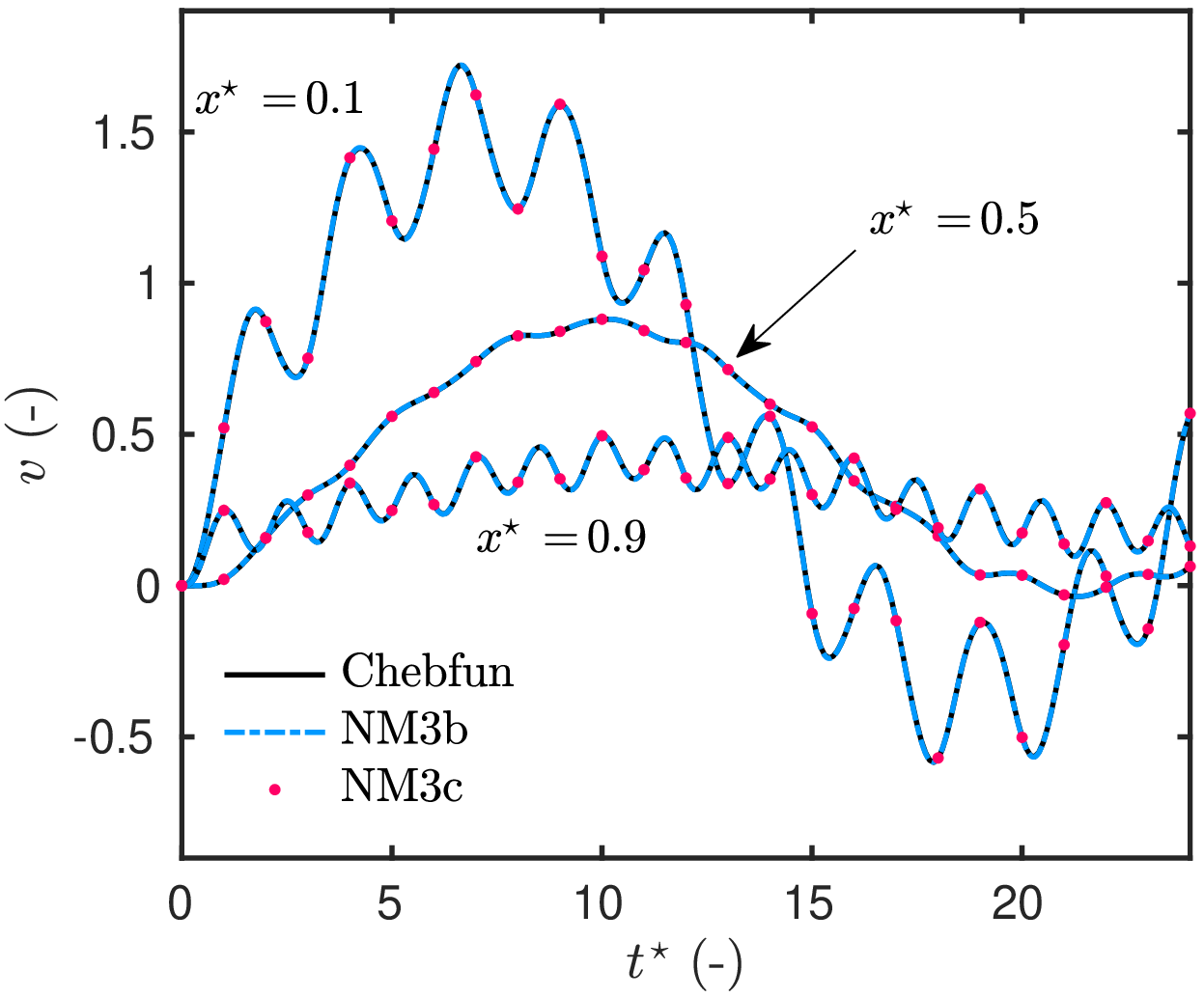}} 
  \hspace{0.2cm}
  \subfigure[\label{fig:Case3_v_fx}]{\includegraphics[width=.45\textwidth]{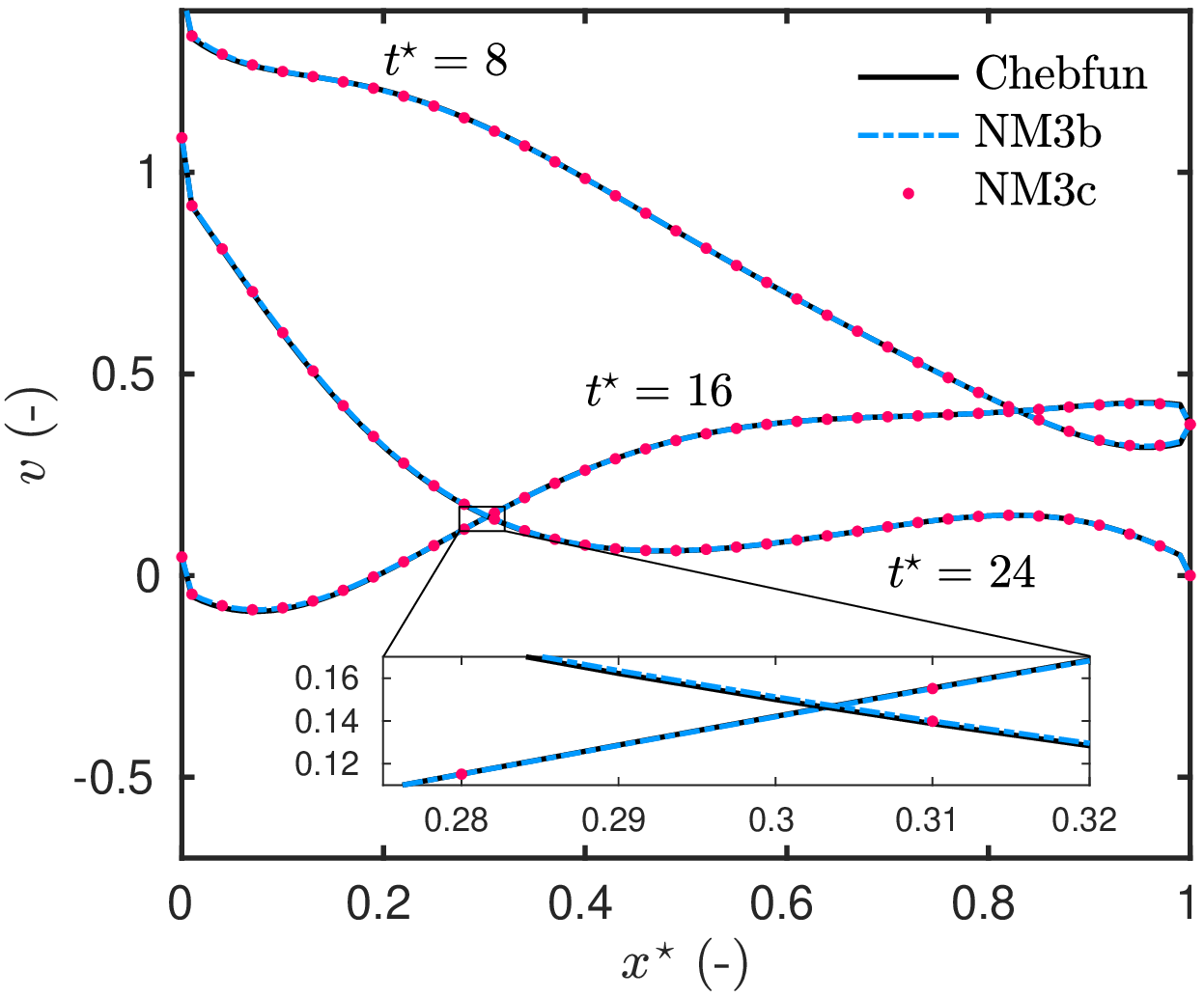}} 
  \\
  \subfigure[\label{fig:Case3_w_ft}]{\includegraphics[width=.45\textwidth]{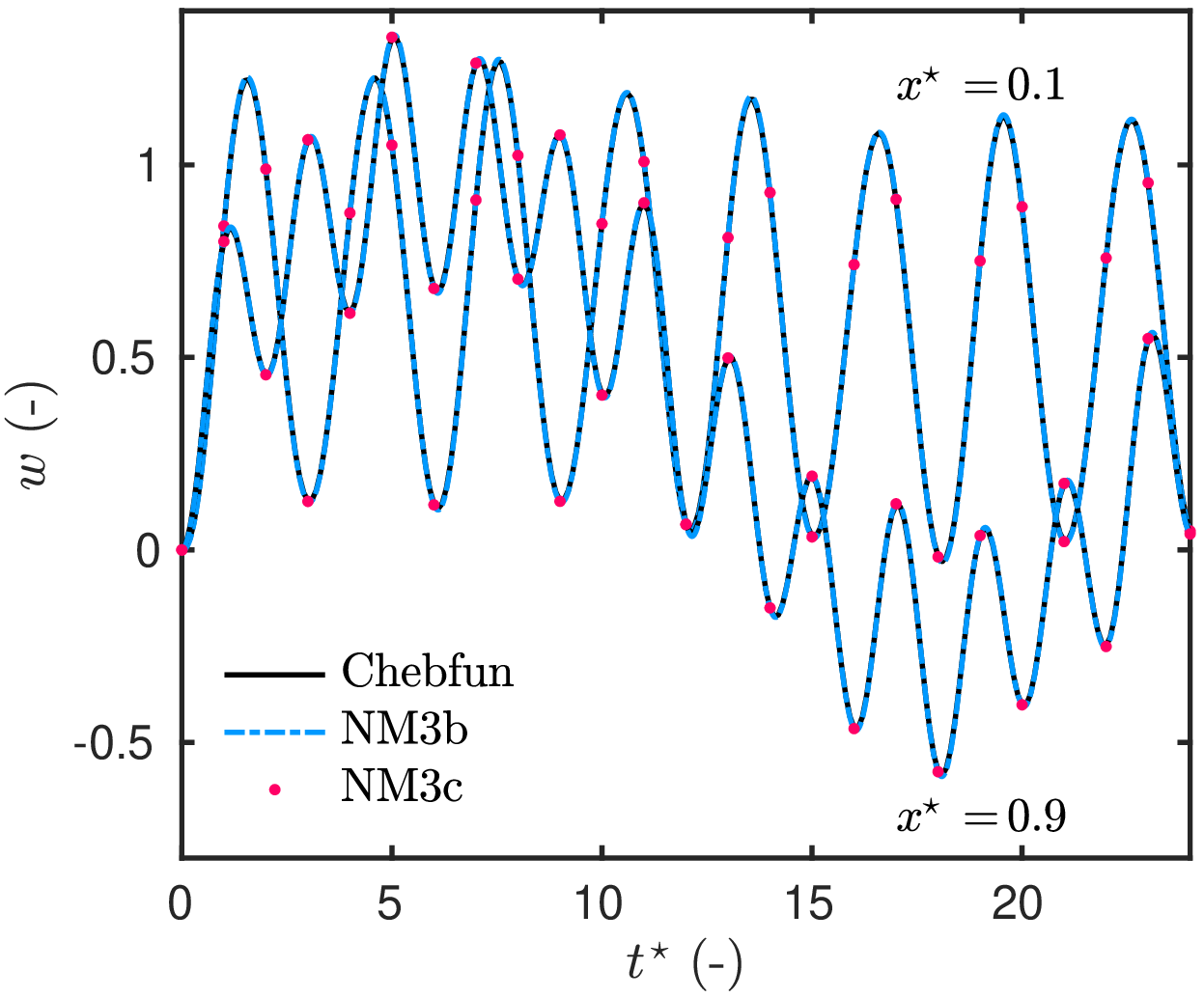}} 
  \hspace{0.2cm}
  \subfigure[\label{fig:Case3_w_fx}]{\includegraphics[width=.45\textwidth]{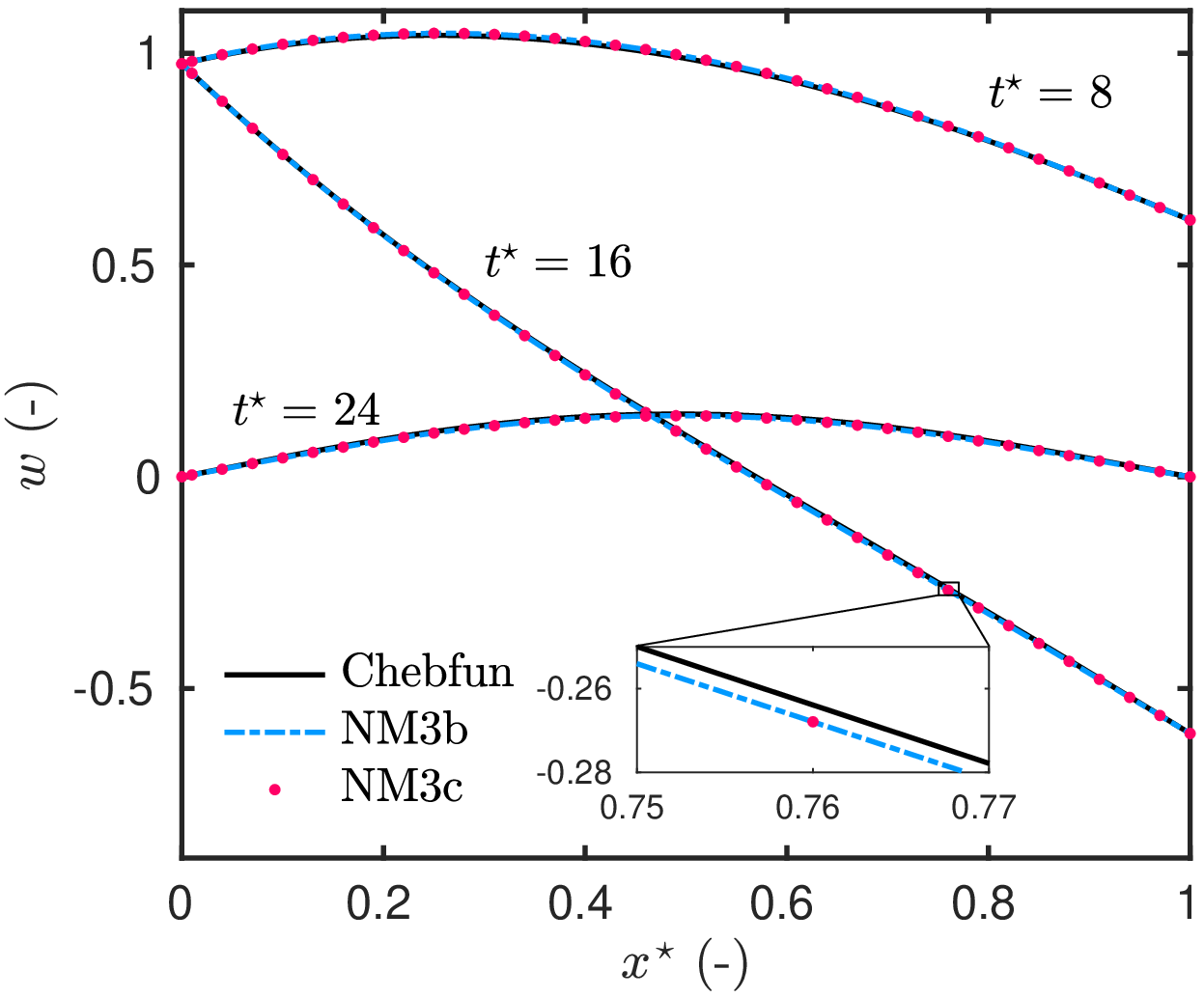}} 
  \\
  \caption{(a, c, e) Time evolution and (b, d, f) profiles at $t^{\,\star} \egal \bigl\{\,8\,,\, 16 \,,\,24 \,\bigr\}$ of the fields $u\,$, $v$ and $w\,$.}
  \label{fig:Case3_uw}
\end{figure}

Figures~\ref{fig:Case3_u_ft}, \ref{fig:Case3_v_ft} and \ref{fig:Case3_w_ft} show the evolution in time of the fields at $x^{\,\star} \ \in \ \bigl\{\,0.1 \,,\, 0.5 \,,\,0.9 \,\bigr\}\,$, which profiles are illustrated in Figures~\ref{fig:Case3_u_fx}, \ref{fig:Case3_v_fx} and \ref{fig:Case3_w_fx}. For the sake of compactness, only the results of NM~$3$ are presented. A perfect agreement is noted between the solutions of the numerical models based on the \SG ~and two--step \RK ~approaches and the \texttt{Chebfun} one. Since the solutions are overlaid, it highlights that the numerical approximations of the numerical models are very small. As noticed in Table~\ref{tb:Case3_cpu_time}, the error with the reference solution is very satisfactory for the three numerical models NM~$3$b, NM~$3$c and NM~$1$. A slight difference can be observed on the error of the field $w$ due to the increase of the time discretisation parameter between NM~$1$ and NM~$3$. The computational time required by the models to compute the solution is reported in Table~\ref{tb:Case3_cpu_time}. For the same level of accuracy, the numerical model NM~$3$b enables to compute the solution almost thirteen times faster than NM~$1$. These results are consistent since the number of operations to perform between two--step \RK ~approaches for $s \egal 2$ and for $s \egal 3$ is multiplied by two, for fixed discretisation parameters. Moreover, for the same level of accuracy, the two--step \RK ~scheme enables to relax the stability condition $\Delta t \egal 10^{\,-4}$ (with $\Delta x^{\,\star} \egal 0.01$) by twenty times with $\Delta t \egal 2 \cdot 10^{\,-3}\,$. This validation case enhances the possibility of using the proposed numerical models for real applications with a very satisfactory accuracy and a reduced computational time compared to more conventional approaches.

\begin{table}
\centering
\caption{Computer run time required by the numerical models to solve the case study (Section~\ref{sec:case3}), with $t_{\,0} \egal 136 \ \unit{s}$, and variation of the error $\varepsilon_{\,\infty}$ of the solutions $u\,$, $v$ and $w$.}
\smallskip
\label{tb:Case3_cpu_time}
\begin{tabular}{c cc c ccc}
\hline
\hline
\textit{Numerical Model (NM)}  
& $\Delta x$
& $\Delta t$
& CPU time
& $\varepsilon_{\,\infty}$ for $u$
& $\varepsilon_{\,\infty}$ for $v$
& $\varepsilon_{\,\infty}$ for $w$\\
\hline
\hline
NM~$1$
& $10^{\,-2}$ 
& $10^{\,-4}$
& $t_{\,0}$ 
& $7.35 \cdot 10^{\,-3}$
& $4.75 \cdot 10^{\,-3}$
& $2.1 \cdot 10^{\,-3}$ \\
NM~$3$b
& $10^{\,-2}$  
& $2 \cdot 10^{\,-3}$
& $0.08 \cdot t_{\,0}$
& $7.36 \cdot 10^{\,-3}$
& $4.76 \cdot 10^{\,-3}$
& $6.48 \cdot 10^{\,-3}$ \\
NM~$3$c
& $10^{\,-2}$ 
& $2 \cdot 10^{\,-3}$
& $0.10 \cdot t_{\,0}$
& $7.35 \cdot 10^{\,-3}$
& $4.76 \cdot 10^{\,-3}$
& $6.48 \cdot 10^{\,-3}$  \\
\hline
\hline
\end{tabular}
\end{table}

%%% ----------------------------------------------------------------------- %%%

\section{Comparison of the numerical predictions with experimental data}
\label{sec:reliability_num_predictions}

Previous section aimed at validating the results of the numerical model with several reference solutions. The proposed numerical model showed a high accuracy with a relaxed stability condition compared to standard approaches. In other words, we verified that the differences due to numerical approximations of the mathematical model by the numerical one are minors. The numerical predictions of the numerical model are now compared with experimental data. The issue is to evaluate the physical approximations introduced by the definition of the mathematical model.

%%% ----------------------------------------------------------------------- %%%

\subsection{Experimental set--up}
\label{sec:exp_case_study}

The experimental data was obtained at the University of Savoie Mont Blanc, laboratory LOCIE (Laboratory of Optimisation of the Conception and Engineering of the Environment) in the frame of the research project HYGRO-BAT \cite{HYGRO-BA2014, Woloszyn2014}. The experimental benchmark provides data for one-dimensional heat and mass transfer in wood fiberboard.

\begin{figure}
  \centering
  \subfigure[\label{fig:Case4_Photo}]{\includegraphics[width=.45\textwidth]{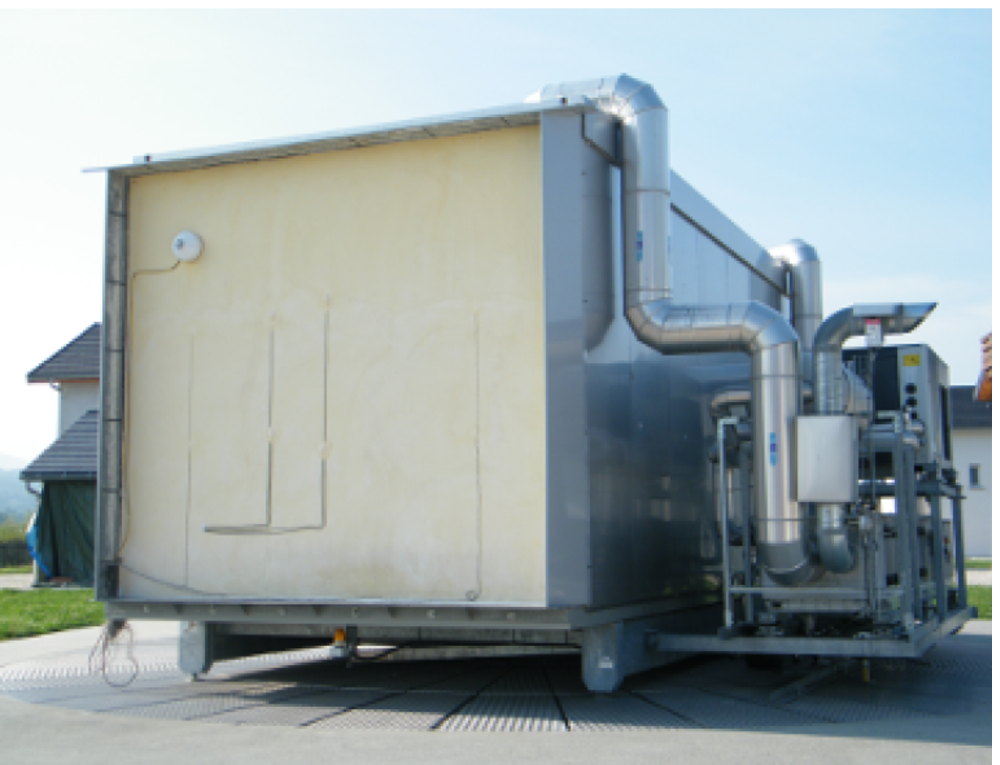}}
  \subfigure[\label{fig:Case4_Paroi}]{\includegraphics[width=.5\textwidth]{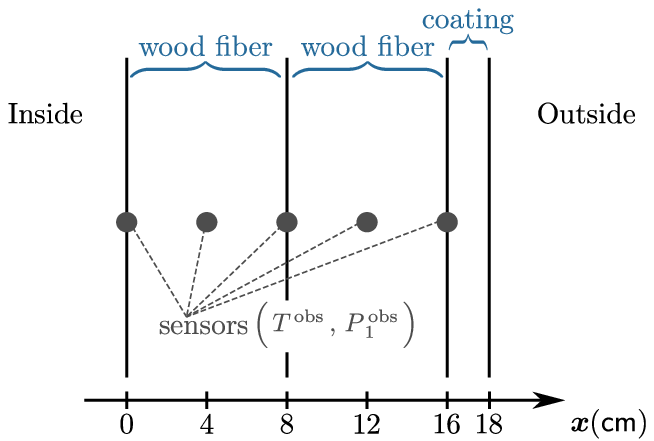}} 
  \caption{(a) Picture of the PASSYS cell and (b) illustration of the wall with the implementation of the sensors.}
  \label{fig:Case4_Passys}
\end{figure}

The site is located in \textsc{Le Bourget-du-Lac}, \textsc{France}. A picture of the PASSYS cell \cite{Jensen1989} is shown in Figure~\ref{fig:Case4_Photo}. The wall is composed of a $2 \ \mathsf{cm}$ outside coating and two $8 \ \mathsf{cm}$ wood fiberboards, as illustrated in Figure~\ref{fig:Case4_Paroi}. As mentioned in Section~\ref{sec:BC_In_IC_cond}, at the interface between the materials, the continuities of the air, moisture and heat fluxes are assumed. The temperature and relative humidity in the cell are controlled by the air handling unit. During the first $7 \ \mathsf{days}$, the inside temperature and relative humidity are set to approximately $24\ \mathsf{^{\,\circ}C}$ and $0.4\,$. After this period, the relative humidity is increased to $0.7\,$, while the temperature is maintained constant. The exterior boundary conditions are imposed by the climate.

Several \texttt{SHT75 Sensirion} sensors are evenly placed inside the material as shown in Figure~\ref{fig:Case4_Paroi}. They provide temperature and relative humidity measurements with an uncertainty $\sigma_{\,T}^{\,\mathrm{meas}} \egal 0.3 \ \mathsf{^{\,\circ}C}$ and $\sigma_{\,\phi}^{\,\mathrm{meas}} \egal 0.018\,$, respectively \cite{Sensirion2019}. The sensors are shifted (in $y$ and $z$ directions) to avoid perturbations of the transfer by themselves. In addition, the sensors at $x \egal \bigl\{\, 4\,,\,12\,\bigr\} \ \mathsf{cm}$ are inserted within the layer by perforating a hole, which is then fulfilled with wood fiber. According to this design, the uncertainty on the position of the sensors scales with $\sigma^{\,\mathrm{pos}} \egal 1 \ \mathsf{cm}$ for the sensors at $x \egal \bigl\{\, 4\,,\,12\,\bigr\} \ \mathsf{cm}$ and $\sigma^{\,\mathrm{pos}} \egal 0.5 \ \mathsf{cm}$ for the others. The difference air pressure between the inside and outside parts of the cell is measured using \texttt{Furness Controls FC0332} sensors with and uncertainty certified to $\sigma_{\,P}^{\,\mathrm{meas}} \egal 0.5 \ \mathsf{\%}\,$. Data measurement is taken with an interval of $5 \ \mathsf{min}\,$.

For the comparison, only the experimental data in the wood fiberboard are considered. It enables to avoid additional uncertainties from the climate-driven boundary conditions, particularly wind-driven rain (that was not measured) and from the unknown surface convective coefficient. Moreover, during the research project, only the material properties of the wood fiber have been fully determined in \cite{Rafidiarison2015, Vololonirina2014}. The properties used for this case are synthesized in Table~\ref{tab:mat_properties_phys_constant} with other physical constants. For the saturation pressure $P_{\,\mathrm{sat}}$, associating the vapor pressure $P_{\,1}$ and the relative humidity $\phi\,$, the following expression are considered:
\begin{subequations}\label{eq:Psat_relation}
  \begin{align}
  & \phi \egal \frac{P_{\,1}}{P_{\,\mathrm{sat}}\,\bigl(\, T \,\bigr)} \,, && P_{\,\mathrm{sat}}\,\bigl(\, T \,\bigr) \egal P_{\,\mathrm{sat}}^{\,\circ} \, \Biggl(\, \frac{T \moins T_{\,A}}{T_{\,B}} \,\Biggr)^{\,\alpha} \,, 
  && \\[4pt]
  & P_{\,\mathrm{sat}}^{\,\circ} \egal 997.3 \ \mathsf{Pa} \,, && T_{\,A} \egal 159.5 \ \mathsf{^{\,\circ}K} \,, \qquad T_{\,B} \egal 120.6 \ \mathsf{^{\,\circ}K} \,, \qquad \alpha \egal 8.275 \,.
  \end{align}
\end{subequations}
The total sequence of experimental data corresponds to $14 \ \mathsf{days}$ and the physical domain is defined for $x \, \in \, \bigl[\, 0 \,,\, L \,\bigr]$ with $L \egal 16 \ \mathsf{cm}\,$.

\begin{table}
\centering
\caption{Material properties of wood fiber and physical constants.}
\label{tab:mat_properties_phys_constant}
\smallskip
\setlength{\extrarowheight}{.3em}
\begin{tabular}[l]{@{} lc}
\hline
\hline
\multicolumn{2}{c}{Material properties of wood fiber.} \\
\hline 
\hline
Sorption curve $\unit{kg/m^{\,3}}$ & $w_{\,12} \egal 70.63 \ \phi^{\,3} - 73.6 \ \phi^{\,2} + 41.05 \ \phi \plus 0.26$ \\
Vapor permeability $\unit{s}$ & $k_{\,1} \egal 3.28 \e{-11} \plus 4.85 \e{-11} \ \phi$\\
Liquid permeability $\unit{s}$ & $k_{\,2} \egal 0$\\
Air permeability $\unit{m^{\,2}}$ & $k_{\,13} \egal 1.1 \e{-13}$\\
Porosity $\unit{-}$ & $\Pi \egal 0.9$ \\
Specific heat $\unit{J/(kg.K)}$ & $c_{\,0} \egal 1103$ \\
Dry--basis specific mass $\unit{kg/m^{\,3}}$ & $\rho_{\,0} \egal 146$ \\
\hline
\hline
\multicolumn{2}{c}{Physical constants} \\
\hline
\hline
Vapor gas constant $\unit{J/(kg.K)}$ & $R_{\,1} \egal 462$ \\
Dry air gas constant  $\unit{J/(kg.K)}$ & $R_{\,3} \egal 287$\\
Latent heat of vaporization $\unit{J/kg}$ & $r_{\,12} \egal 2.5 \e{6}$\\
Vapor specific heat $\unit{J/(kg.K)}$ & $c_{\,1} \egal 1870$ \\
Liquid specific heat $\unit{J/(kg.K)}$ & $c_{\,2} \egal 4180$ \\
Dry air specific heat $\unit{J/(kg.K)}$ & $c_{\,3} \egal 1006$ \\
Dynamic viscosity $\unit{Pa.s}$ & $\mu  \egal 1.8 \e{-5}$ \\
\hline
\hline
\end{tabular}
\end{table}

\begin{figure}
  \centering
  \subfigure[\label{fig:Case4_BC_Pv}]{\includegraphics[width=.45\textwidth]{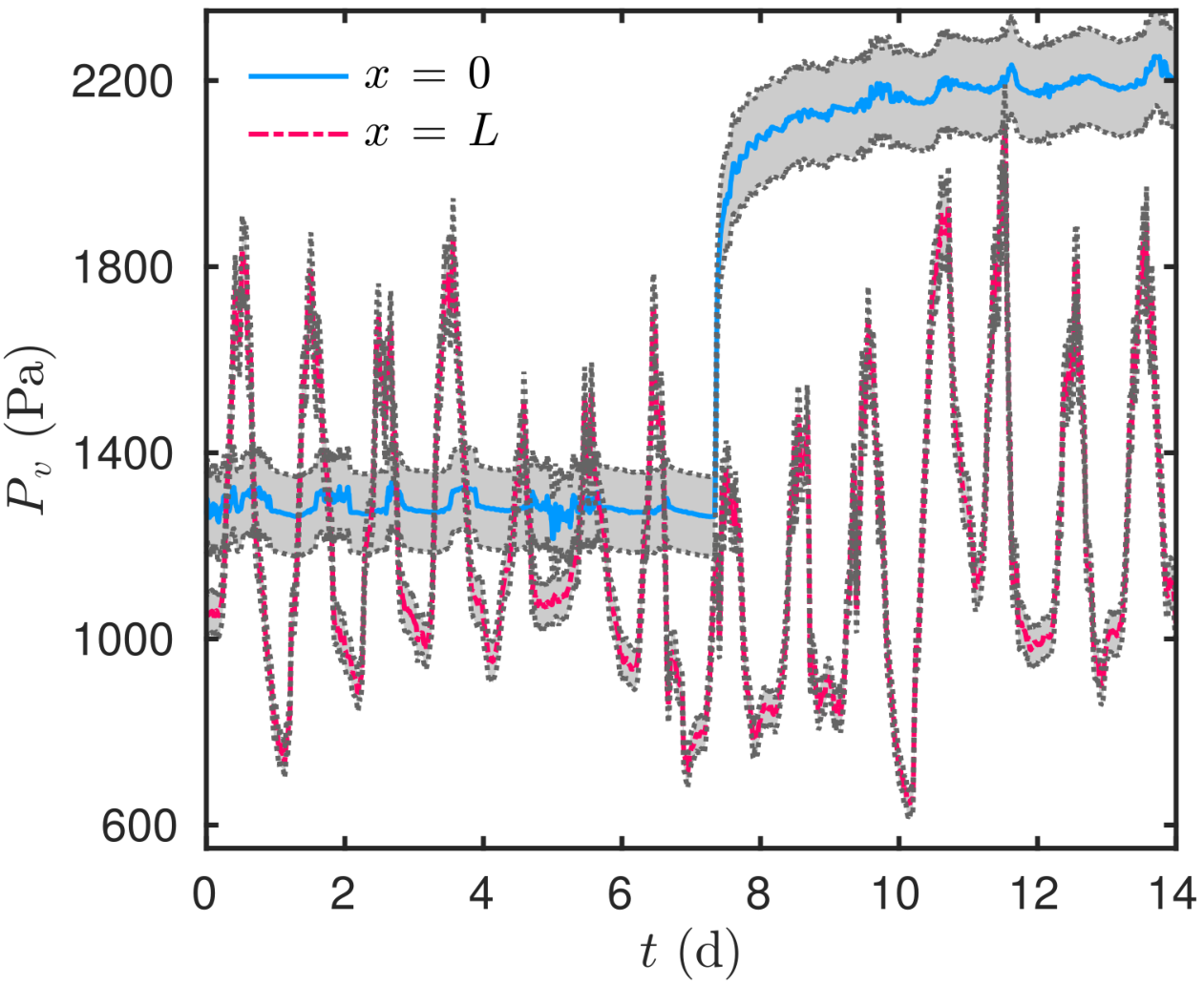}}
  \subfigure[\label{fig:Case4_BC_T}]{\includegraphics[width=.45\textwidth]{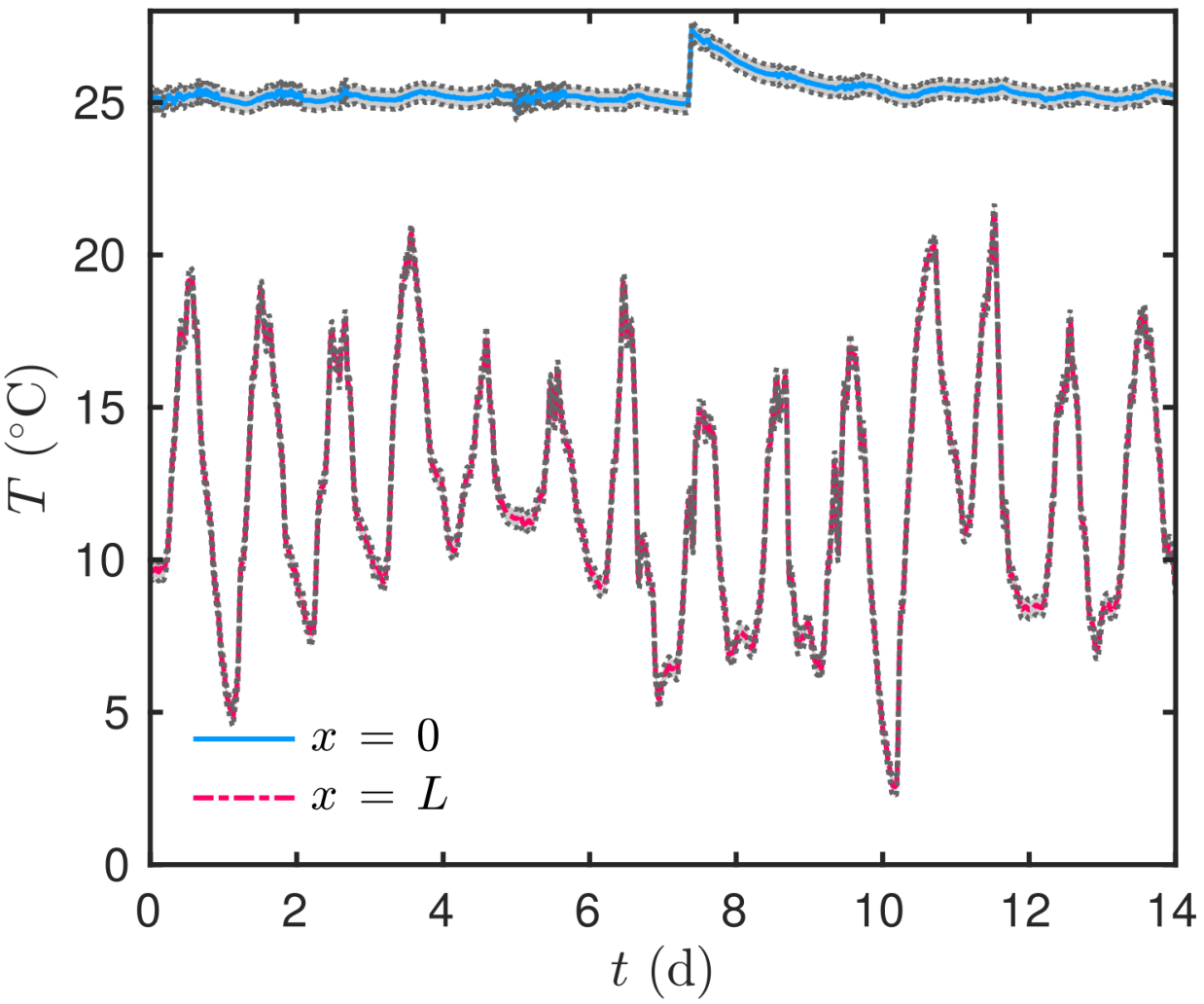}}
  \subfigure[\label{fig:Case4_BC_P}]{\includegraphics[width=.45\textwidth]{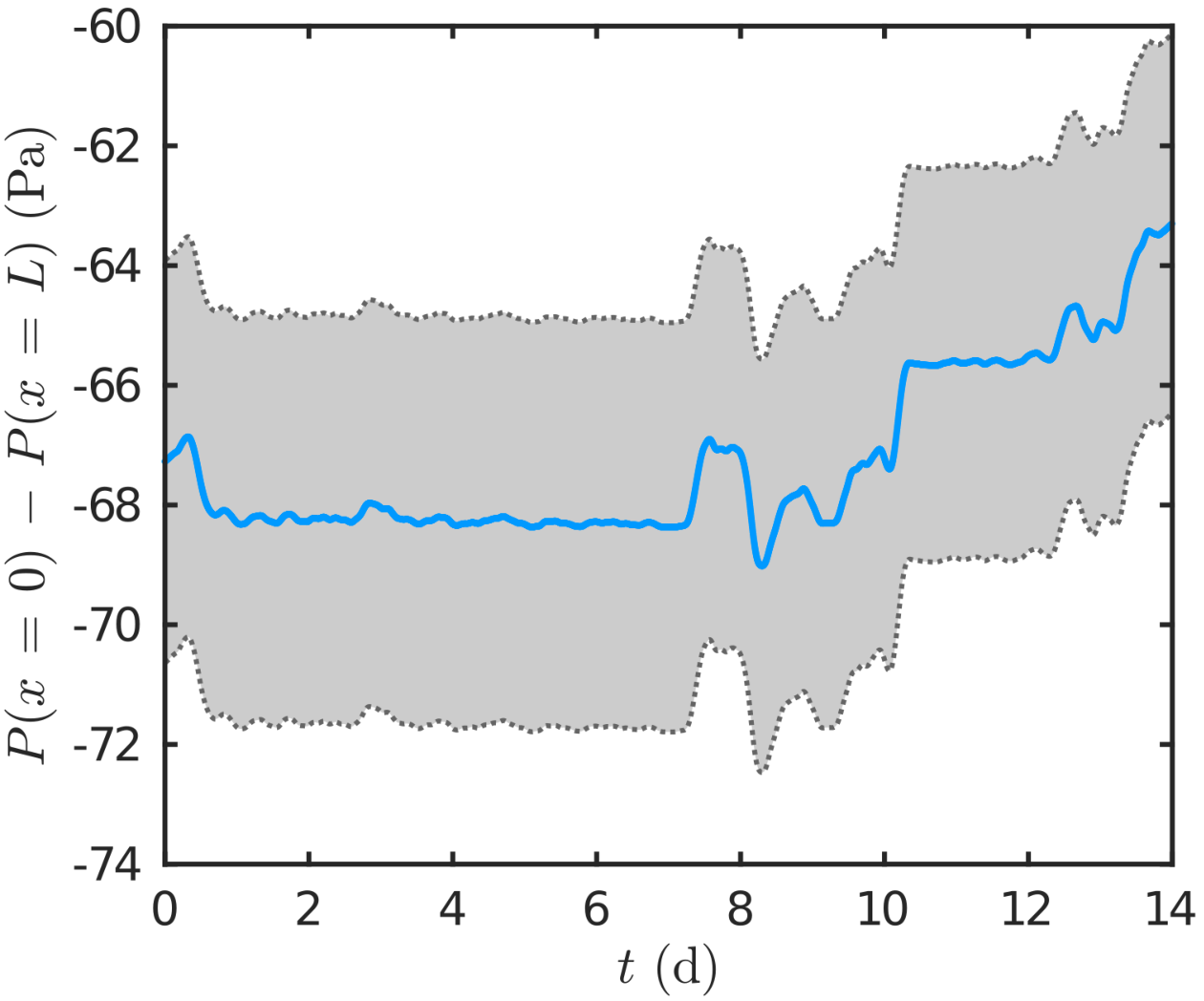}} 
  \caption{Measured boundary conditions used for the numerical model.}
  \label{fig:Case4_BC}
\end{figure}

Accordingly, for the three fields, \textsc{Dirichlet} boundary conditions are considered. Measurements provided by sensors at $x \egal 0 \ \mathsf{cm}$ and $x \egal 16 \ \mathsf{cm}$ are used to prescribe both temperature and vapor pressure at the boundaries. For the air pressure $P\,$, at $x \egal 0 \ \mathsf{cm}\,$, the pressure is imposed using measurements from a differential sensor and considering the reference pressure of $1 \ \mathsf{bar}\,$. At $x \egal 16 \ \mathsf{cm}\,$, the air pressure is assumed as constant in time and equal to the initial condition. This assumption is equivalent to consider a null air velocity at the interface between the coating and the wood fiber ($x \egal 16 \ \mathsf{cm}$). This hypothesis is very likely since the air permeability of the coating is $100$ times lower than wood fiber one. The boundary conditions with their measurement uncertainties are shown in Figure~\ref{fig:Case4_BC}. The increase of vapor pressure on the inside part of the PASSYS cell can be noticed. The maximal uncertainty due to sensor measurements reaches $100 \ \mathsf{Pa}$ for the vapor pressure and  $3 \ \mathsf{Pa}$ for the air pressure difference.

\begin{figure}
  \centering
  \subfigure[\label{fig:Case4_IC_Pv}]{\includegraphics[width=.45\textwidth]{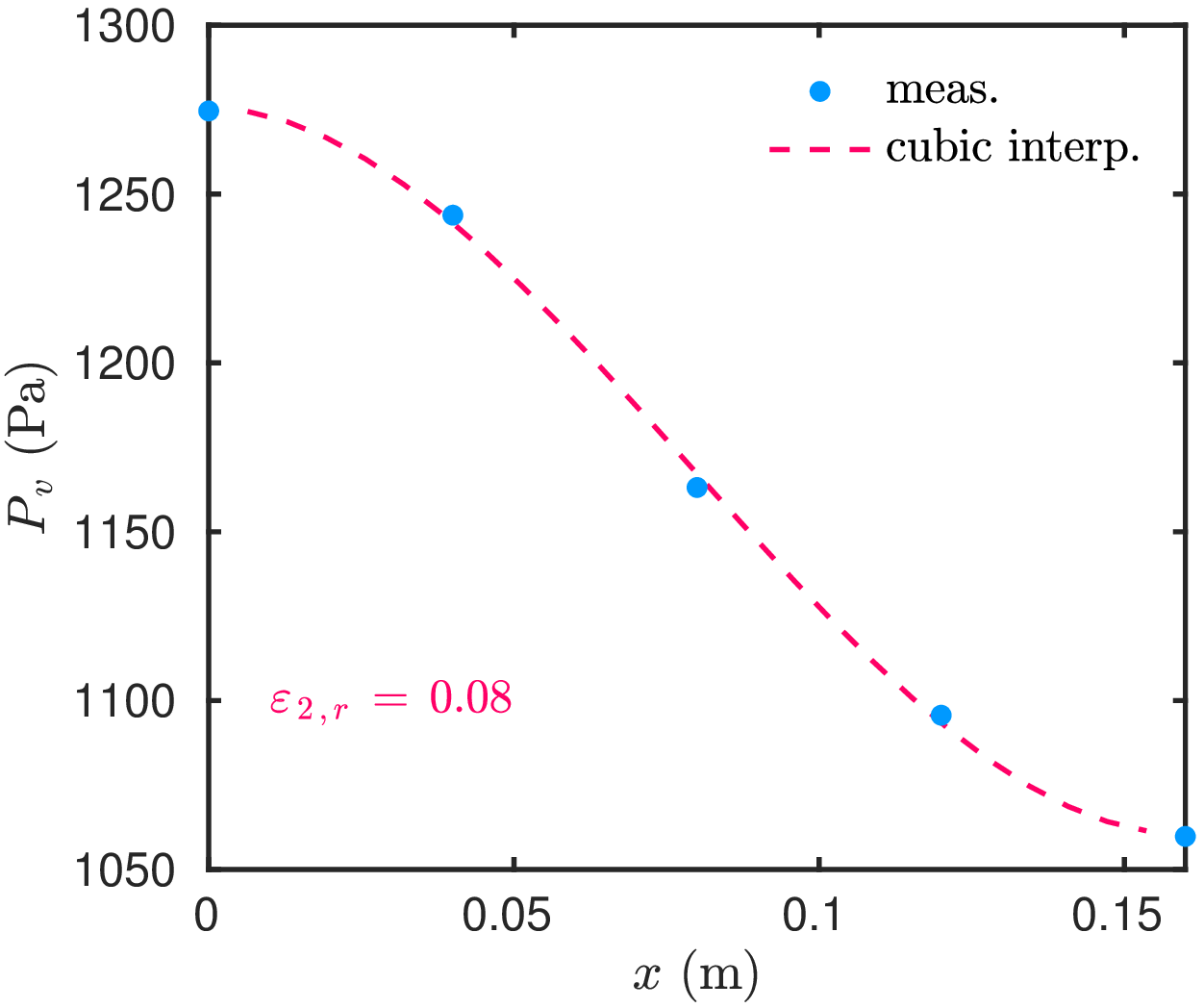}}
  \subfigure[\label{fig:Case4_IC_T}]{\includegraphics[width=.45\textwidth]{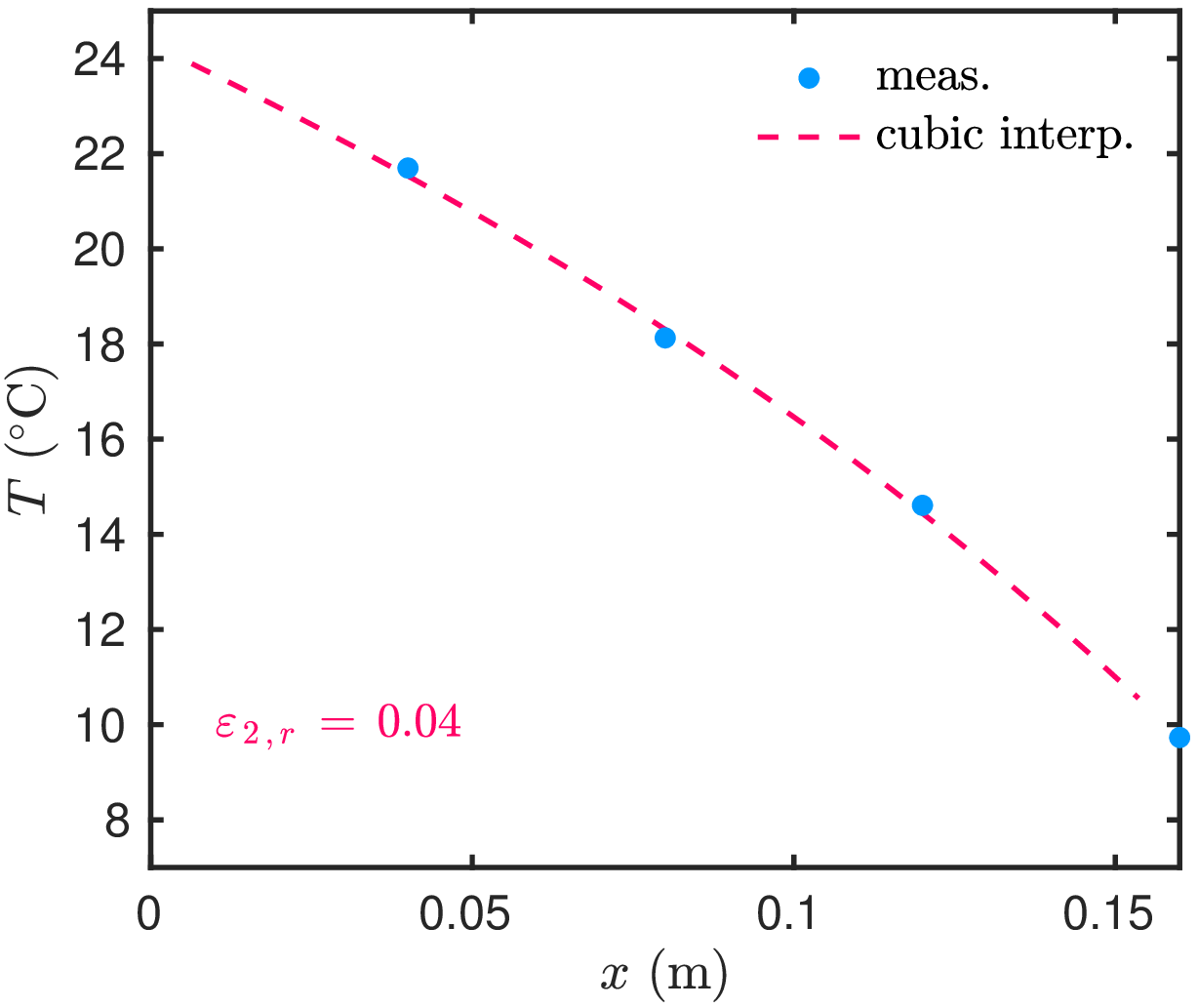}}
  \subfigure[\label{fig:Case4_IC_P}]{\includegraphics[width=.45\textwidth]{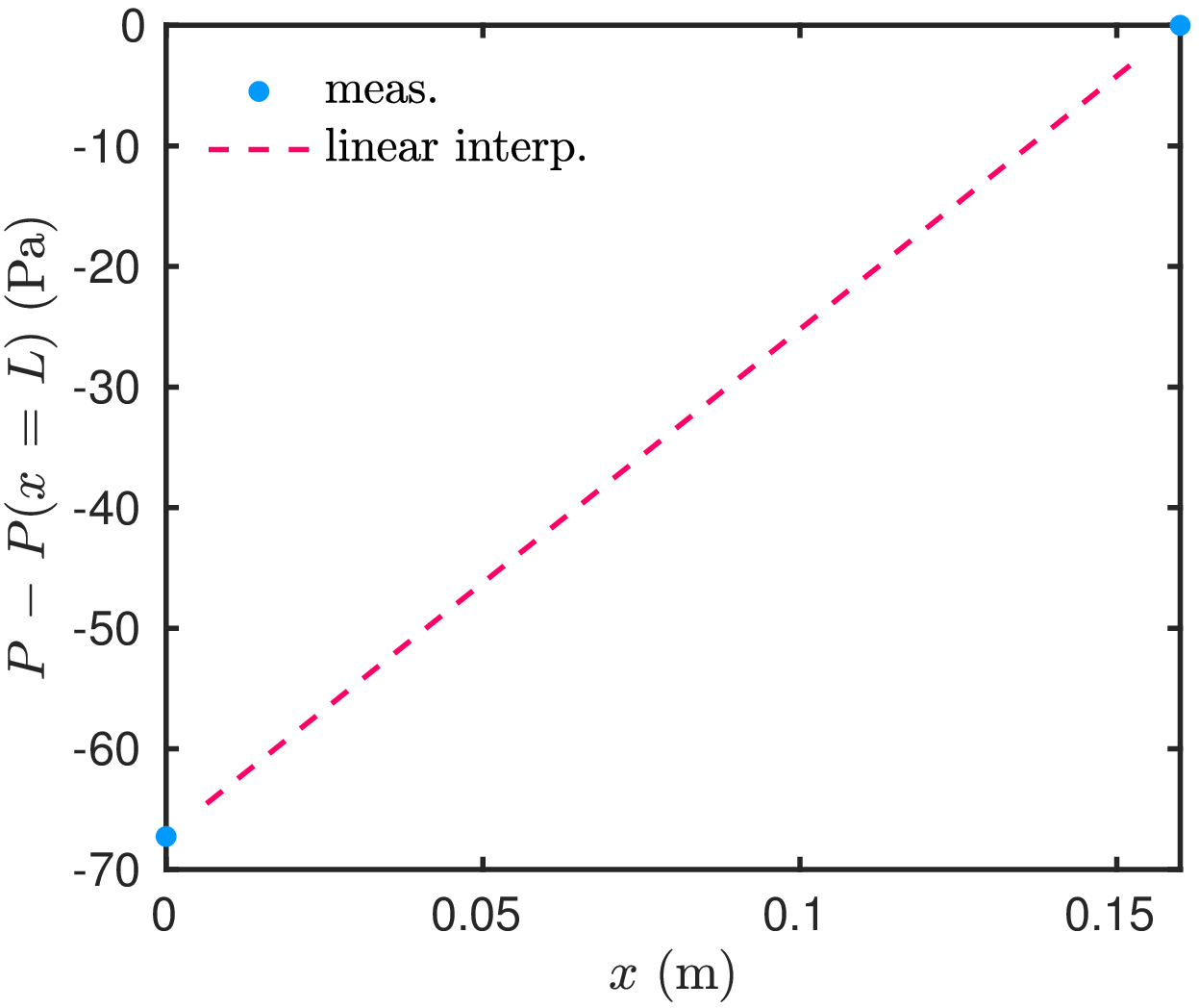}} 
  \caption{Measured and interpolated initial conditions used for the numerical model.}
  \label{fig:Case4_IC}
\end{figure}

It should be noted there was a long period (around $6$ months) between the installation of the sensors and the presented set of experimental data. However, there are no evidences that gradients of the temperature and the vapor pressure are well established within the wall. An open question to perform numerical predictions with the proposed model is the initial condition in the wall. Here, we used the third--order polynomial interpolation of the measurements at $x \egal \bigl\{\,0,\,4,\,8,\,12,\,16\,\bigr\}\, \mathsf{cm}\,$, at the first time instant. For the air pressure, the linear interpolation is supposed, assuming an established gradient of $P\,$. The initial condition of the numerical model are shown in Figure~\ref{fig:Case4_IC}. An absolute error $\varepsilon_{\,2}$ of $5.21 \ \mathsf{Pa}$ and $0.29 \ \mathsf{^{\,\circ}C}$ and a relative error  error $\varepsilon_{\,2_,,r}$ of $0.08$ and $0.04$ are obtained between the experimental data and the interpolated initial profiles of vapor pressure and temperature, respectively. Some complementary information on the experimental design is given in \cite{Gasparin2018}.

%%% ----------------------------------------------------------------------- %%%

\subsection{Discussion on some hypotheses regarding the mathematical model}

Before comparing the numerical predictions with the experimental observations, some possible hypotheses on the formulation of the mathematical model are discussed. The analysis is based on the relative error $\varepsilon_{\,2\,,r}$ defined in Section~\ref{sec:efficiency_of_num_models} and its probability density functions computed using the kernel smoothing function for the extrapolation \cite{Hill1985}.

\begin{figure}
  \centering
  \subfigure[\label{fig:Case4_hyp_pdf_eps2r_1}]{\includegraphics[width=.45\textwidth]{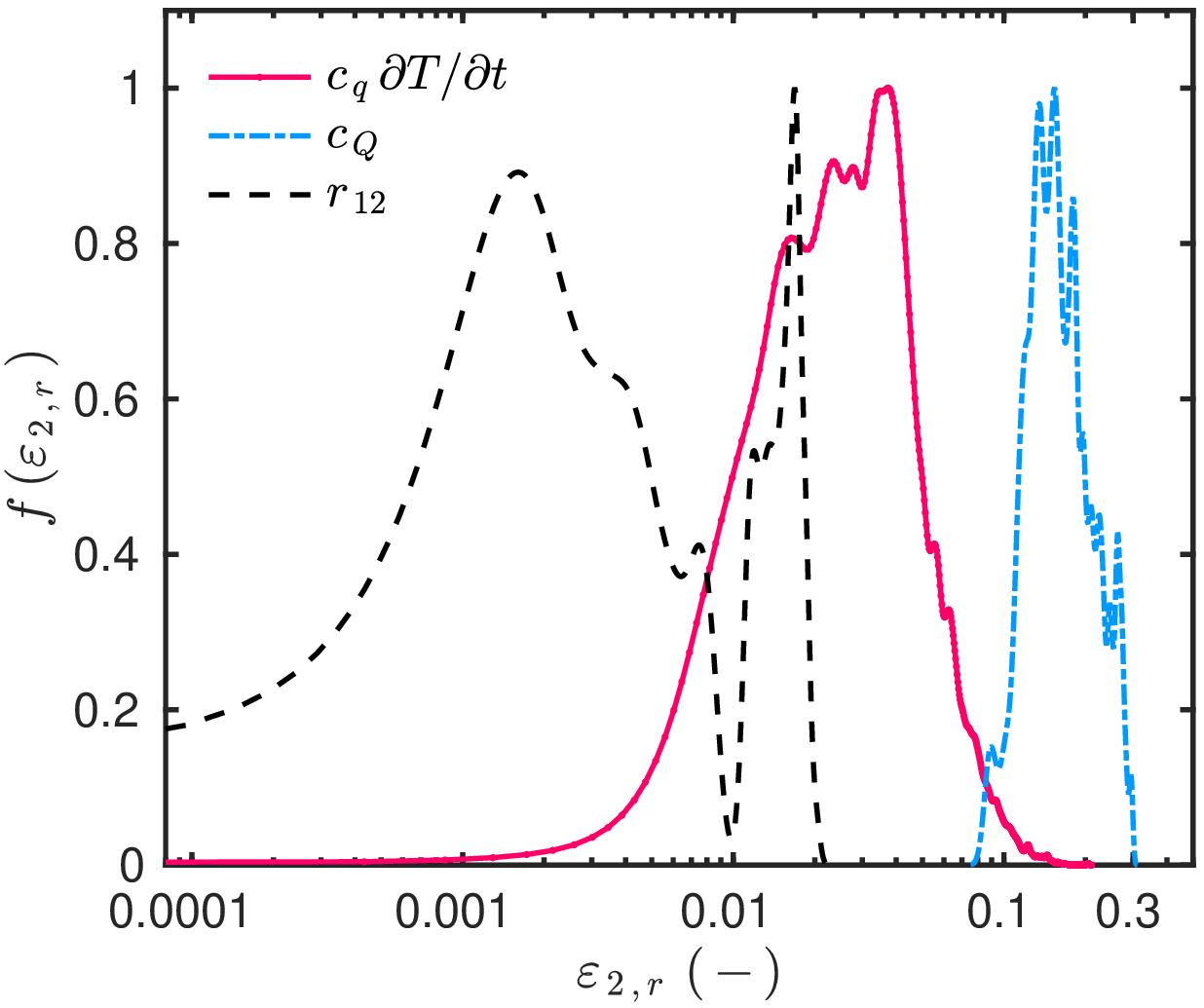}}
  \subfigure[\label{fig:Case4_hyp_pdf_eps2r_2}]{\includegraphics[width=.45\textwidth]{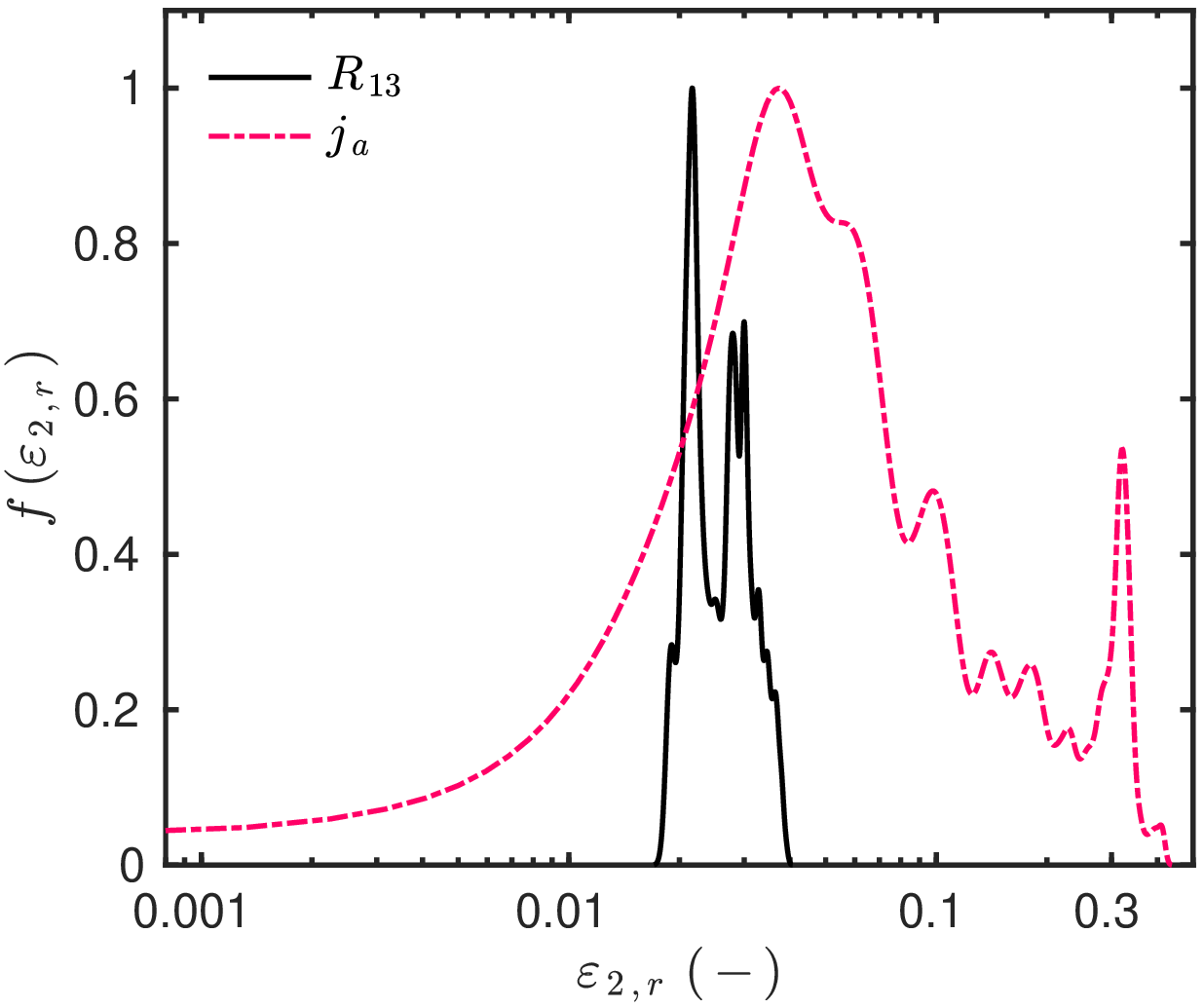}}
  \caption{Probability density function of the relative error for some hypothesis in the formulation of the mathematical model (a) for the latent heat of vaporization $r_{\,12}\,$, for the volumetric heat capacity $c_{\,q}\,$ and for the transient term in the right hand side of the heat Equation~\eqref{eq:phys_dim_H} and (b) for the for the constant gas law $R_{\,13}\,$  and the gradient of enthalpy.}
  \label{fig:Case4_hyp_pdf_eps2r}
\end{figure}

First of all, the investigations concern the constant gas law $R_{\,13}\,$. The exact definition is provided in Eq.~\eqref{eq:definition_R13}. Since in building physics application the dry air pressure is higher than the vapor pressure $P_{\,3} \, \gg \, P_{\,1}\,$, the constant gas law can be approximated by $R_{\,13} \, \approx \, R_{\,3}\,$. The probability of the relative error between both expressions is shown in Figure~\ref{fig:Case4_hyp_pdf_eps2r}. One may conclude that for this case study an error lower than $\sim \,4\,\%$ is committed using the approximation of $R_{\,13}\,$.

Another approximation may be taken in the expression of the latent heat of vaporization $r_{\,12}$. The expression provided by \textsc{Kelvin}'s law in Eq.~\eqref{eq:definition_r12} can be approximated by the vaporization latent heat at the reference temperature $r_{\,12} \, \approx \, r_{\,12}^{\,\circ}\,$.  
As shown in Figure~\ref{fig:Case4_hyp_pdf_eps2r}, the relative error regarding the effect of temperature on $r_{\,12}$ is lower than $10^{\,-2}\,$. Thus, the approximation might be also satisfactory for this case study.

In Eq.~\eqref{eq:phys_dim_H}, the volumetric specific heat $c_{\,q}$ is defined by summing up the capacity of the dry material and that of each constituting phase (liquid, vapor and dry air). Commonly in literature, as for instance in \cite{Belleudy2015}, the thermal capacity is approximated by the sum of the one for the dry--basis material and the one for the liquid phase $c_{\,q} \, \approx \, c_{\,0} \, \rho_{\,0} \plus c_{\,2} \, \w_{\,12}\,$. It should be remarked that in the approximation, the liquid water content is provided by the water sorption curve $w_{\,12}\,$ . In the exact expression, the liquid water content is given by the expression of $w_{\,2}$ in Eq.~\eqref{eq:liquid_volumetric_mass}. As noticed in Figure~\ref{fig:Case4_hyp_pdf_eps2r}, the relative error reaches $\sim \, 30\,\%$ indicating that the approximation may not provide reliable results for this case study.

The significance of the two transient terms $\displaystyle \biggl(\, - \ r_{\,12} \, c_{\,qv} \, \pd{P_{\,1}}{t} \plus r_{\,12} \, c_{\,qs} \, \pd{\sigma}{t} \,\biggr) $ in the right--hand side of the heat Equation~\eqref{eq:phys_dim_H} is studied. The relative error on the right--hand side is computed for both cases. As shown in Figure~\ref{fig:Case4_hyp_pdf_eps2r}, the maximum error reaches $\sim \,21\,\%$ with an average value of $\sim \,4\,\%\,$. Moreover, the probability of the relative error has an important standard deviation. It indicated that the approximation may impact locally the reliability of the numerical predictions. But overall, the average prediction of the temperature may be satisfactory.

\begin{figure}
  \centering
  \subfigure[\label{fig:Case4_hyp_I1}]{\includegraphics[width=.45\textwidth]{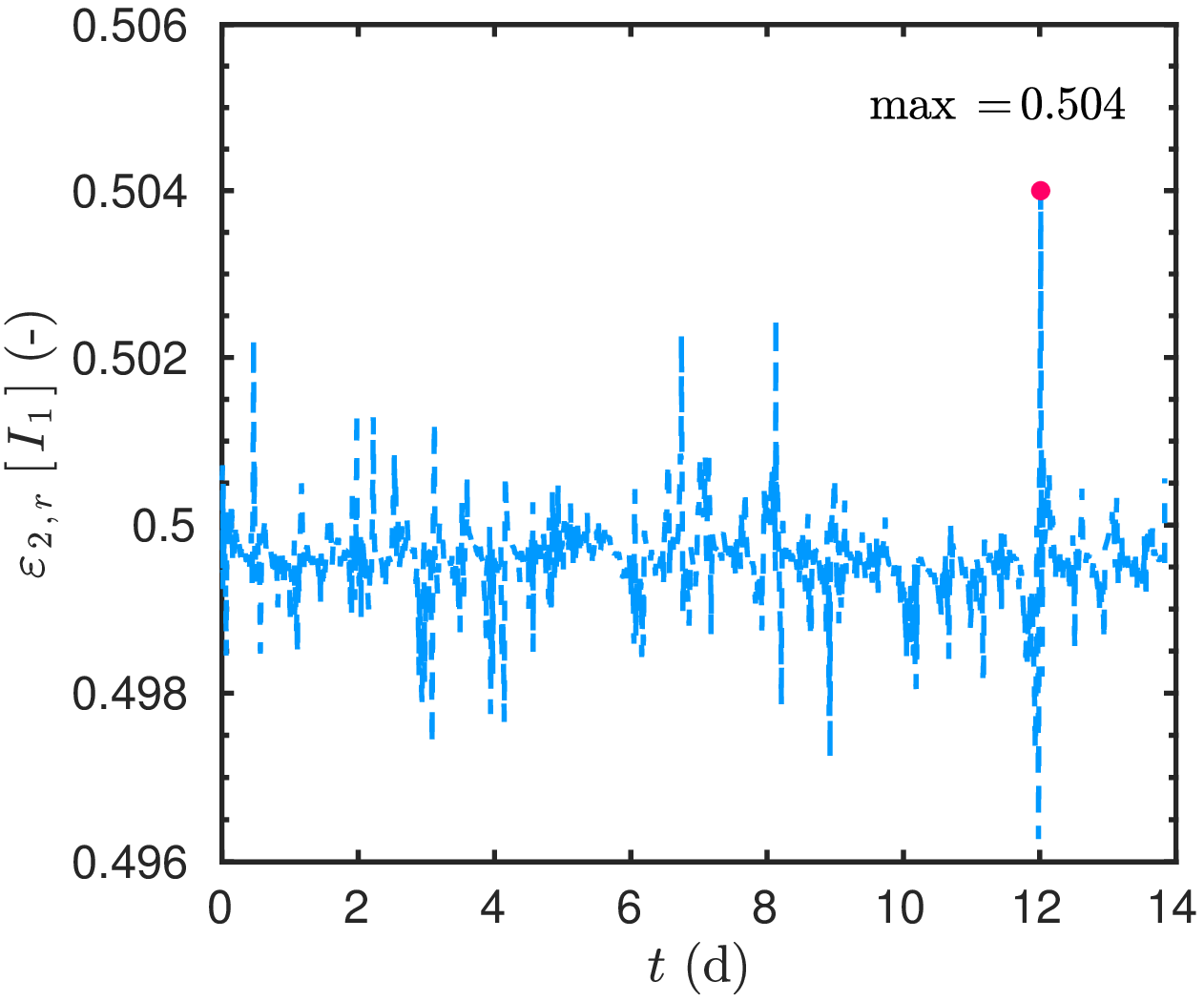}}
  \subfigure[\label{fig:Case4_I1_ft}]{\includegraphics[width=.45\textwidth]{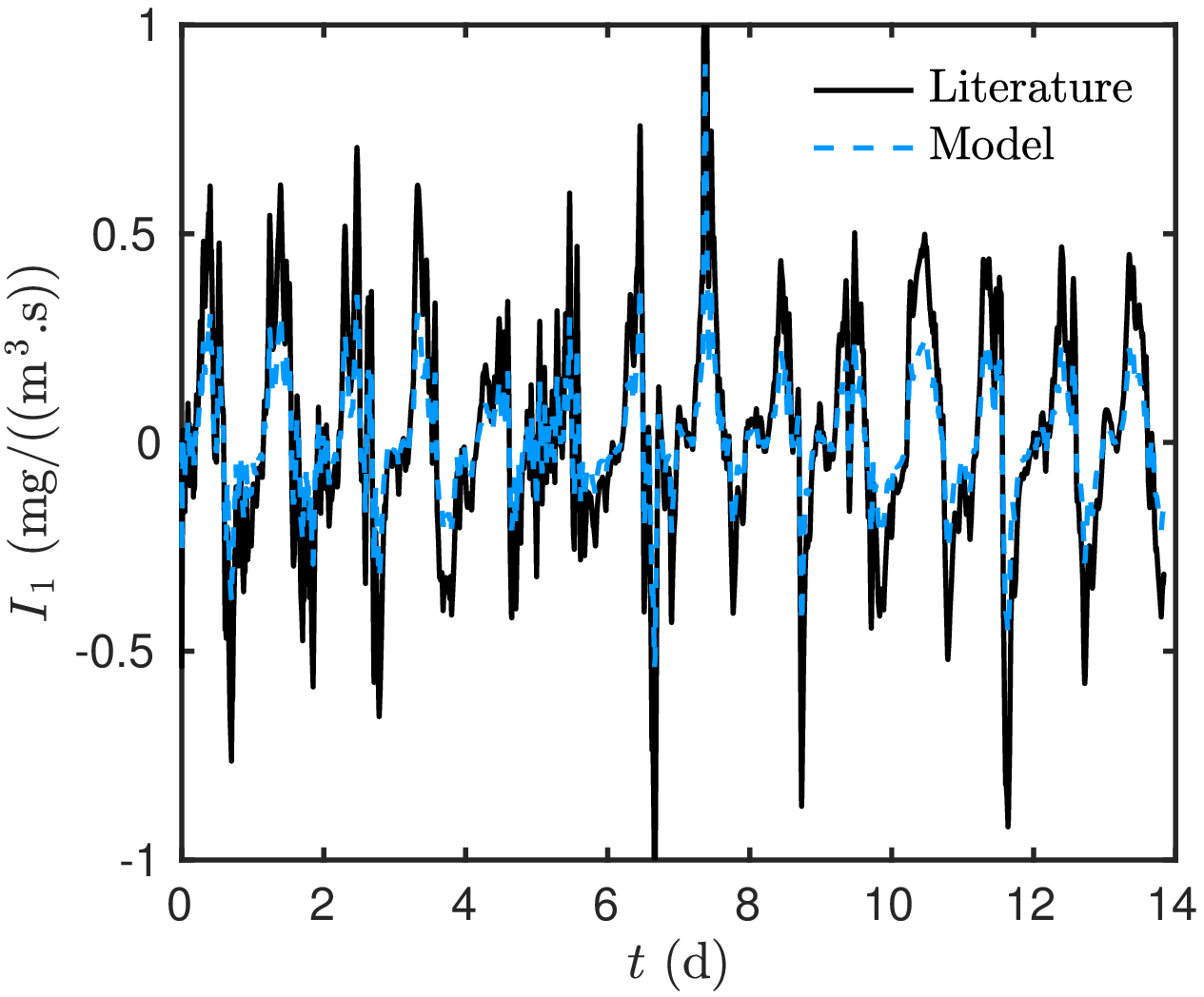}}
  \caption{Importance of the approximation of the volumetric vapor source $I_{\,1}\,$.}
\end{figure}

An important hypothesis in the mathematical model concerns the expression of the volumetric vapor source as presented in Section~\ref{sec:source_I2}. In \cite{Luikov1966}, \textsc{Luikov} neglects the time variation of the vapor mass. The importance of this approximation is evaluated by comparing both literature expressions:
\begin{align*}
  && \text{Exact expression:} & \quad I_{\,1}   \, = \, \pd{\w_{\,1}}{t} \plus \div \, j_{\,c\,,\,1}  \,, \\[3pt]
  && \text{Approximation from the literature:} & \quad I_{\,1}  \, \approx \, \div \, j_{\,c\,,\,1} \,.
\end{align*}
The relative error is shown in Figure~\ref{fig:Case4_hyp_I1} and is of the order of $50\,\%\,$. Therefore, the importance of the transient term is non-negligible for this case. Important discrepancies are remarked in Figure~\ref{fig:Case4_I1_ft} between the two approximations.

The last hypothesis involves the gradient of enthalpy in the heat Equation~\eqref{eq:phys_dim_H}. As mentioned in Section~\ref{sec:extension_nonlinear}, the term $\sum_{i=1}^{3} \,\grad \, \bigl(\, c_{\,i} \, T \,\bigr) \scal \boldsymbol{j}_{\,c\,,\,i} $ is included into the advective heat flux by freezing the convective flux at each time iteration. The importance of this term is investigated by comparing the two following expressions:
\begin{align*}
  && \text{Exact expression:} & \quad j_{\,a}  \, = \, a_{\,q} \, T \plus \sum_{i=1}^{3} \, c_{\,i} \, j_{\,c\,,\,i} \, T \,, \\[3pt]
  && \text{Approximation from the literature:} & \quad j_{\,a} \, \approx \, a_{\,q} \, T  \,.
\end{align*}
In Figure~\ref{fig:Case4_hyp_pdf_eps2r_2}, a relative error of $40\,\%$ on the advective heat flux is committed with this approximation. The term with the gradient of enthalpy is particularly important when the vapor pressure increases within the wall, \ie, for $t \, \in \, \bigl[\, 8 \,,\,14 \,\bigr] \ \mathsf{days}$.

%%% ----------------------------------------------------------------------- %%%

\subsection{Results of the comparison and discussion}

If the experimental data was already used in previous works \cite{Rouchier2017, HYGRO-BA2014, Woloszyn2014}, only the uncertainties of the sensor measurement were presented. Here, we propose to evaluate the propagation of uncertainties due to sensor measurement and sensor location. The total uncertainty on measurement of temperature $T$ and vapor pressure $P_{\,1}$ are denoted by $\sigmaT$ and $\sigmaPv\,$, respectively. At the measurement point $x_{\,0} \egal \bigl\{\, 4 \,,\, 8 \,,\, 12\,\bigr\} \ \mathsf{cm}$, they are computed by:
\begin{align*}
  \sigmaT^{\,2} \,(\,x_{\,0} \,,\,t \,) & \egal \Bigl(\, \sigmaT^{\,\mathrm{meas}} \,\Bigr)^{\,2} \plus \Biggl(\, \pd{T}{x} \,\biggl|_{\,x \egal x_{\,0}} \; \sigma^{\,\mathrm{pos}} \,\Biggr)^{\,2}  \,, \\[4pt]
  \sigmaPv^{\,2} \,(\,x_{\,0} \,,\,t \,) & \egal \Bigl(\,  \sigmaPv^{\,\mathrm{meas}} \,\Bigr)^{\,2} \plus \Biggl(\, \pd{P_{\,1}}{x}\,\biggl|_{\,x \egal x_{\,0}} \;\sigma^{\,\mathrm{pos}} \,\Biggr)^{\,2}\,,
\end{align*}
where, as mentioned in Section~\ref{sec:exp_case_study}, $\sigma^{\,\mathrm{pos}}$ is the sensor position uncertainty. The quantities $\sigmaT^{\,\mathrm{meas}}$ and $\sigmaPv^{\,\mathrm{meas}}$ are the sensor measurement uncertainties. For the temperature, $\sigmaT^{\,\mathrm{meas}}$  is directly given by the sensor manufacturer. For the vapor pressure, $\sigmaPv^{\,\mathrm{meas}}$ is evaluated using Eq.~\eqref{eq:Psat_relation}:
\begin{align*}
  \sigmaPv^{\,\mathrm{meas}} \egal P_{\,\mathrm{sat}} \, \bigl(\, T \,\bigr) \; \sigma_{\,\phi}^{\,\mathrm{meas}} \plus \phi \; \frac{P_{\,\mathrm{sat}}\,\bigl(\, T \,\bigr)}{T \moins T_{\,A}} \; \sigmaT^{\,\mathrm{meas}} \,,
\end{align*}
where $\sigma_{\,\phi}^{\,\mathrm{meas}} $ is directly given by the sensor manufacturer. The quantities $\displaystyle \pd{T}{x}\,\biggl|_{\,x \egal x_{\,0}} $ and $\displaystyle \pd{P_{\,1}}{x}\,\biggl|_{\,x \egal x_{\,0}} $ are evaluated using the results of the numerical model at the points of probes $x_{\,0} \egal \bigl\{\, 4 \,,\, 8 \,,\, 12\,\bigr\} \ \mathsf{cm}$.

Using the results obtained in Section~\ref{sec:validation_numerical_model} on the efficiency of the proposed numerical models in terms of accuracy and reduced computational time, the solution of the present case study is solved using NM~$3$c. The discretisation parameters are set to $\Delta x^{\,\star} \egal 4 \e{\,-2}$ and $\Delta t^{\,\star} \egal 5 \, \e{-4}\,$, corresponding to $\Delta x \egal 6.4 \ \mathsf{mm}$ and $\Delta t \egal 1.8 \ \mathsf{s}$  in physical dimensions. According to the numerical values of the parameters, we have the following stability conditions:
\begin{align*}
  & \text{Eq.}~\eqref{eq:phys_dim_A} \ : \ \Delta t \ \leqslant \  0.6 \ \mathsf{s}  \,, \\
  & \text{Eq.}~\eqref{eq:phys_dim_H} \ : \ \Delta t \ \leqslant \ 1.8  \ \mathsf{s}  \,, \\
  & \text{Eq.}~\eqref{eq:phys_dim_A} \ : \ \Delta t \ \leqslant \ 0.03 \ \mathsf{s}  \,.
\end{align*}
Thus, the most restrictive conditions are due to to Eq.~\eqref{eq:phys_dim_A}. It can be remarked that the stability conditions are relaxed by a factor of $\sim \, 50$ due to the efficiency of the proposed innovative method. It enables to save important computational efforts. The model needs $117 \ \mathsf{min}$ to compute the solution with the given discretisation parameters. It corresponds to a ratio CPU time / time horizon of simulation of $8.3 \ \mathsf{min./day}\,$ for the proposed numerical model and a ratio of $139.3 \ \mathsf{min./day}$ for the standard approach based on \Eu ~explicit scheme with central finite-differences. In other words, the NM~$3$c is sixteen faster to predict the physical phenomena during occurring during one day. It is also important to remark that with the discretisation parameter $\Delta x^{\,\star} \egal 4 \e{\,-2}\,$, the solution is computed only at $24$ points in the discrete space domain. The fields at the points of observation are computed using the exact interpolation equations for solutions $u\,$, $v\,$ and $w$ as detailed in Section~\ref{sec:spat_discretisation}.

\begin{figure}
  \centering
  \subfigure[$x \egal 4 \ \mathsf{cm}$ \label{fig:Case4_Pvx4_ft}]{\includegraphics[width=.45\textwidth]{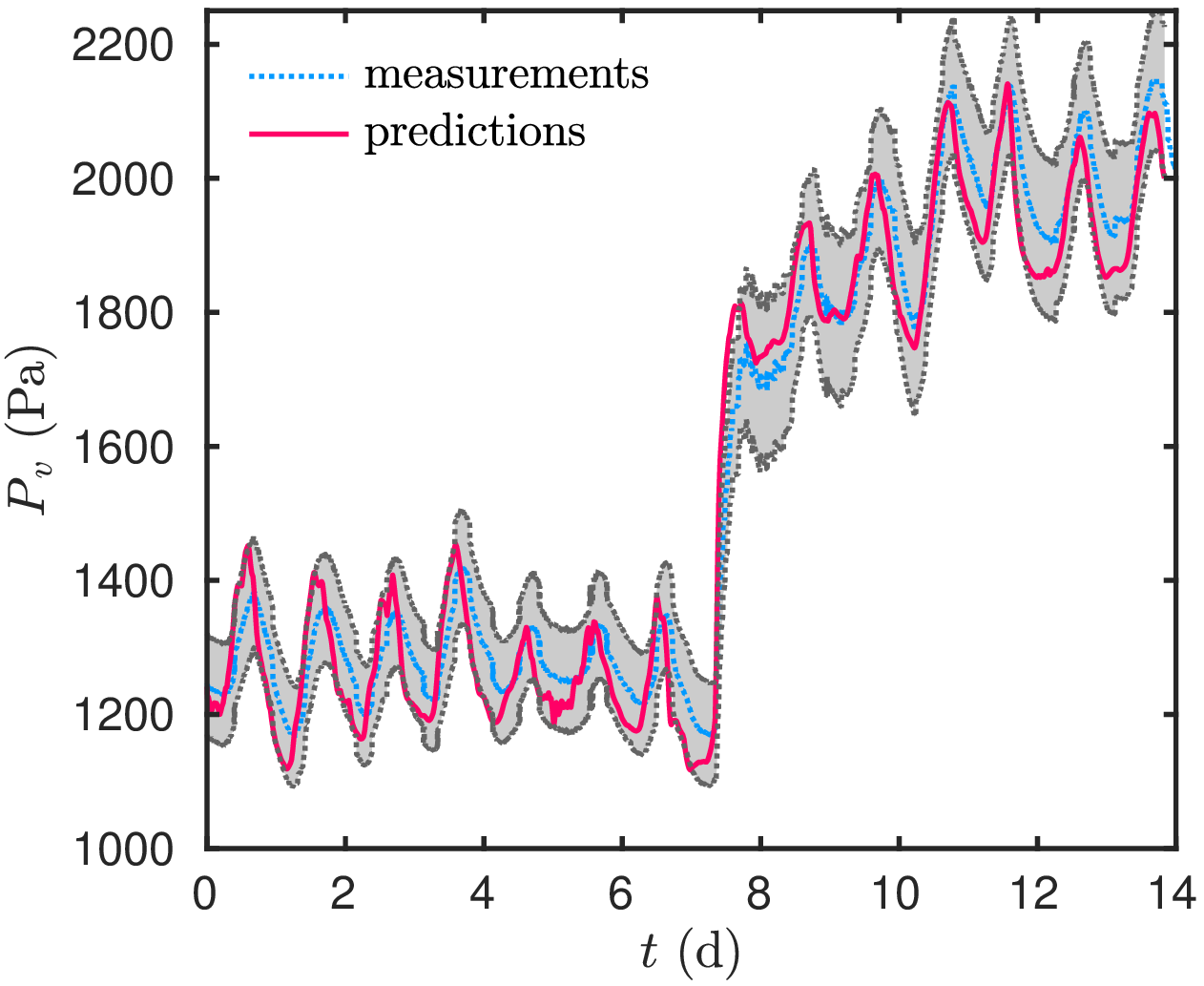}}
  \subfigure[$x \egal 4 \ \mathsf{cm}$ \label{fig:Case4_Tx4_ft}]{\includegraphics[width=.45\textwidth]{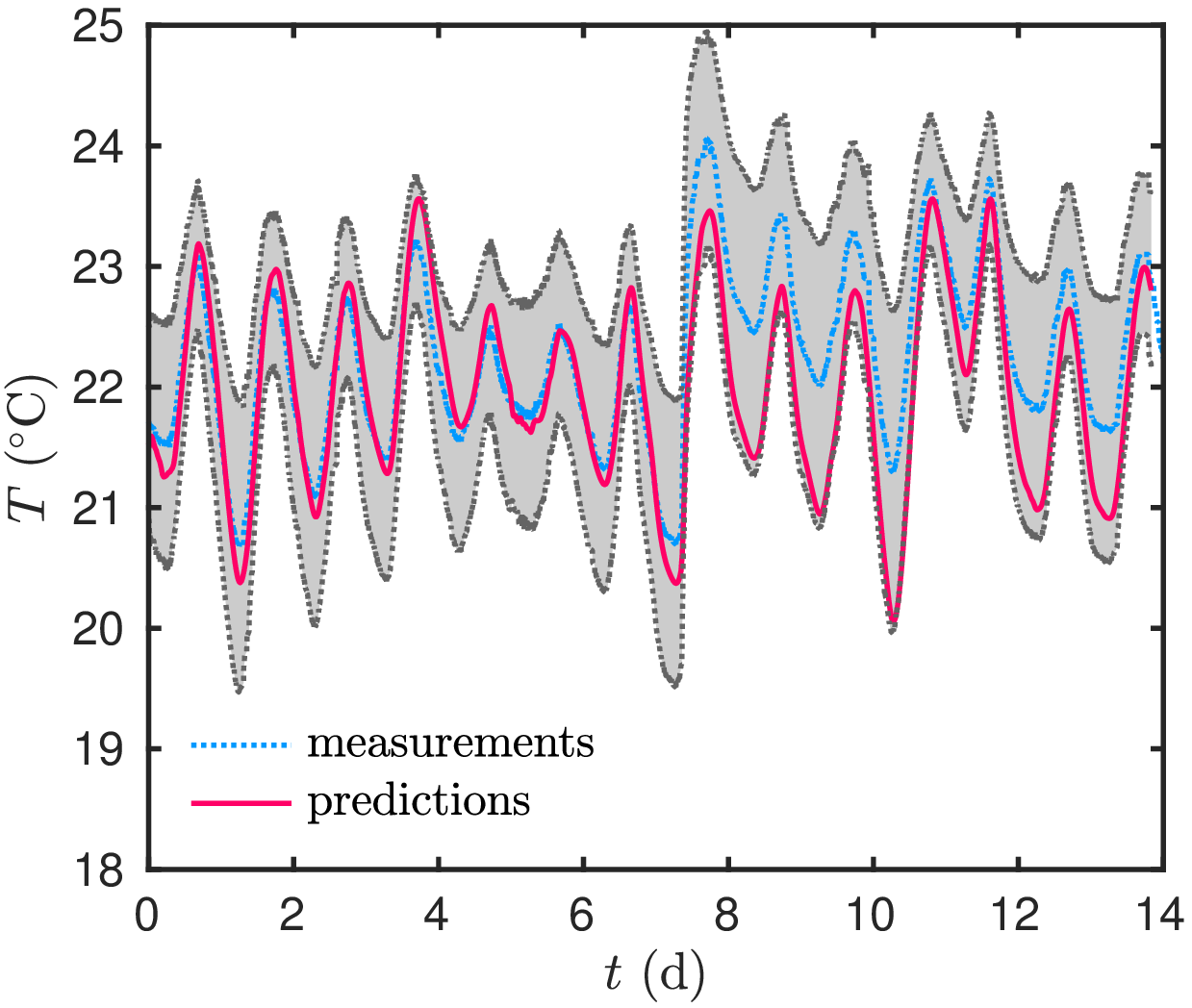}}
  \subfigure[$x \egal 8 \ \mathsf{cm}$ \label{fig:Case4_Pvx8_ft}]{\includegraphics[width=.45\textwidth]{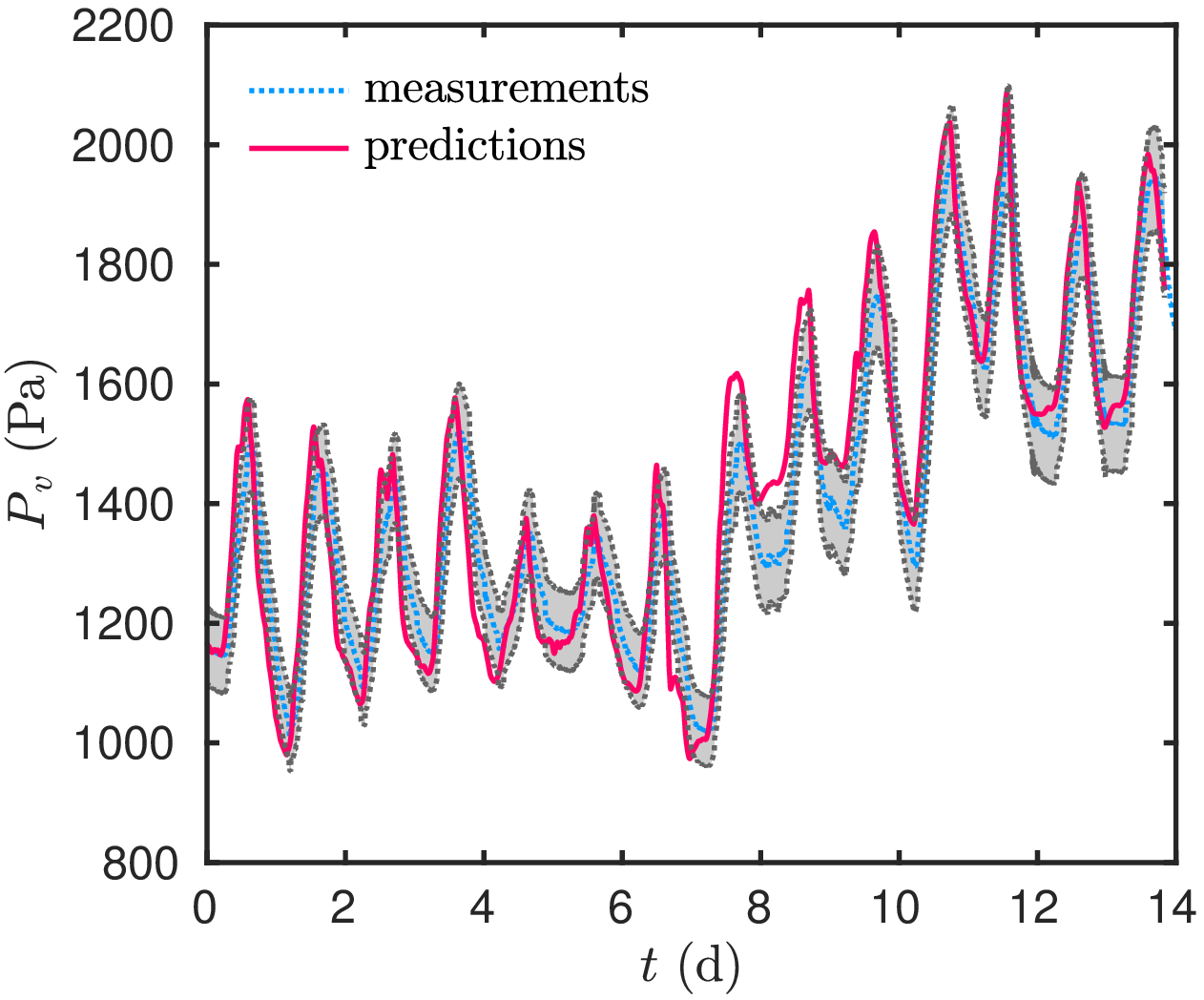}}
  \subfigure[$x \egal 8 \ \mathsf{cm}$ \label{fig:Case4_Tx8_ft}]{\includegraphics[width=.45\textwidth]{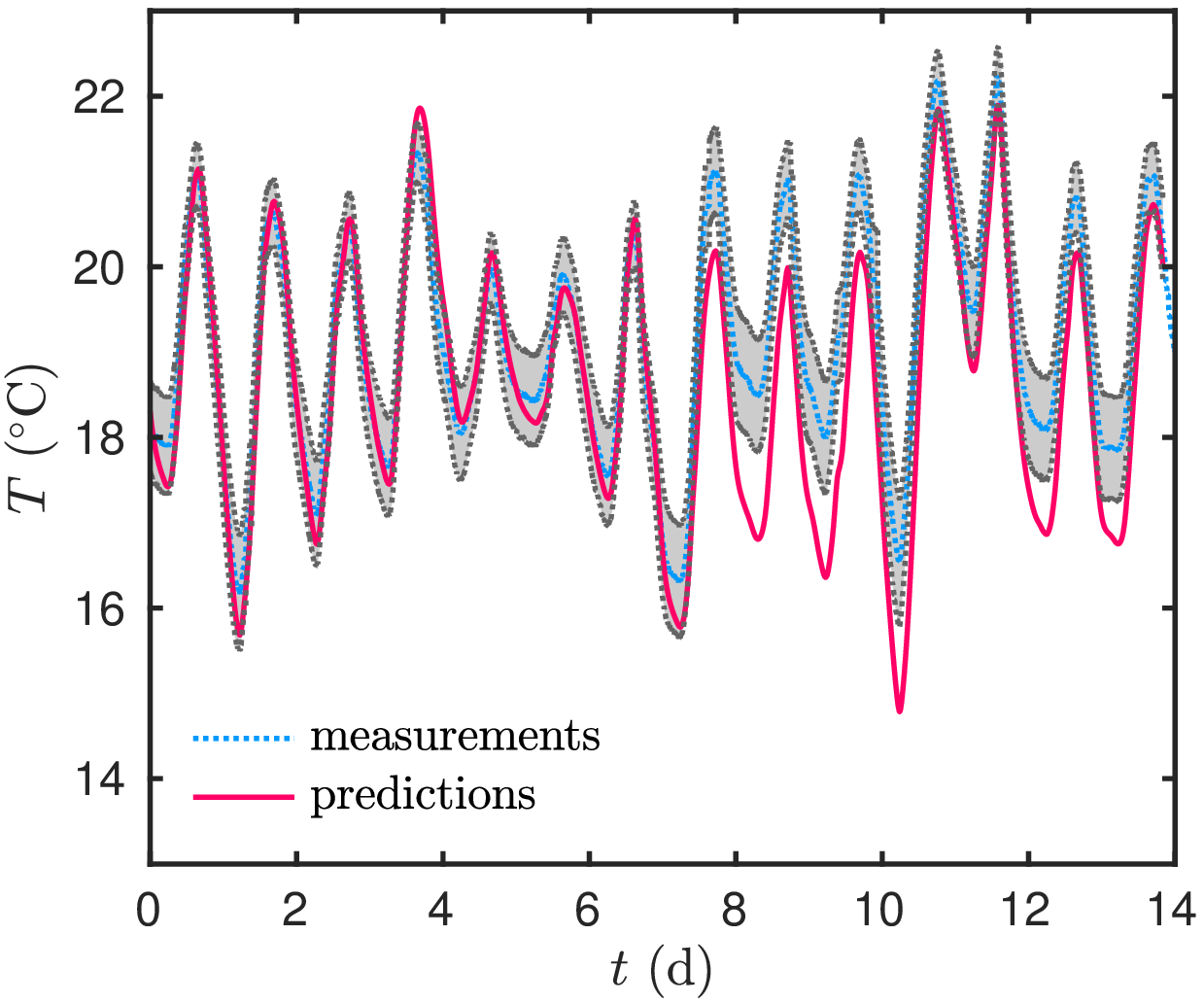}}
  \subfigure[$x \egal 12 \ \mathsf{cm}$\label{fig:Case4_Pvx12_ft}]{\includegraphics[width=.45\textwidth]{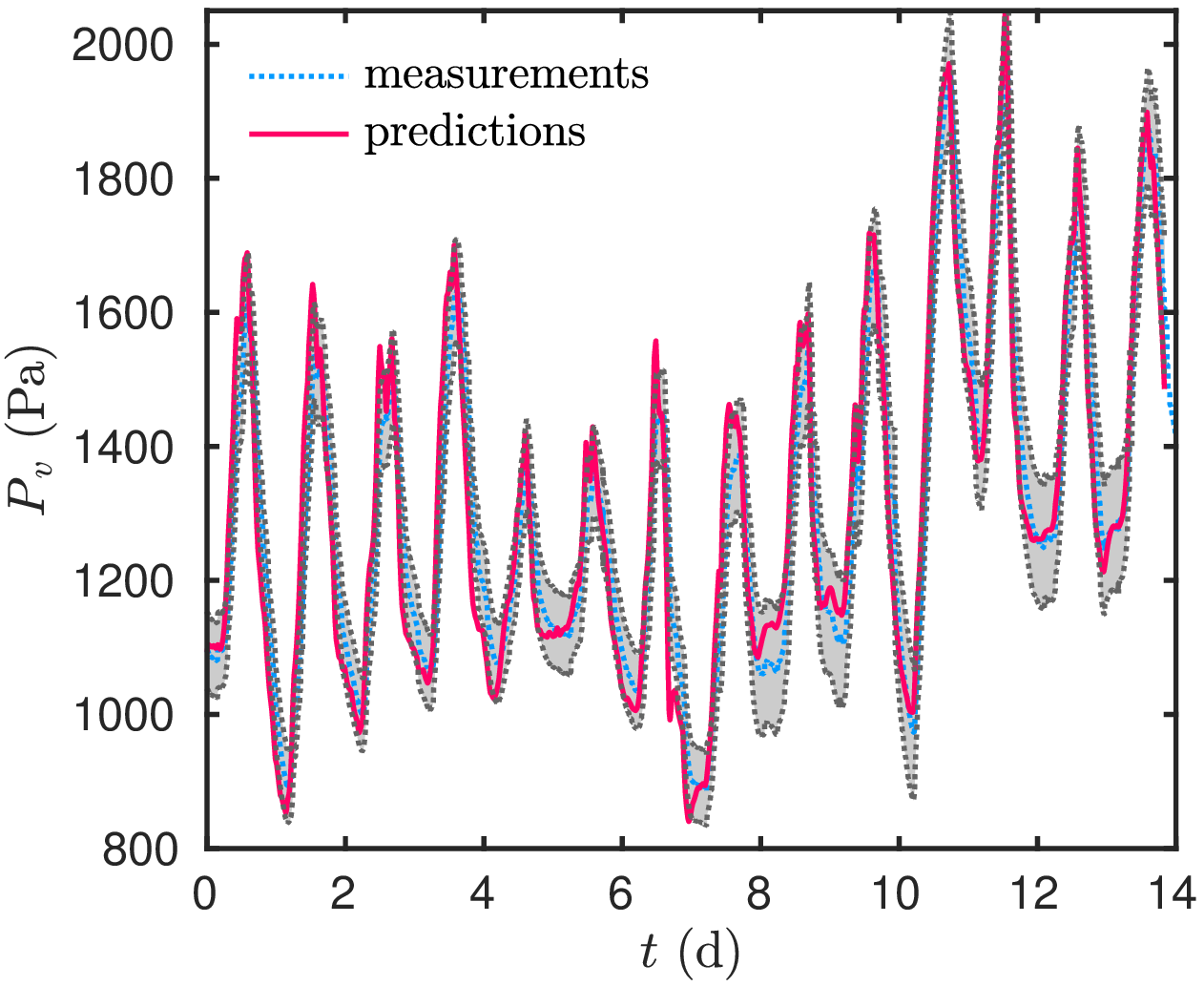}}
  \subfigure[$x \egal 12 \ \mathsf{cm}$ \label{fig:Case4_Tx12_ft}]{\includegraphics[width=.45\textwidth]{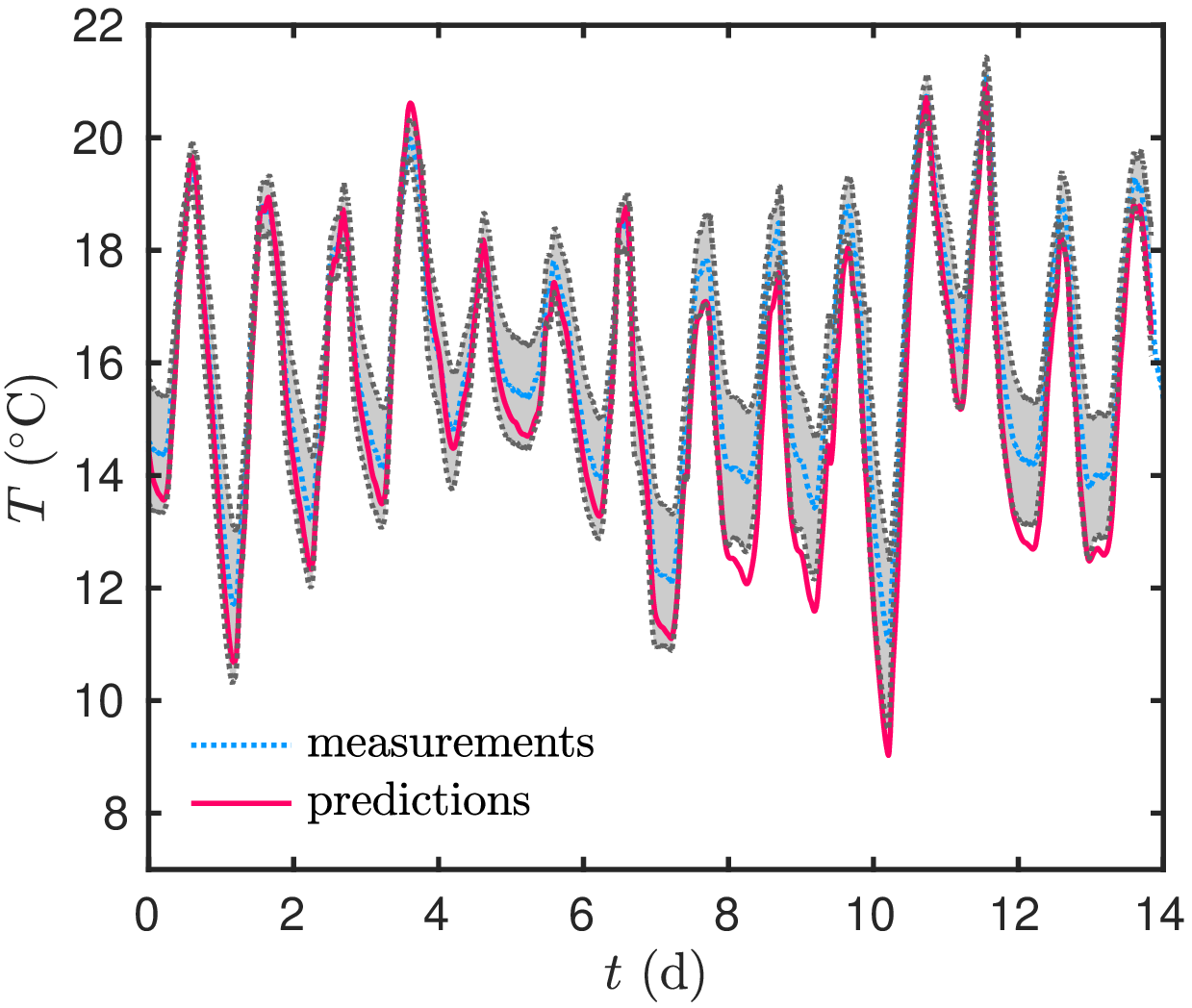}} 
  \caption{Comparison of the experimental observations with the numerical predictions for  vapor pressure (a, c, e) and  temperature (b, d, f). The grey shadows represent the measurement uncertainties.}
  \label{fig:Case4_Pv_T_xm_ft}
\end{figure}

\begin{figure}
  \centering
  \subfigure[\label{fig:Case4_epsu_ft}]{\includegraphics[width=.45\textwidth]{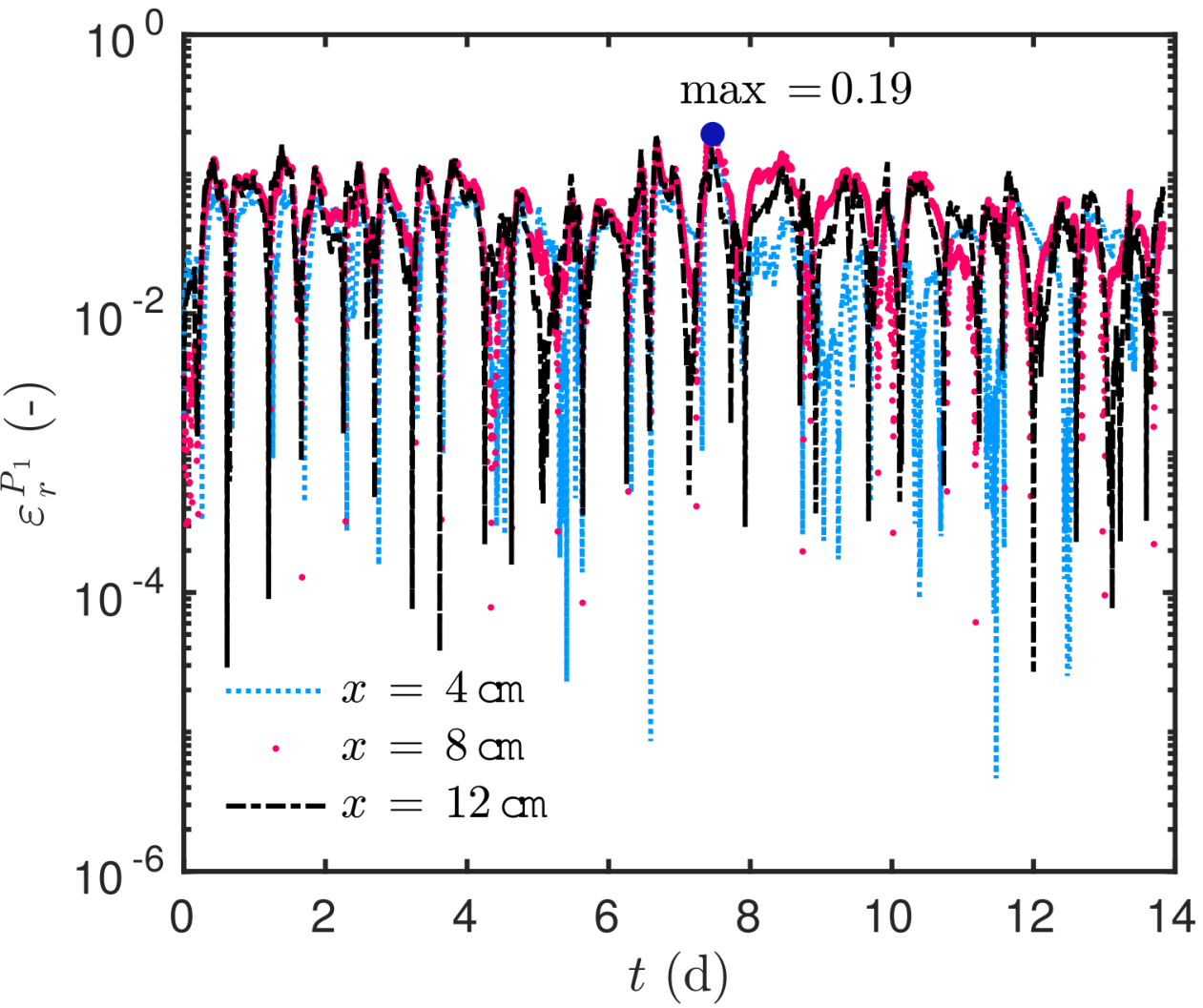}}
  \subfigure[\label{fig:Case4_epsv_ft}]{\includegraphics[width=.45\textwidth]{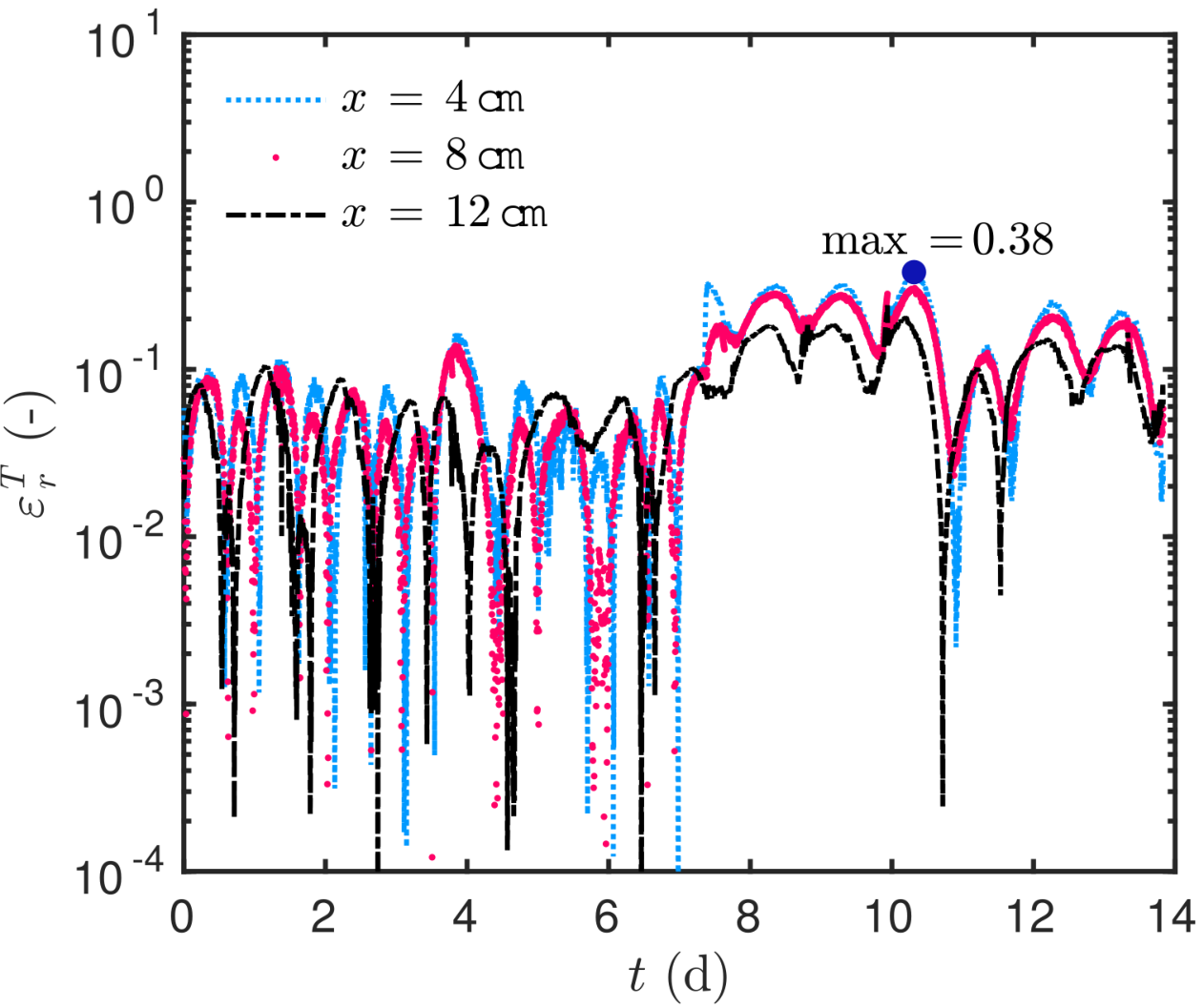}} 
  \caption{Relative error between the experimental observations and the numerical predictions for vapor pressure (a) and temperature (b).}
  \label{fig:Case4_eps_ft}
\end{figure}

Figure~\ref{fig:Case4_Pv_T_xm_ft} gives an overview of the comparison of the experimental observations at $x \egal \bigl\{\, 4 \,,\, 8 \,,\, 12\,\bigr\} \ \mathsf{cm}$ with the numerical predictions. The numerical predictions remain within the uncertainty range of the experimental observations. The relative error is shown in Figures~\ref{fig:Case4_epsu_ft} and \ref{fig:Case4_epsv_ft} for vapor pressure and temperature, respectively. One may deduce there is an overall satisfactory agreement among the observations and predictions. Some moderate discrepancies can be noticed in the temperature field for $t \, \in \, \bigl[\, 8 \,,\, 12 \,\bigr] \ \mathsf{days}$, particularly at $x \egal 8 \ \mathsf{cm}\,$. It corresponds to the increase of the vapor pressure. It highlights that there might be uncertainties in the material properties given in Table~\ref{tab:mat_properties_phys_constant} and/or that some physical phenomena are not considered in the actual model. It may arise from the interface at $x \egal 8 \ \mathsf{cm}$ between the two layers of wood fiberboard. In the literature \cite{DeFreitas1996, Guimaraes2018}, some works question this hypothesis of flux continuity at the interface, particularly for mass transfer. During the construction of the tested wall, the intention was to obtain a good contact between the two layers. Although the experiments were designed to avoid perturbation due to the contact resistance, some uncertainties sources might still exist.

\begin{figure}
  \centering
  \subfigure[\label{fig:Case4_P_ft}]{\includegraphics[width=.45\textwidth]{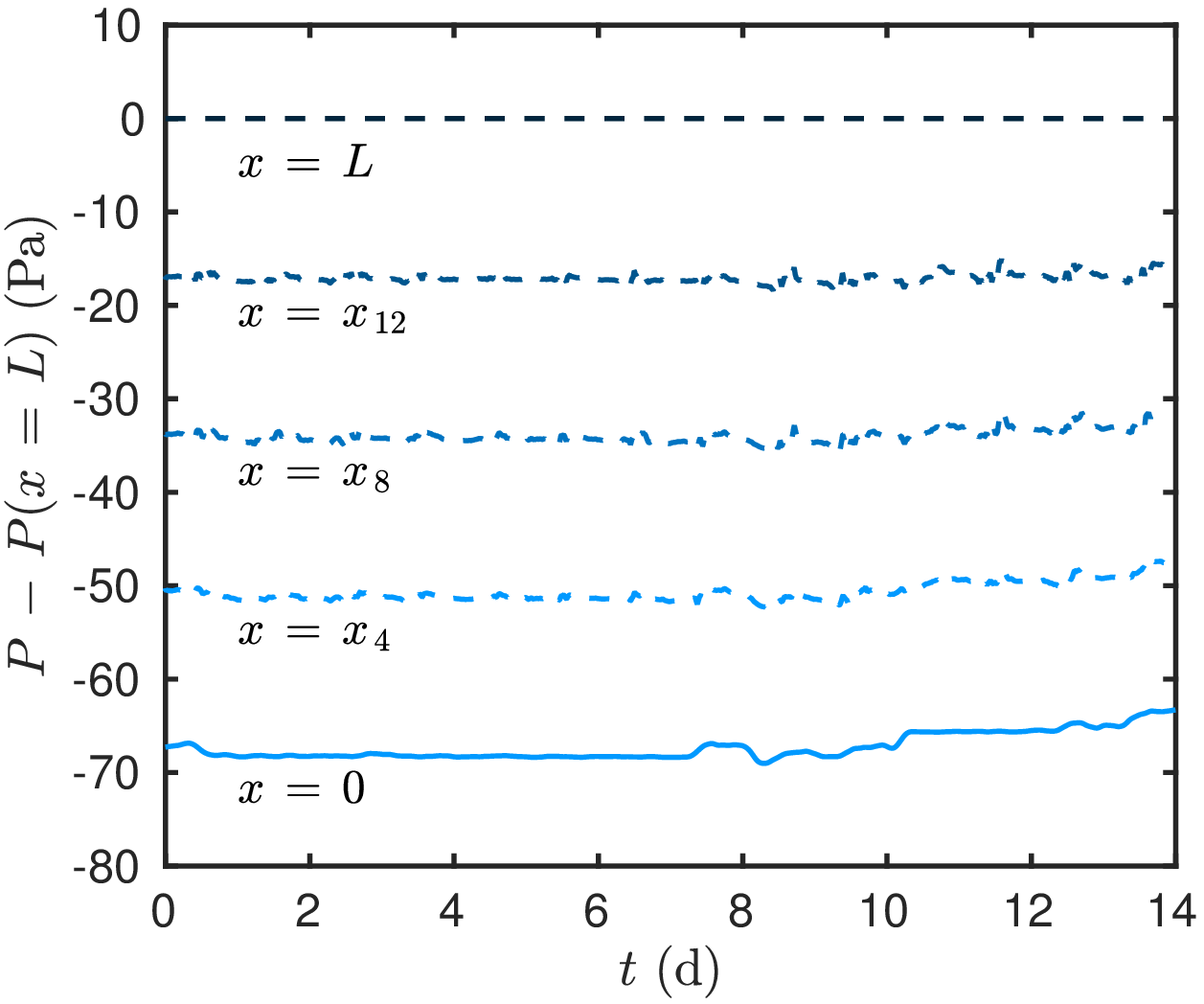}}
  \subfigure[\label{fig:Case4_vit_ft}]{\includegraphics[width=.45\textwidth]{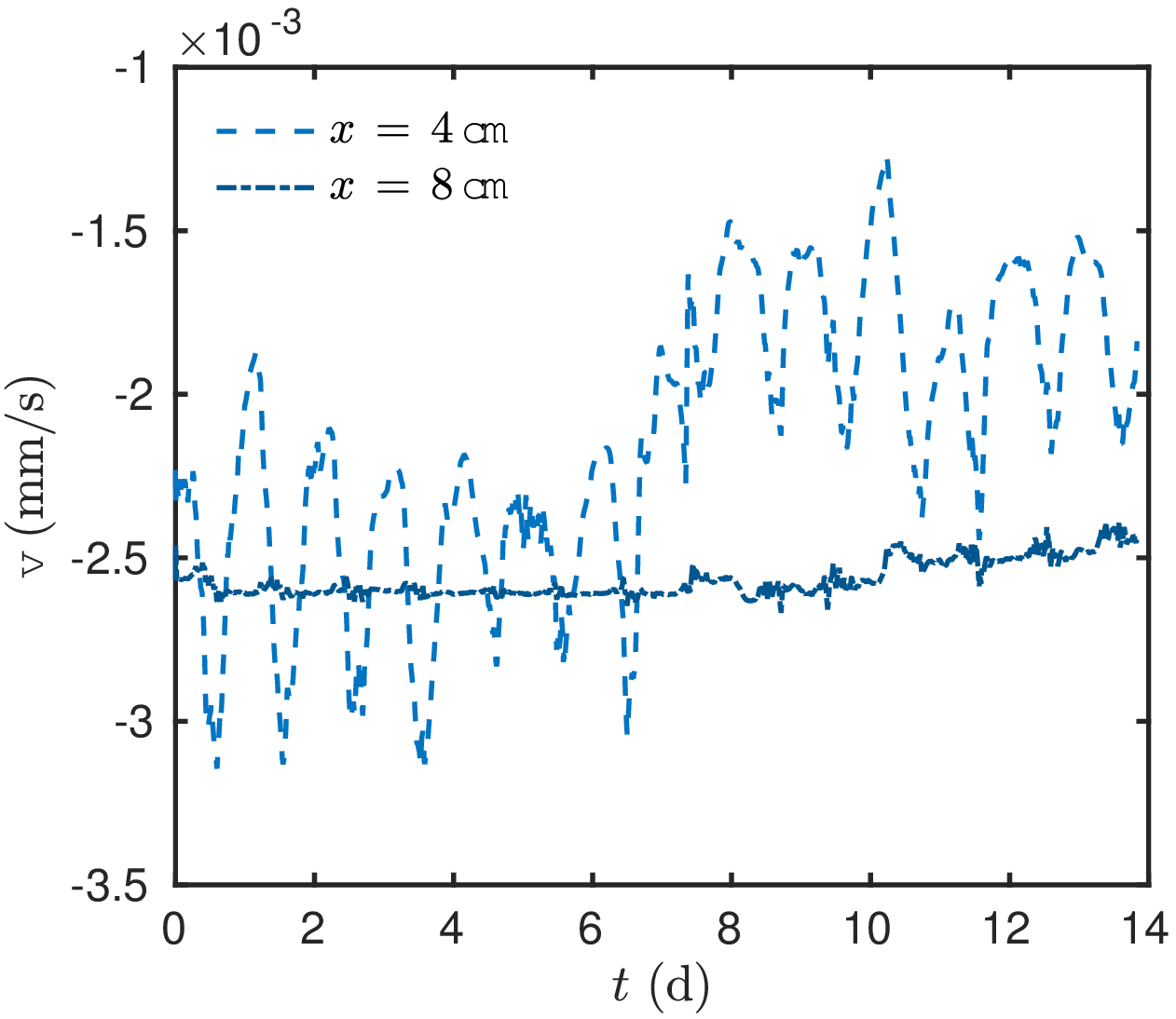}} 
  \caption{Variation of the differential pressure (a) and of the mass average velocity (b), both obtained with the numerical predictions (no experiments available).}
  \label{fig:Case4_P_vit_ft}
\end{figure}

The difference air pressure $P\,(\,x\,,\,t\,) \moins P\,(\,L\,,\,t\,)$ and the mass average velocity inside the material are depicted in Figure~\ref{fig:Case4_P_vit_ft}. The model enables to compute a velocity varying with time and space. It provides a more complex representation of the physical phenomena than the model proposed in \cite{Berger2017a, Berger2018a} where the air velocity is assumed as a constant in the whole material. Unfortunately, no experimental observations of air pressure were carried out for this case study so the reliability of the predictions cannot be evaluated. The gradient of the pressure is well established and it remains almost invariant in time. The air velocity is negative, indicating that the air flux is directed from the outside to the inside of the cell. In addition, at $x \egal 8 \ \mathsf{cm}\,$, interface between the two layers of wood fiber, the velocity is almost constant. As mentioned before, to explain the discrepancies on the temperature at the interface, this hypothesis may be reevaluated. The introduction of sensors inside the wall, with electronic connection passing through the layers, may disturb the air leakage at this interface.

\begin{figure}
  \centering
  \subfigure[\label{fig:Case4_jMadv_ft}]{\includegraphics[width=.45\textwidth]{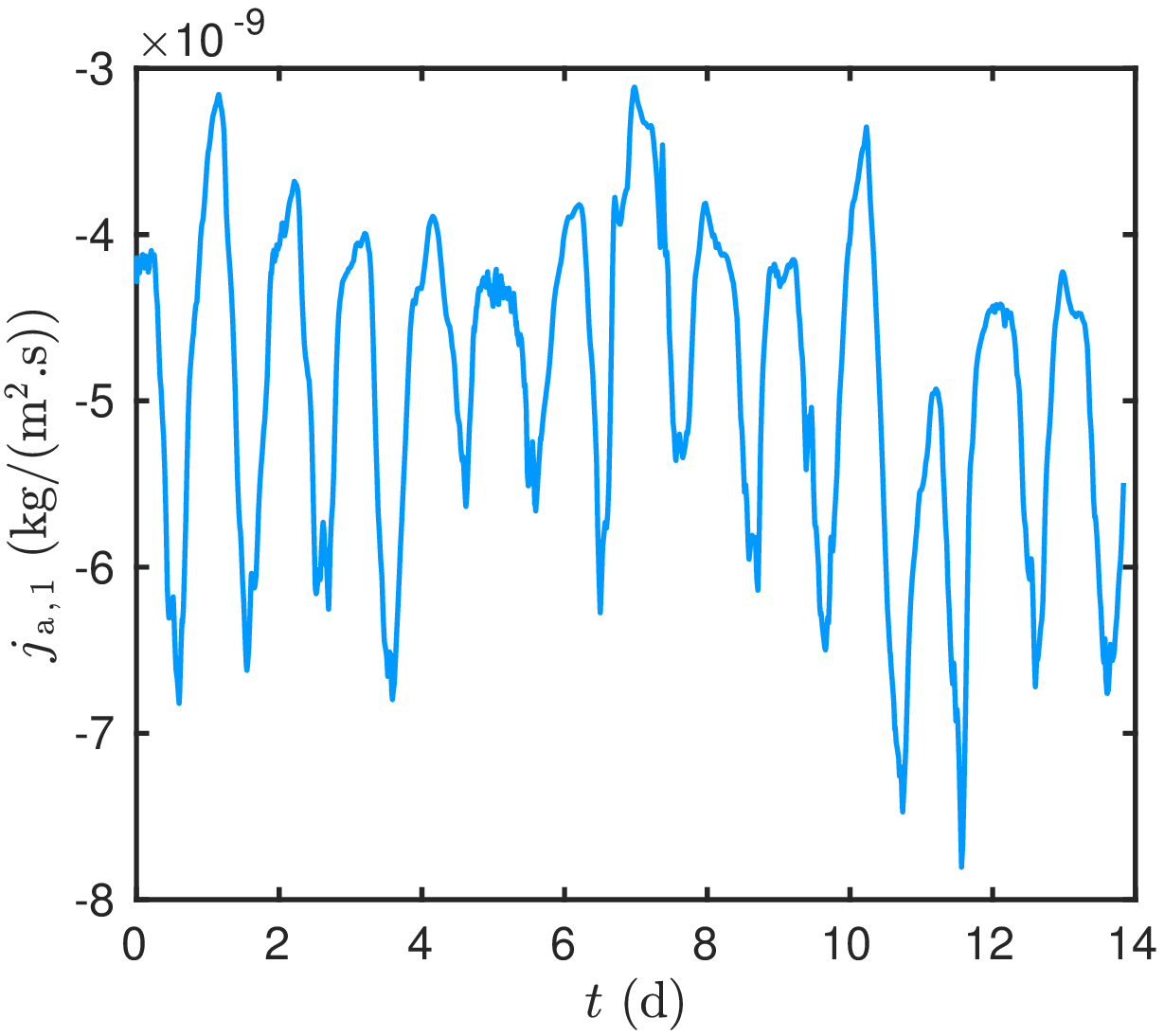}}
  \subfigure[\label{fig:Case4_jMdiff_ft}]{\includegraphics[width=.45\textwidth]{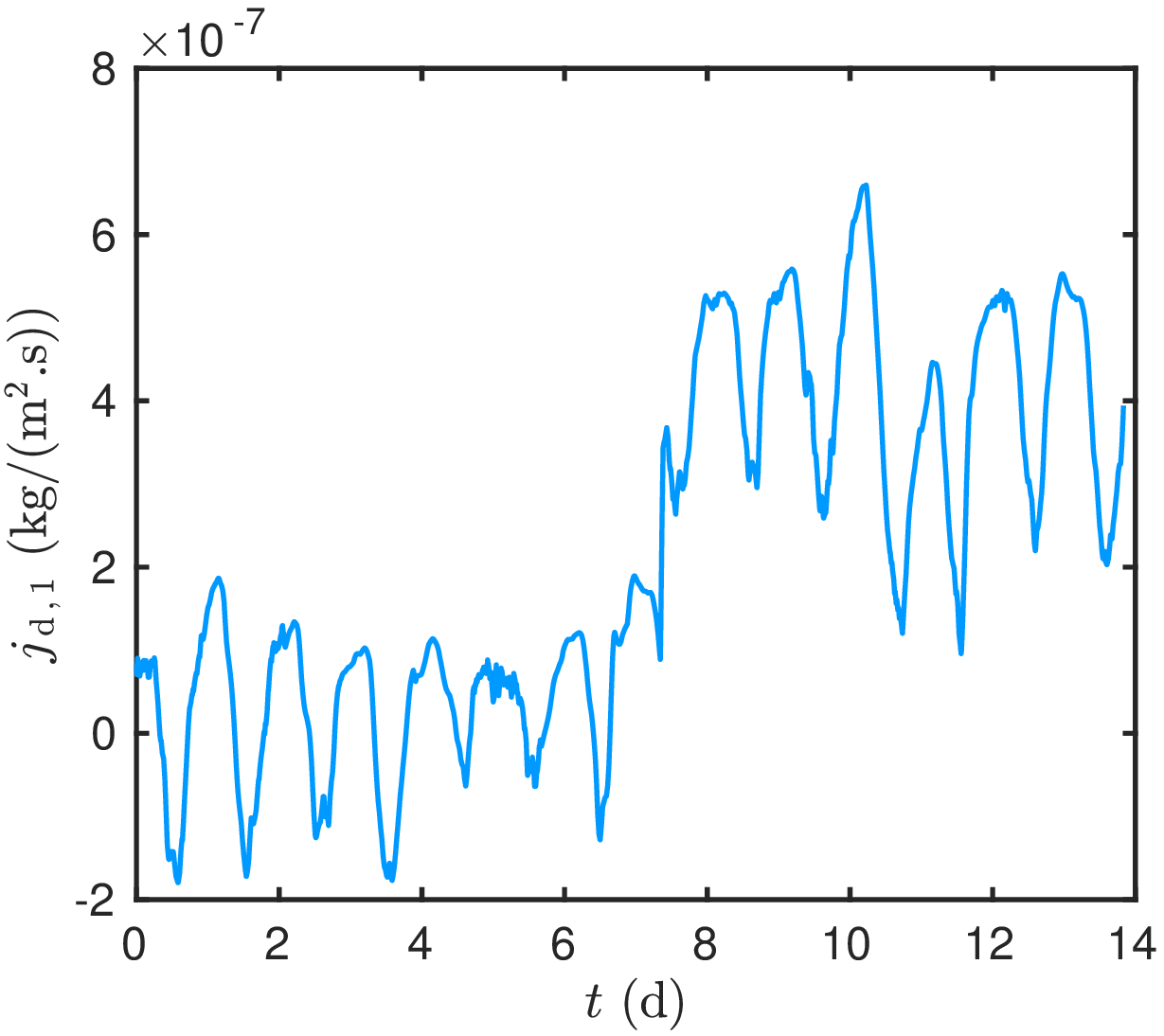}}
  \subfigure[\label{fig:Case4_jQadv_ft}]{\includegraphics[width=.45\textwidth]{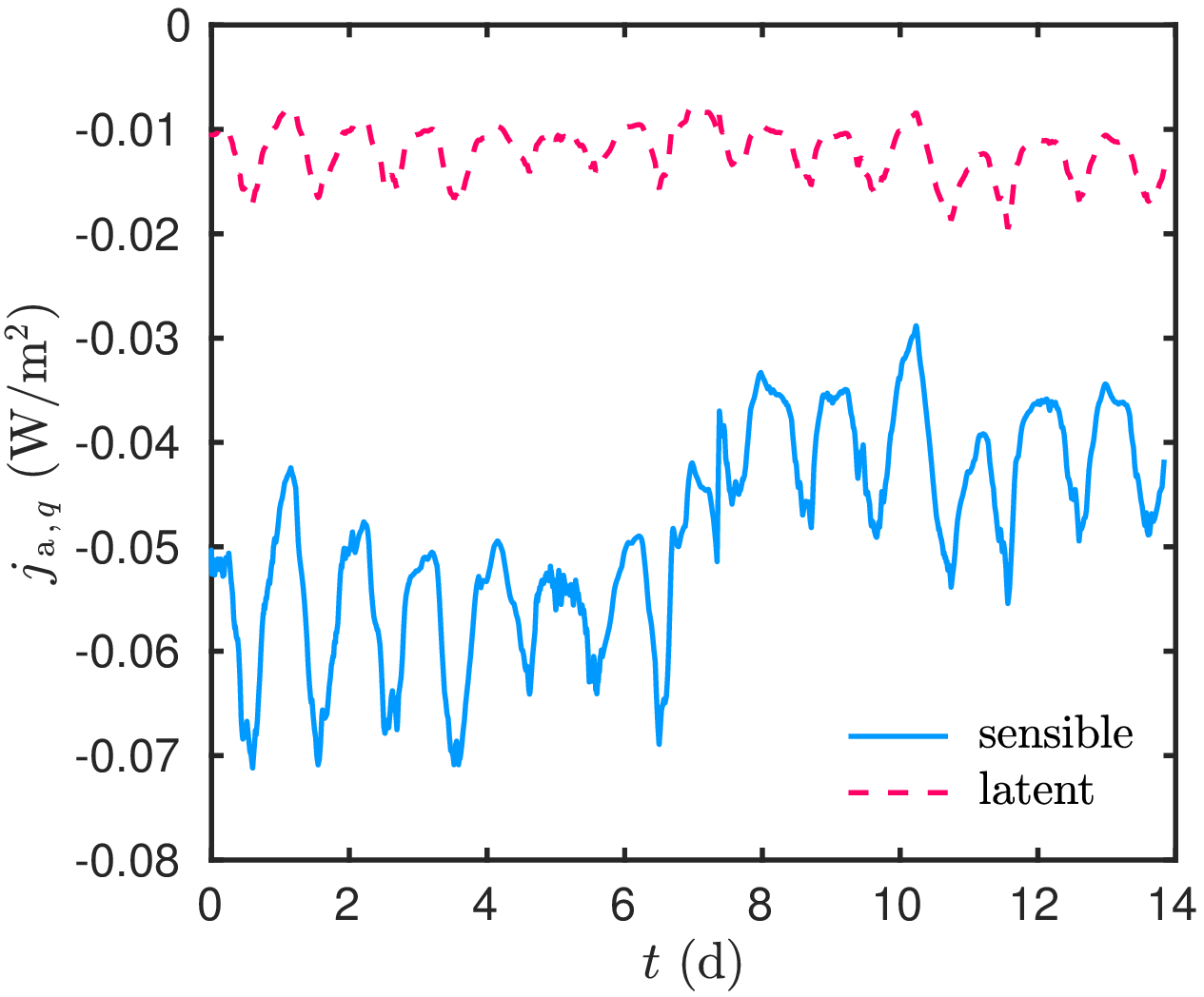}}
  \subfigure[\label{fig:Case4_jQdiff_ft}]{\includegraphics[width=.45\textwidth]{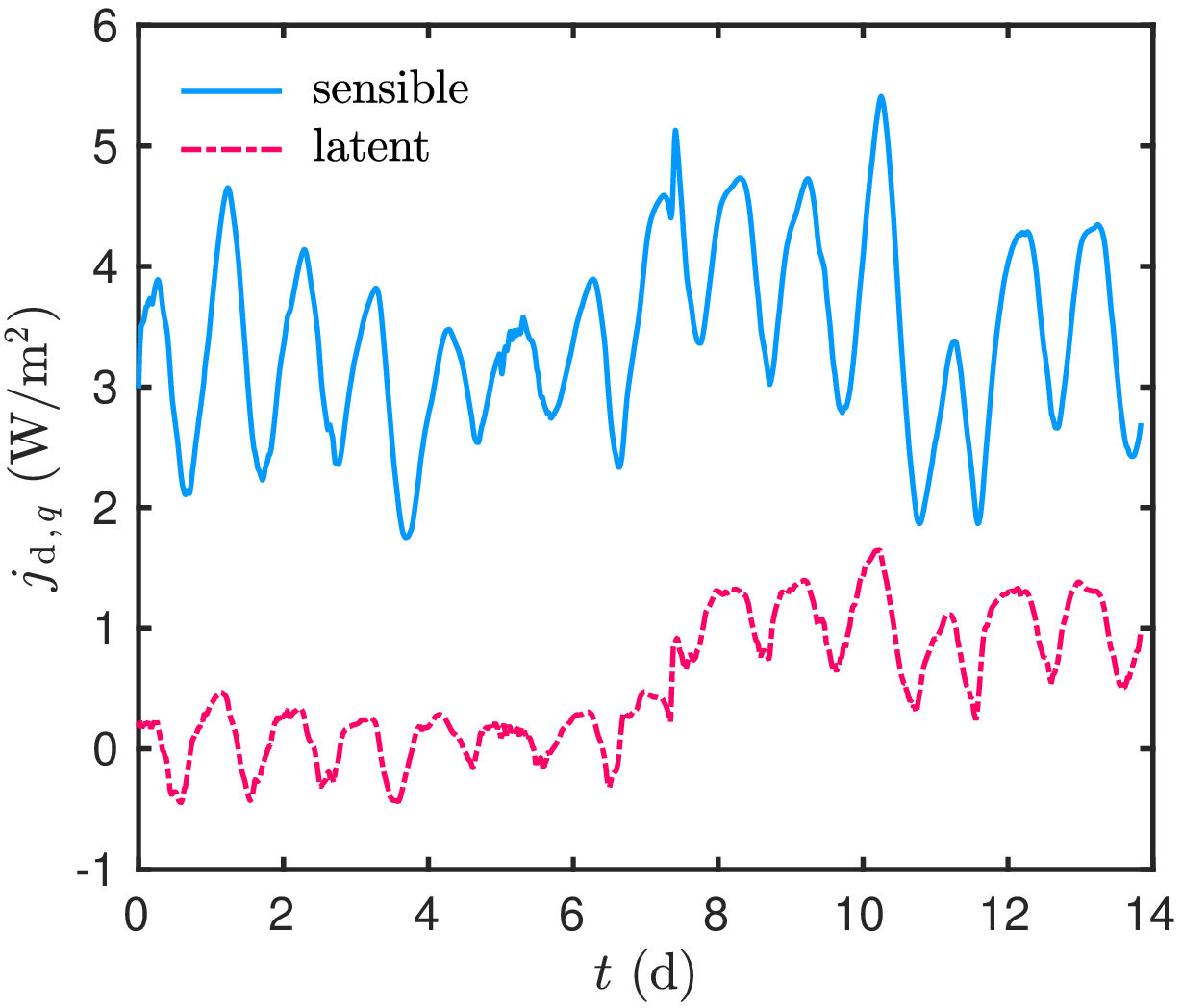}} 
  \caption{Time evolution of the mass flux by advection (a) and  diffusion \emph{(b)}. Time evolution of the heat flux by advection (c) and diffusion (d).}
\end{figure}

The mass flux driven by advection and diffusion are shown in Figures~\ref{fig:Case4_jMadv_ft} and \ref{fig:Case4_jMdiff_ft}, respectively. The advection driven flux is around hundred times lower than the diffusion one. Similar observations can be made for the heat flux in Figures~\ref{fig:Case4_jQadv_ft} and \ref{fig:Case4_jQdiff_ft}. The air transfer has a reduced impact on the heat and mass ones. It is due to a relatively low air permeability of the material $k_{\,13} \egal 1.1 \e{-13} \ \mathsf{m^{\,2}}\,$. In addition, the boundary conditions for the air pressure are almost invariant in time as can be seen in Figure~\ref{fig:Case4_BC_P}. In Figures~\ref{fig:Case4_jQadv_ft} and \ref{fig:Case4_jQdiff_ft}, it can also be remarked that the latent heat flux is smaller than the sensible one.

\begin{figure}
  \centering
  \subfigure[$x \egal 4 \ \mathsf{cm}$ \label{fig:Case4_diff_model_Pvx4}]{\includegraphics[width=.45\textwidth]{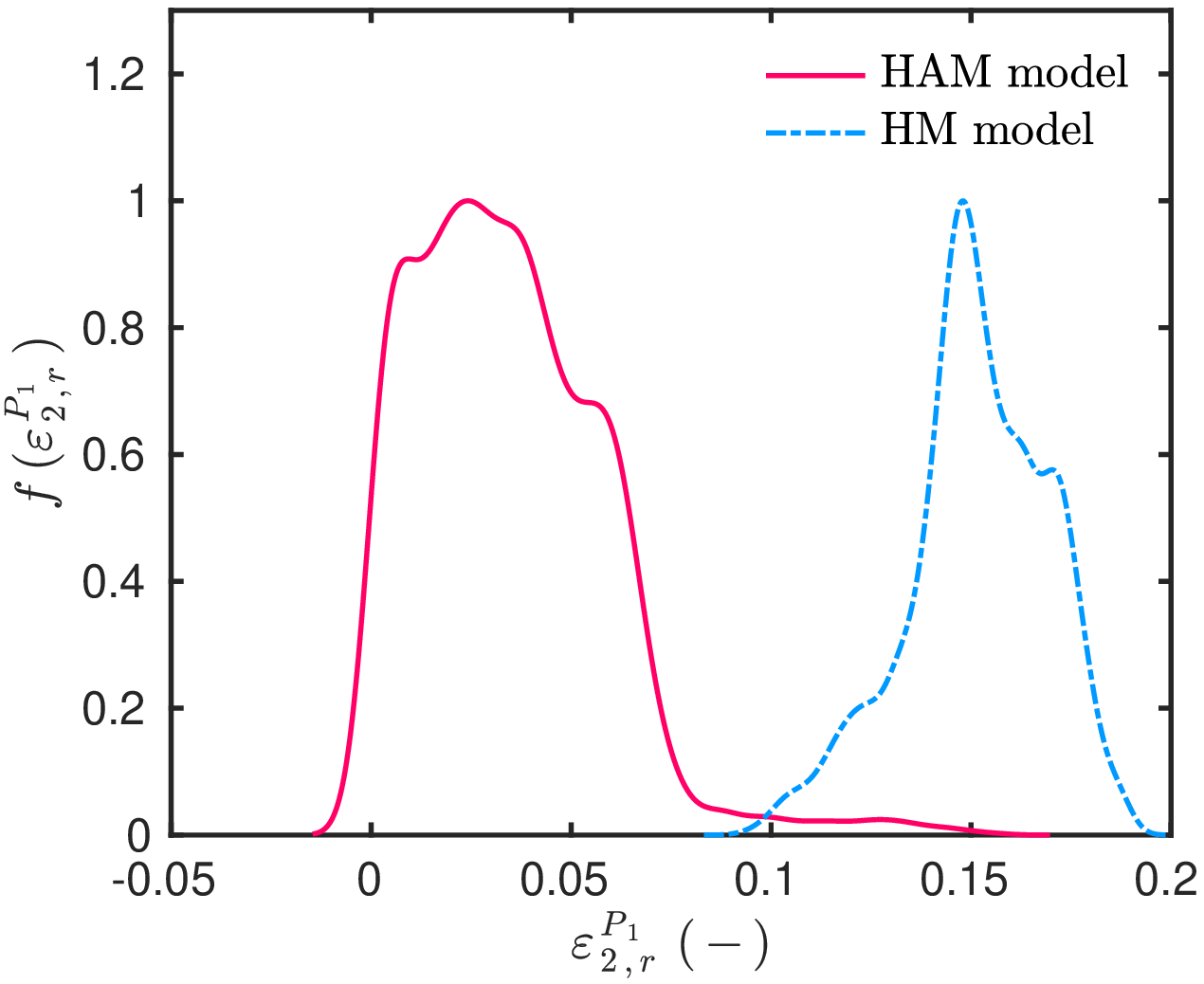}}
  \subfigure[$x \egal 4 \ \mathsf{cm}$ \label{fig:Case4_diff_model_Tx4}]{\includegraphics[width=.45\textwidth]{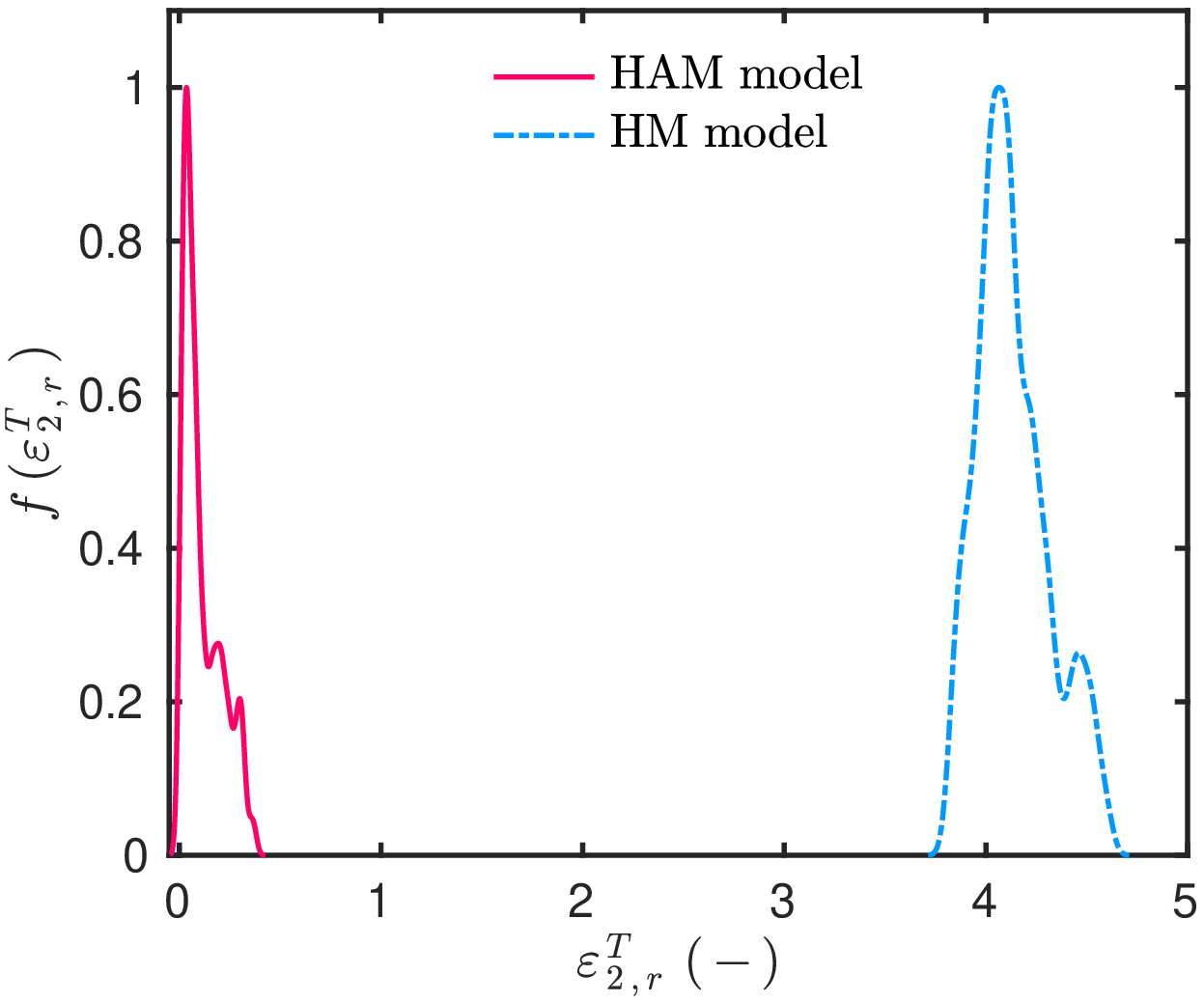}}
  \subfigure[$x \egal 8 \ \mathsf{cm}$\label{fig:Case4_diff_model_Pvx8}]{\includegraphics[width=.45\textwidth]{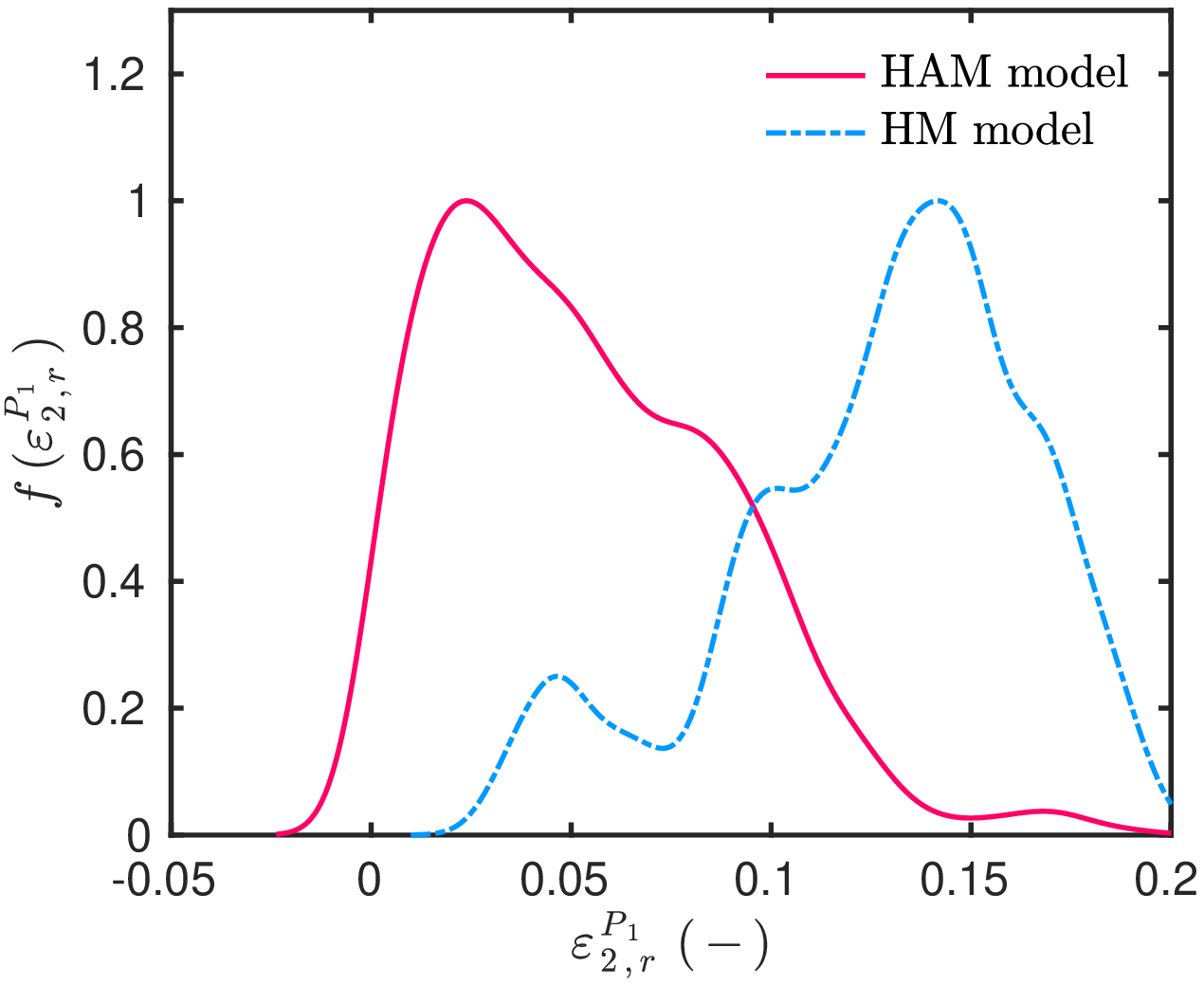}}
  \subfigure[$x \egal 8 \ \mathsf{cm}$\label{fig:Case4_diff_model_Tx8}]{\includegraphics[width=.45\textwidth]{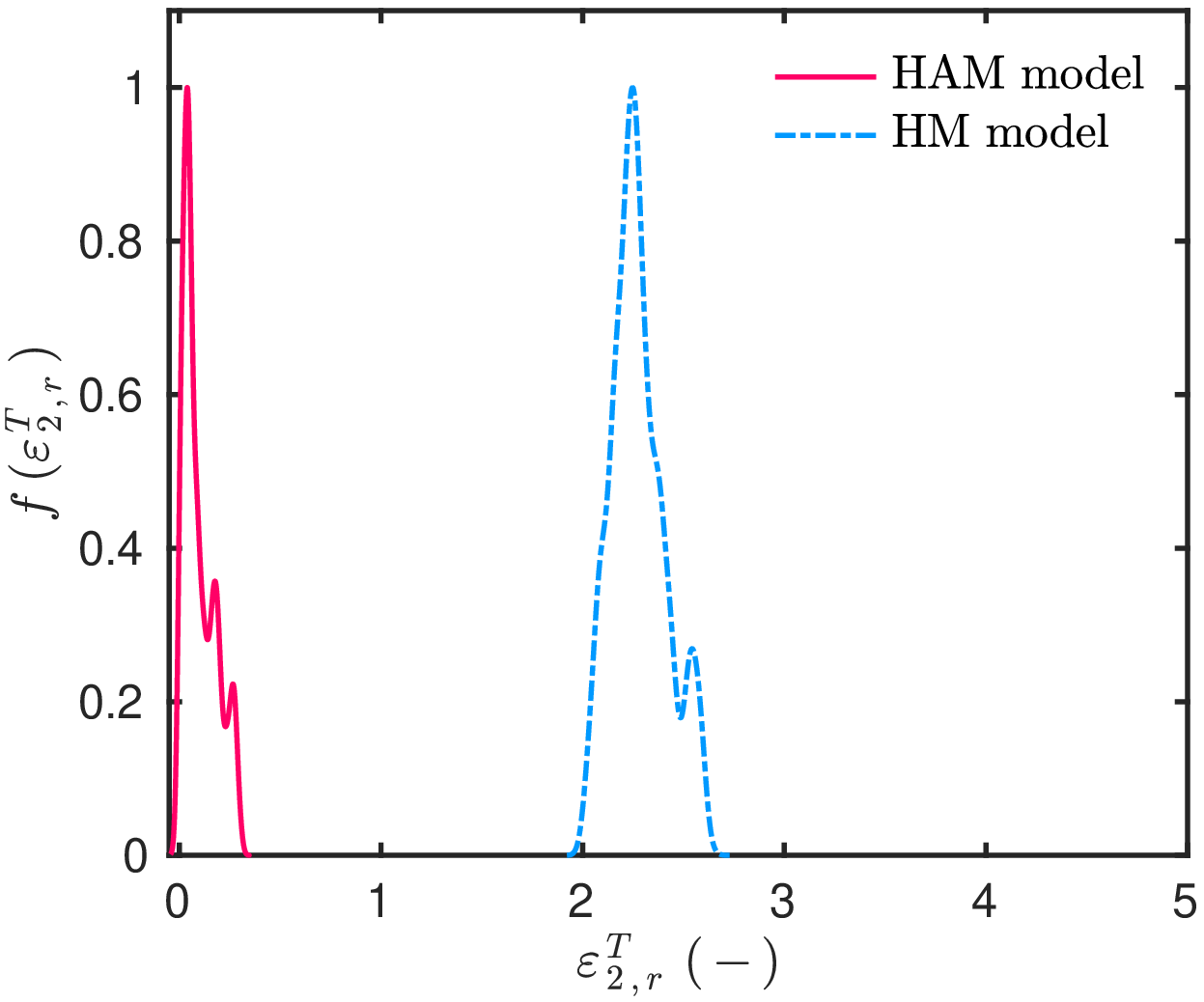}}
  \subfigure[$x \egal 12 \ \mathsf{cm}$\label{fig:Case4_diff_model_Pvx12}]{\includegraphics[width=.45\textwidth]{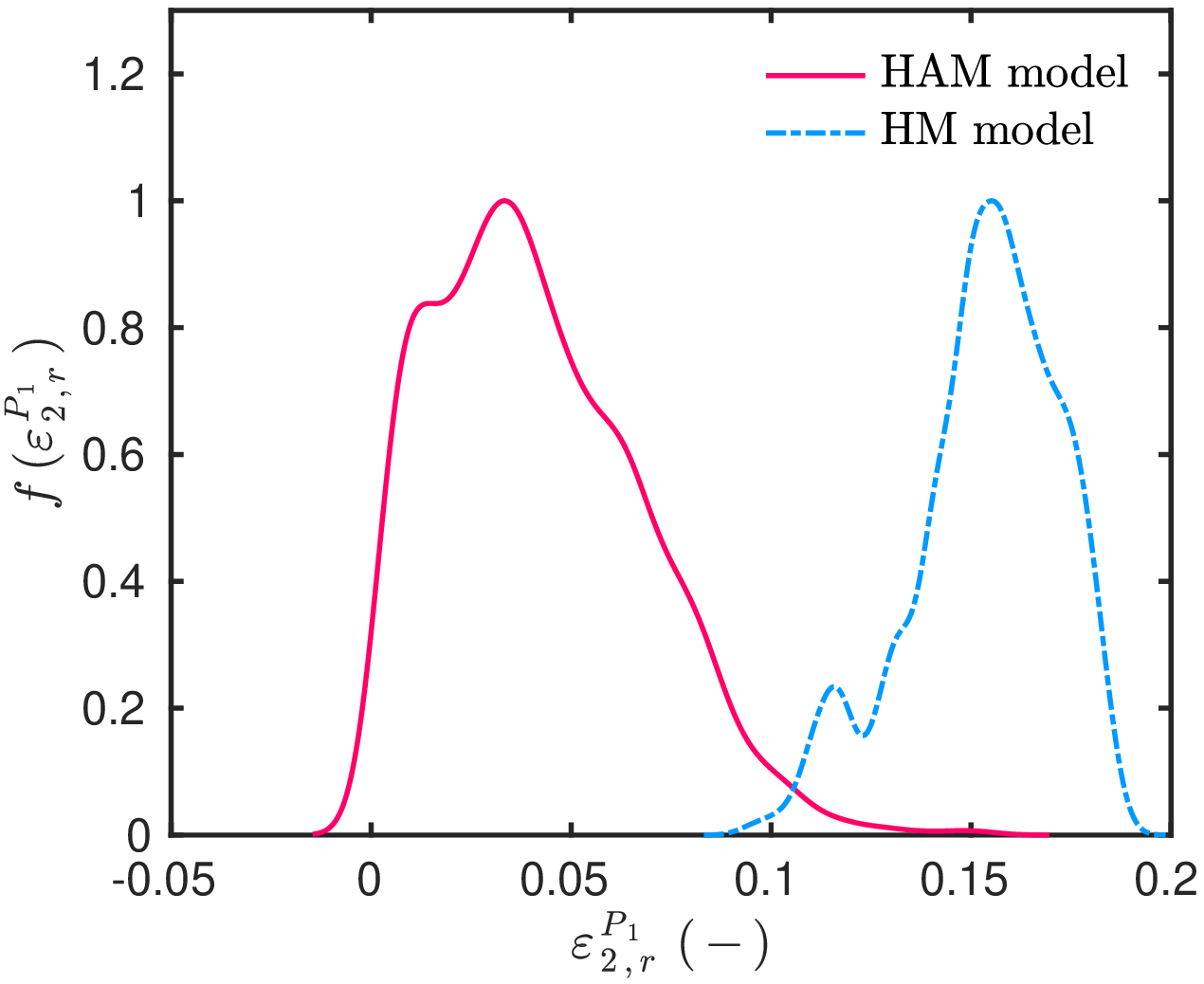}}
  \subfigure[$x \egal 12 \ \mathsf{cm}$\label{fig:Case4_diff_model_Tx12}]{\includegraphics[width=.45\textwidth]{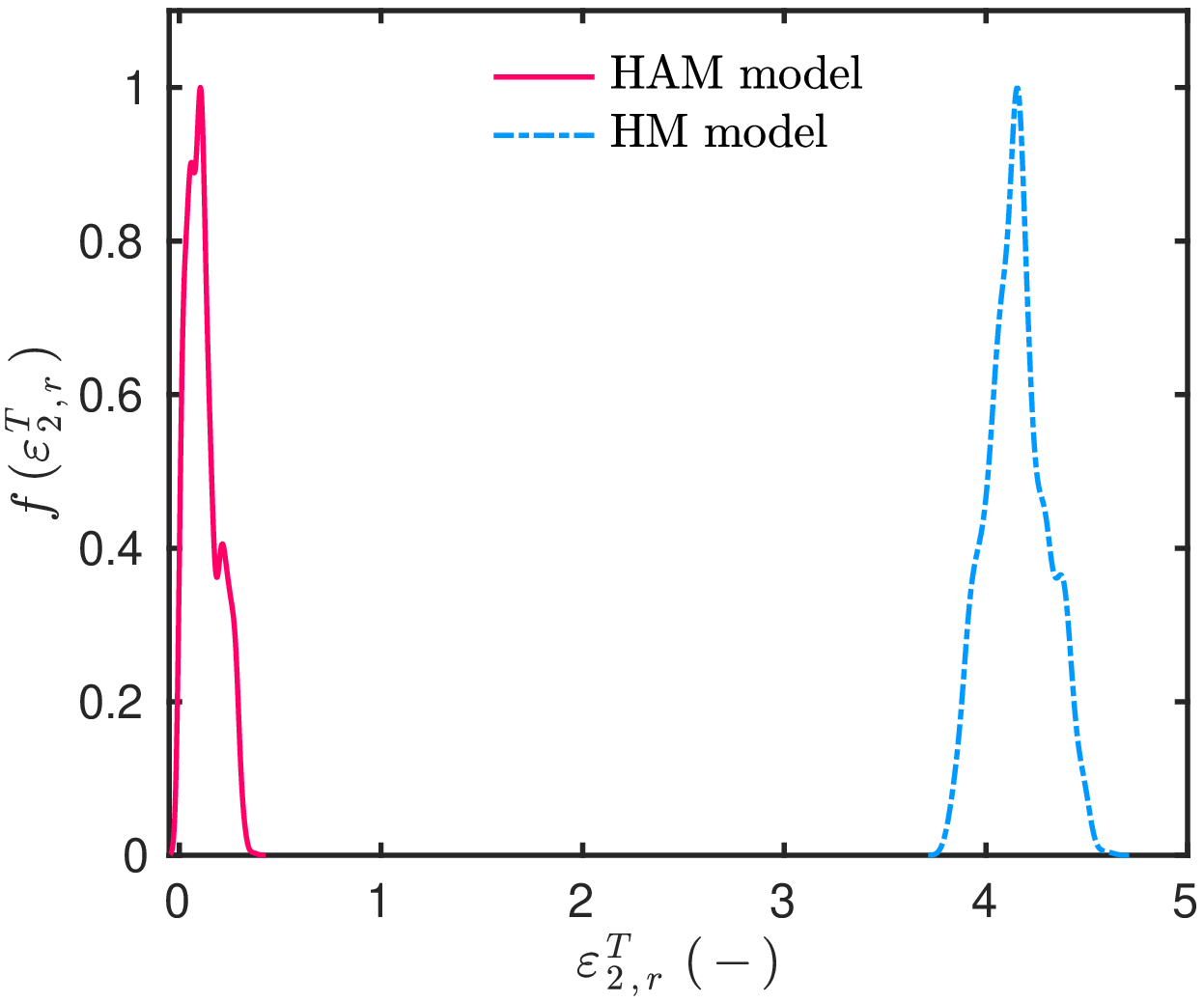}} 
  \caption{Comparison of the probability density function of the relative error between the experimental observations and the numerical predictions for vapor pressure (a,c,e) and temperature (b,d,f), obtained with the proposed model and the purely diffusive one.}
  \label{fig:Case4_diff_model}
\end{figure}

If the reliability of the proposed model is very satisfactory, most approaches from the literature do not consider the air transport equation~\eqref{eq:phys_dim_A}. For instance, in \cite{Gasparin2018} the reliability is proven for the same case study but considering a model with only heat and moisture transport by diffusion processes. A natural question raises on the importance of considering air transport to better represent the whole physical phenomenon. To answer this question, the numerical predictions from the model in \cite{Gasparin2018}, denoted as HM (Heat and Moisture) model are compared to the one obtained with the proposed model, denoted as HAM (Heat, Air and Moisture) model. Figure~\ref{fig:Case4_diff_model} shows the probability density function of the relative error for each model between the numerical predictions and the experimental observations. If both models admit a very satisfactory reliability, the fidelity is slightly better for the proposed model. Particularly, the relative error of the predictions of temperature scales with $0.3 \%$ and with $4\%$ for the HAM and HM models, respectively. The difference between the predictions of the models is reduced for the observations at the middle of the wall, \ie $x \egal 8 \ \mathsf{cm}\,$.

%%% ----------------------------------------------------------------------- %%%

\section{Conclusion}

Within the context of predicting the impact of air transport on the combined heat and moisture transfer in porous building materials, this article proposes a new model with an efficient numerical scheme. In contrast to earlier proposed models in literature, the approach takes into account the transient effects of air convection without being constrained by the high numerical cost induced by the very small characteristic times of the physical phenomena. After describing in details the achievements of the mathematical model to describe the physical phenomena in Section~\ref{sec:definition_physical_model}, the numerical model is defined in Section~\ref{sec:numerical_model}. A system of three partial differential equations is obtained, being two of them of advective--diffusive nature, while the third one is purely diffusive. The numerical model is based on the \SG ~numerical scheme for the spatial discretisation of the two coupled advection--diffusion equations. It has important numerical properties as well-balanced and asymptotically preserving, ensuring the accuracy of the computed solutions. The scheme is explicit in time avoiding costly sub--iterations at each time step to handle the nonlinearities of the problem.  In addition, as demonstrated in \cite{Gosse2016}, the CFL stability condition of the scheme  $\Delta t \, \simeq \, \Delta x$ is relaxed compared to standard finite--difference approaches, enabling to save extra computational efforts. For the time discretisation, a two--step \RK ~approach is employed. With this scheme, the stability condition $\Delta t \, \sim  \, \Delta x$ is even more relaxed. For the exclusive diffusion equation, the explicit and unconditionally stable \DF ~scheme is used. It circumvents the strong restrictions on the stability of the numerical model coming from the very small time characteristics of the air transfer. Finally, an efficient numerical scheme is proposed to compute an accurate solution. The computational efforts are saved since the restrictions on the choice of the time discretisation steps are weakened by the use of robust numerical schemes.

The features of the numerical model are enhanced in Section~\ref{sec:validation_numerical_model} with one case study, where all coefficients are nonlinear and the \textsc{Robin}-type boundary conditions are applied. The reference solution is computed using a pseudo--spectral approach. Results show a very accurate solution with a reduced computational time.

After validating the efficiency of the numerical model, Section~\ref{sec:reliability_num_predictions} intends to evaluate the reliability of the numerical predictions by comparing them to experimental observations. The latter are achieved by measuring temperature and vapor pressure in a wall composed of two layers of wood fiberboard. The internal boundary conditions are controlled, with a sudden isothermal increase on the vapor pressure imposed by an air handling unit. For the external side, the boundary conditions are driven by climate variations. There is a very satisfactory agreement between the numerical predictions and the experimental observations. The results indicate a high reliability of the numerical model to represent the physical phenomena. The stability condition of the problem is reduced by an order of $50$ with the innovative numerical model, providing a reduction of the computational effort by $16$ compared to standard approach based on \Eu ~explicit scheme with central finite differences.

Further research should be carried out on the characterization of the conditions at the interface of two materials. As noticed in Section~\ref{sec:reliability_num_predictions}, some small discrepancies were observed at the interface between two layers. Some works in the literature suggest how to modify the conditions in the model \cite{DeFreitas1996, Guimaraes2018} but the resistances at the interfaces between porous materials is still an open question. To conclude, we can say the air transport has been considered important since the HAMSTAD project (2002) \cite{Hagentoft2004}, but a little research has been presented in the literature regarding the importance of advection in porous material and on the calculation of the velocity and pressure profiles. Therefore, to reinforce the research in the field, it is important to promote on one hand detailed experimental research to include, \eg, pressure differences along the thickness as well as microscopic simulations using methods such as the Lattice \textsc{Boltzmann} model including energy and momentum balances and phase change terms.

%%% ----------------------------------------------------------------------- %%%

\section*{Nomenclature}

\begin{tabular}{|cll|}
\hline
\multicolumn{3}{|c|}{\emph{Latin letters}} \\
$a_{\,q}$ & heat advection coefficient & $[\mathsf{W/(m^{\,2}\cdot K)}]$ \\
$a_{\,v}$ & vapor advection coefficient & $[\mathsf{s/m}]$ \\
$c$ & specific heat & $[\mathsf{J/(kg\cdot K)}]$ \\
$c_{\,at}$ & storage coefficient & $[\mathsf{kg \cdot Pa/(J\cdot K)}]$ \\
$c_{\,m}$ & moisture storage capacity & $[\mathsf{kg/(m^3\cdot Pa)}]$ \\
$c_{\,q}$ & volumetric heat capacity & $[\mathsf{J/(m^{\,3}\cdot K)}]$ \\
$c_{\,qs}\,,\,c_{\,as}$ & storage coefficient & $[\mathsf{kg\cdot Pa/J}]$ \\
$c_{\,qv}\,,\,c_{\,a}\,,\,c_{\,av}$ & storage coefficient & $[\mathsf{kg/J}]$ \\
$g$ & liquid flux &  $[\mathsf{kg/(s\cdot m^{\,2})}]$ \\
$h$ & specific enthalpy  & $[\mathsf{J/kg}]$ \\
$I$ & volumetric capacity of source/sink & $[\mathsf{kg/(m^{\,3}.s)}]$\\
$j$ & mass flow &  $[\mathsf{kg/(s\cdot m^{\,2})}]$ \\
$j_{\,q}$ & heat flow & $[\mathsf{W/m^{\,2}}]$ \\
$k_{\,a}$ & permeability coefficient & $[\mathsf{s^{\,2}/m^{\,2}}]$ \\
$k_{\,1}\,,\,k_{\,2}\,,\,k_{\,v}\,,\,k_{\,m}$ & vapor, liquid, moisture permeability & $[\mathsf{s}]$ \\
$k_{\,q}$ & thermal conductivity & $[\mathsf{W/(m\cdot K)}]$ \\
$k_{\,13}$ & intrinsic permeability & $[\mathsf{m^{\,2}}]$ \\
$L$ & length & $[\mathsf{m}]$ \\
$m$ & mass & $[\mathsf{kg}]$ \\
$P\,,\,P_{\,1}\,,\,P_{\,2}\,,\,P_{\,3}$ & pressure & $[\mathsf{Pa}]$ \\
$R_{\,1}\,,\,R_{\,3}\,,\,R_{\,13}$ & specie gaz constant & $[\mathsf{J/(kg\cdot K)}]$ \\
$r_{\,12}$ & latent heat of evaporation & $[\mathsf{J/kg}]$ \\
$T$ & temperature & $[\mathsf{K}]$ \\
$t$ & time coordinate & $[\mathsf{s}]$ \\
$x$ & space coordinate & $[\mathsf{m}]$ \\
$V$ & volume & $[\mathsf{m^{\,3}}]$ \\
$v$ & mass average velocity & $[\mathsf{m/s}]$ \\
$\w$ & volumetric concentration & $[\mathsf{kg/m^{\,3}}]$ \\
\hline
\end{tabular}

%\bigskip

\begin{tabular}{|cll|}
\hline
\multicolumn{3}{|c|}{\emph{Greek letters}} \\
$\alpha_{\,m}$ & surface vapour transfer coefficient & $[\mathsf{s/m}]$ \\
$\alpha_{\,q}$ & surface heat transfer coefficient & $[\mathsf{W/(m^2 \cdot K)}]$ \\
$\phi$ & relative humidity & $[-]$ \\
$\rho$ & specific mass & $[\mathsf{kg/m^3}]$ \\
$\Pi$ & material porosity & $[\mathsf{-}]$ \\
$\sigma$ & saturation rate & $[\mathsf{-}]$ \\
$\mu$ & dynamic viscosity & $[\mathsf{Pa \cdot s}]$ \\
\hline
\end{tabular}

%%% ----------------------------------------------------------------------- %%%

\section*{Acknowledgments}

The authors acknowledge the Junior Chair Research program ``Building performance assessment, evaluation and enhancement'' from the University of Savoie Mont Blanc in collaboration with The French Atomic and Alternative Energy Center (CEA) and Scientific and Technical Center for Buildings (CSTB). The authors thanks the grants from the Ministry of Education and Science of the Republic of Kazakhstan and the visiting professor grants from the University of Savoie Mont Blanc. The authors also acklowledge the French and Brazilian agencies for their financial supports through the project CAPES--COFECUB, as weel as the CNPQ of the Brazilian  Ministry of Education and of the Ministry of Science, Technology and Innovation, respectively, for co-funding.

%%% ----------------------------------------------------------------------- %%%

%%% Bibliography

\bigskip


\bigskip\bigskip
\addcontentsline{toc}{section}{References}

\begin{thebibliography}{10}

\bibitem{Abahri2016a}
K.~Abahri, R.~Bennacer, and R.~Belarbi.
\newblock {Sensitivity analyses of convective and diffusive driving potentials
  on combined heat air and mass transfer in hygroscopic materials}.
\newblock {\em Numerical Heat Transfer, Part A: Applications},
  69(10):1079--1091, may 2016.

\bibitem{HYGRO-BA2014}
{ANR Project HYGRO-BAT}.
\newblock {Vers une m{\'{e}}thode de conception HYGRO-thermique des BATiments
  performants}, 2014.

\bibitem{Belleudy2015}
C.~Belleudy, A.~Kayello, M.~Woloszyn, and H.~Ge.
\newblock {Experimental and numerical investigations of the effects of air
  leakage on temperature and moisture fields in porous insulation}.
\newblock {\em Building and Environment}, 94:457--466, dec 2015.

\bibitem{Belleudy2016}
C.~Belleudy, M.~Woloszyn, M.~Chhay, and M.~Cosnier.
\newblock {A 2D model for coupled heat, air, and moisture transfer through
  porous media in contact with air channels}.
\newblock {\em Int. J. Heat Mass Transfer}, 95:453--465, apr 2016.

\bibitem{Berger2017a}
J.~Berger, S.~Gasparin, D.~Dutykh, and N.~Mendes.
\newblock {Accurate numerical simulation of moisture front in porous material}.
\newblock {\em Building and Environment}, 118:211--224, jun 2017.

\bibitem{Berger2018a}
J.~Berger, S.~Gasparin, D.~Dutykh, and N.~Mendes.
\newblock {On the Solution of Coupled Heat and Moisture Transport in Porous
  Material}.
\newblock {\em Transport in Porous Media}, 121(3):665--702, feb 2018.

\bibitem{Berger2015a}
J.~Berger, S.~Guernouti, M.~Woloszyn, and C.~Buhe.
\newblock {Factors governing the development of moisture disorders for
  integration into building performance simulation}.
\newblock {\em J. Building Eng.}, 3:1--15, sep 2015.

\bibitem{Chollom2003}
J.~Chollom and Z.~Jackiewicz.
\newblock {Construction of two-step Runge-Kutta methods with large regions of
  absolute stability}.
\newblock {\em J. Comp. Appl. Math.}, 157(1):125--137, aug 2003.

\bibitem{DeFreitas1996}
V.~P. {De Freitas}, V.~Abrantes, and P.~Crausse.
\newblock {Moisture migration in building walls - Analysis of the interface
  phenomena}.
\newblock {\em Building and Environment}, 31(2):99--108, mar 1996.

\bibitem{Desta2011}
T.~Z. Desta, J.~Langmans, and S.~Roels.
\newblock {Experimental data set for validation of heat, air and moisture
  transport models of building envelopes}.
\newblock {\em Building and Environment}, 46(5):1038--1046, may 2011.

\bibitem{DosSantos2009a}
G.~H. dos Santos and N.~Mendes.
\newblock {Heat, air and moisture transfer through hollow porous blocks}.
\newblock {\em Int. J. Heat Mass Transfer}, 52(9-10):2390--2398, apr 2009.

\bibitem{Driscoll2014}
T.~A. Driscoll, N.~Hale, and L.~N. Trefethen.
\newblock {Chebfun Guide}.
\newblock {\em Pafnuty Publications}, Oxford, 2014.

\bibitem{DuFort1953}
E.~C. DuFort and S.~P. Frankel.
\newblock {Stability Conditions in the Numerical Treatment of Parabolic
  Differential Equations}.
\newblock {\em Mathematical Tables and Other Aids to Computation}, 7(43):135,
  jul 1953.

\bibitem{Gasparin2018}
S.~Gasparin, J.~Berger, D.~Dutykh, and N.~Mendes.
\newblock {An adaptive simulation of nonlinear heat and moisture transfer as a
  boundary value problem}.
\newblock {\em International Journal of Thermal Sciences}, 133:120--139, nov
  2018.

\bibitem{Gasparin2017b}
S.~Gasparin, J.~Berger, D.~Dutykh, and N.~Mendes.
\newblock {An improved explicit scheme for whole-building hygrothermal
  simulation}.
\newblock {\em Building Simulation}, 11(3):465--481, jun 2018.

\bibitem{Gasparin2017}
S.~Gasparin, J.~Berger, D.~Dutykh, and N.~Mendes.
\newblock {Stable explicit schemes for simulation of nonlinear moisture
  transfer in porous materials}.
\newblock {\em J. Building Perf. Simul.}, 11(2):129--144, 2018.

\bibitem{Gosse2013}
L.~Gosse.
\newblock {\em {Computing Qualitatively Correct Approximations of Balance Laws:
  Exponential-Fit, Well-Balanced and Asymptotic-Preserving}}, volume~2 of {\em
  SIMAI Springer Series}.
\newblock Springer Milan, Milano, 1 edition, 2013.

\bibitem{Gosse2017}
L.~Gosse.
\newblock {Viscous equations treated with L-splines and Steklov-Poincare
  operator in two dimensions}.
\newblock In L.~Gosse and R.~Natalini, editors, {\em Innovative Algorithms and
  Analysis}. Springer International Publishing, Milano, 2017.

\bibitem{Gosse2016}
L.~Gosse.
\newblock {L-Splines and Viscosity Limits for Well-Balanced Schemes Acting on
  Linear Parabolic Equations}.
\newblock {\em Acta Applicandae Mathematicae}, 153(1):101--124, feb 2018.

\bibitem{Guimaraes2018}
A.~S. Guimaraes, J.~M. P.~Q. Delgado, A.~C. Azevedo, and V.~P. de~Freitas.
\newblock {Interface influence on moisture transport in buildings}.
\newblock {\em Construction and Building Materials}, 162:480--488, feb 2018.

\bibitem{Hagentoft2004}
C.-E. Hagentoft, A.~S. Kalagasidis, B.~Adl-Zarrabi, S.~Roels, J.~Carmeliet,
  H.~Hens, J.~Grunewald, M.~Funk, R.~Becker, D.~Shamir, O.~Adan, H.~Brocken,
  K.~Kumaran, and R.~Djebbar.
\newblock {Assessment Method of Numerical Prediction Models for Combined Heat,
  Air and Moisture Transfer in Building Components: Benchmarks for
  One-dimensional Cases}.
\newblock {\em J. Building Phys.}, 27(4):327--352, apr 2004.

\bibitem{Hill1985}
P.~D. Hill.
\newblock {Kernel estimation of a distribution function}.
\newblock {\em Communications in Statistics - Theory and Methods},
  14(3):605--620, jan 1985.

\bibitem{Hindmarsh2005}
A.~C. Hindmarsh, P.~N. Brown, K.~E. Grant, S.~L. Lee, R.~Serban, D.~E.
  Shumaker, and C.~S. Woodward.
\newblock {SUNDIALS}.
\newblock {\em ACM Transactions on Mathematical Software}, 31(3):363--396, sep
  2005.

\bibitem{Jackiewicz1991}
Z.~Jackiewicz, R.~Renaut, and A.~Feldstein.
\newblock {Two-Step Runge-Kutta Methods}.
\newblock {\em SIAM J. Num. Anal.}, 28(4):1165--1182, aug 1991.

\bibitem{Jensen1989}
S.~O. Jensen.
\newblock {The PASSYS Project Phase 1: Subgroup Model Validation and
  Development : Final Report 1986-1989}.
\newblock Technical report, Commission of the European Communities:
  Directorate-General for Science, Research, and Development, Brussels,
  Belgium, 1989.

\bibitem{Kalamees2010}
T.~Kalamees and J.~Kurnitski.
\newblock {Moisture Convection Performance of External Walls and Roofs}.
\newblock {\em Journal of Building Physics}, 33(3):225--247, jan 2010.

\bibitem{Khanduri1998}
A.~C. Khanduri, T.~Stathopoulos, and C.~B{\'{e}}dard.
\newblock {Wind-induced interference effects on buildings - a review of the
  state-of-the-art}.
\newblock {\em Engineering Structures}, 20(7):617--630, jul 1998.

\bibitem{Kreiss1992}
H.~O. Kreiss and G.~Scherer.
\newblock {Method of lines for hyperbolic equations}.
\newblock {\em SIAM Journal on Numerical Analysis}, 29:640--646, 1992.

\bibitem{Kuenzel1995}
H.~M. Kuenzel and J.~Kiessl.
\newblock {\em {Simultaneous heat and moisture transport in building
  components: one- and two-dimensional calculation using simple parameters}}.
\newblock IRB Verlag, Stuttgart, 1995.

\bibitem{Langmans2012}
J.~Langmans, A.~Nicolai, R.~Klein, and S.~Roels.
\newblock {A quasi-steady state implementation of air convection in a transient
  heat and moisture building component model}.
\newblock {\em Building and Environment}, 58:208--218, dec 2012.

\bibitem{Luikov1966}
A.~V. Luikov.
\newblock {\em {Heat and mass transfer in capillary-porous bodies}}.
\newblock Pergamon Press, New York, 1966.

\bibitem{Mendes2017}
N.~Mendes, M.~Chhay, J.~Berger, and D.~Dutykh.
\newblock {\em {Numerical methods for diffusion phenomena in building
  physics}}.
\newblock PUCPRess, Curitiba, Parana, 1 edition, 2017.

\bibitem{Mnasri2017}
F.~Mnasri, K.~Abahri, G.~El, R.~Bennacer, and S.~Gabsi.
\newblock {Numerical analysis of heat, air, and moisture transfers in a wooden
  building material}.
\newblock {\em Thermal Science}, 21(2):785--795, 2017.

\bibitem{Rafidiarison2015}
H.~Rafidiarison, R.~R{\'{e}}mond, and E.~Mougel.
\newblock {Dataset for validating 1-D heat and mass transfer models within
  building walls with hygroscopic materials}.
\newblock {\em Building and Environment}, 89:356--368, jul 2015.

\bibitem{Rouchier2017}
S.~Rouchier, T.~Busser, M.~Pailha, A.~Piot, and M.~Woloszyn.
\newblock {Hygric characterization of wood fiber insulation under uncertainty
  with dynamic measurements and Markov Chain Monte-Carlo algorithm}.
\newblock {\em Building and Environment}, 114:129--139, mar 2017.

\bibitem{Scharfetter1969}
D.~L. Scharfetter and H.~K. Gummel.
\newblock {Large-signal analysis of a silicon Read diode oscillator}.
\newblock {\em IEEE Transactions on Electron Devices}, 16(1):64--77, jan 1969.

\bibitem{Sensirion2019}
Sensirion.
\newblock {Digital Humidity Sensor SHT7x (RH/T)}, 2019.

\bibitem{Soderlind2006a}
G.~S{\"{o}}derlind and L.~Wang.
\newblock {Evaluating numerical ODE/DAE methods, algorithms and software}.
\newblock {\em J. Comp. Appl. Math.}, 185(2):244--260, jan 2006.

\bibitem{Soudani2016}
L.~Soudani, A.~Fabbri, J.-C. Morel, M.~Woloszyn, P.-A. Chabriac, H.~Wong, and
  A.-C. Grillet.
\newblock {Assessment of the validity of some common assumptions in
  hygrothermal modeling of earth based materials}.
\newblock {\em Energy and Buildings}, 116:498--511, mar 2016.

\bibitem{Tariku2010}
F.~Tariku, K.~Kumaran, and P.~Fazio.
\newblock {Transient model for coupled heat, air and moisture transfer through
  multilayered porous media}.
\newblock {\em Int. J. Heat Mass Transfer}, 53(15-16):3035--3044, jul 2010.

\bibitem{Taylor1970}
P.~J. Taylor.
\newblock {The stability of the Du Fort-Frankel method for the diffusion
  equation with boundary conditions involving space derivatives}.
\newblock {\em The Computer Journal}, 13(1):92--97, jan 1970.

\bibitem{Vololonirina2014}
O.~Vololonirina, M.~Coutand, and B.~Perrin.
\newblock {Characterization of hygrothermal properties of wood-based products -
  Impact of moisture content and temperature}.
\newblock {\em Construction and Building Materials}, 63:223--233, jul 2014.

\bibitem{Wang2017}
L.~Wang and H.~Ge.
\newblock {Effect of air leakage on the hygrothermal performance of highly
  insulated wood frame walls: Comparison of air leakage modelling methods}.
\newblock {\em Building and Environment}, 123:363--377, oct 2017.

\bibitem{Whitaker1986}
S.~Whitaker.
\newblock {Flow in porous media I: A theoretical derivation of Darcy's law}.
\newblock {\em Transport in Porous Media}, 1(1):3--25, 1986.

\bibitem{Whitaker1986a}
S.~Whitaker.
\newblock {Flow in porous media II: The governing equations for immiscible,
  two-phase flow}.
\newblock {\em Transport in Porous Media}, 1(2):105--125, 1986.

\bibitem{Woloszyn2014}
M.~Woloszyn, N.~{Le Pierr{\`{e}}s}, Y.~Kedowid{\'{e}}, J.~Virgone, A.~Trabelsi,
  Z.~Slimani, E.~Mougel, P.~Reymond, H.~Rafidiarison, P.~Perr{\'{e}},
  F.~Pierre, R.~Belarbi, N.~Issaadi, K.~Abahri, T.~Bejat, A.~Piot, E.~Wurtz,
  T.~Duforestel, M.~Colmet-Daage, B.~Perrin, M.~Coutand, O.~Vololonirina, W.~W.
  Jomaa, J.-S. Lauffer, P.~Thiriet, R.~Diss, N.~R{\'{e}}mond, and O.~Legrand.
\newblock {Vers une m{\'{e}}thode de conception HYGRO-thermique des BATiments
  performants : d{\'{e}}marche du projet HYGRO-BAT}.
\newblock In {\em IBPSA France}, page~8, 2014.

\end{thebibliography}
\end{document}